\documentclass[a4paper,11pt]{article}
\usepackage{subfiles}
\pdfoutput=1 

\usepackage{amsmath,amsfonts,amssymb,graphicx,color}
\usepackage{longtable,booktabs}
\usepackage{placeins}

\AtBeginDocument{
\heavyrulewidth=.08em
\lightrulewidth=.05em
\cmidrulewidth=.03em
\belowrulesep=.65ex
\belowbottomsep=0pt
\aboverulesep=.4ex
\abovetopsep=0pt
\cmidrulesep=\doublerulesep
\cmidrulekern=.5em
\defaultaddspace=.5em
}

\usepackage{geometry}
\usepackage{setspace}
\usepackage{changepage}  
\usepackage{xspace}

\usepackage{jcappub} 
                     
\usepackage{lineno}   

\usepackage{orcidlink}
\usepackage[T1]{fontenc} 
                     
\usepackage{multirow}      
\usepackage[bf]{caption}

\usepackage{subcaption}

\usepackage{blkarray, bigstrut}
\usepackage{xparse}

\usepackage{lipsum}  
\usepackage{graphbox}   
\usepackage{stackengine}
\usepackage{tabularx}
\usepackage{array}
                     
\usepackage{graphicx}
\usepackage{dcolumn}
\usepackage{hyperref}
\hypersetup{colorlinks}
\usepackage[T1]{fontenc} 
\usepackage[normalem]{ulem}
\usepackage[utf8]{inputenc}

\usepackage[nameinlink,noabbrev]{cleveref}
\crefname{section}{Section}{Sections}
\crefname{table}{Table}{Tables}
\crefname{appendix}{Appendix}{Appendices}
\Crefname{figure}{Figure}{Figures}
\Crefname{equation}{Equation}{Equations}
\Crefname{section}{Section}{Sections}
\Crefname{table}{Table}{Tables}

\usepackage{booktabs}
\usepackage{float} 

\usepackage{soul}

\usepackage{bm}
\let\vec\bm

\newcommand{\code}[1]{\textsf{#1}}

\newcommand{\Dk}[1]{\frac{d^3#1}{(2\pi)^3}}

\newcommand{\ve}[1]{{\text{\bf #1}}}

\newcommand{\vk}{\vec k}
\newcommand{\vp}{\vec p}

\newcommand{\vx}{\vec x}

\newcommand{\vu}{\vec u}
\newcommand{\vhn}{\hat{\vec n}}

\newcommand{\ikk}{\underset{\vk_{12}= \vk}{\int}}

\newcommand{\dD}{\delta_\text{D}}

\newcommand{\folps}{\textsc{Folps}}
\newcommand{\folpsrepo}{\url{https://github.com/cosmodesi/FolpsD}}

\newcommand{\ihMpc}{\,h\text{Mpc}^{-1}}

\usepackage{float} 

\usepackage{booktabs}

\newcommand{\revised}[1]{#1}

\title{FolpsD: combining EFT and phenomenological approaches for joint power spectrum and bispectrum analyses}

\emailAdd{prakharb@umich.edu, aviles@icf.unam.mx}
\affiliation{Affiliations are in Appendix \ref{sec:affiliations}}

\author[1,2]{{P.~Bansal}\orcidlink{0009-0000-7309-4341},}
\author[3,4]{{A.~Aviles}\orcidlink{0000-0001-5998-3986},}
\author[3]{{H.~E.~Noriega}\orcidlink{0000-0002-3397-3998},}
\author[5]{{C.~Guandalin},}
\author[3]{{I.~Garzon},}
\author[6,4]{{G.~Niz}\orcidlink{0000-0002-1544-8946},}
\author[5]{{M.~S.~Wang}\orcidlink{0000-0002-2652-4043},}
\author[1,2]{{U.~Andrade}\orcidlink{0000-0002-4118-8236},}
\author[5]{{F.~Beutler}\orcidlink{0000-0003-0467-5438},}
\author[7]{{A.~de~Mattia}\orcidlink{0000-0003-0920-2947},}
\author[6,3]{{D.~Gonzalez}\orcidlink{0009-0009-6485-640X},}
\author[8,9,10]{{J.~Hou}\orcidlink{0000-0001-6083-1947},}
\author[11,2]{{D.~Huterer}\orcidlink{0000-0001-6558-0112},}
\author[12,13]{{E.~Paillas}\orcidlink{0000-0002-4637-2868},}
\author[5]{{M. Pellejero Ibanez}\orcidlink{0000-0003-4680-7275},}
\author[14]{{J.~Aguilar},}
\author[15]{{S.~Ahlen}\orcidlink{0000-0001-6098-7247},}
\author[16,17]{{D.~Bianchi}\orcidlink{0000-0001-9712-0006},}
\author[18]{{D.~Brooks},}
\author[14]{{T.~Claybaugh},}
\author[14]{{A.~Cuceu}\orcidlink{0000-0002-2169-0595},}
\author[19]{{A.~de la Macorra}\orcidlink{0000-0002-1769-1640},}
\author[20,21]{{B.~Dey}\orcidlink{0000-0002-5665-7912},}
\author[18]{{P.~Doel},}
\author[14,22]{{S.~Ferraro}\orcidlink{0000-0003-4992-7854},}
\author[23,24]{{A.~Font-Ribera}\orcidlink{0000-0002-3033-7312},}
\author[25,26]{{J.~E.~Forero-Romero}\orcidlink{0000-0002-2890-3725},}
\author[27,28,29]{{E.~Gaztañaga}\orcidlink{0000-0001-9632-0815},}
\author[30]{{S.~Gontcho A Gontcho}\orcidlink{0000-0003-3142-233X},}
\author[31]{{G.~Gutierrez},}
\author[32]{{C.~Hahn}\orcidlink{0000-0003-1197-0902},}
\author[33,7]{{H.~K.~Herrera-Alcantar}\orcidlink{0000-0002-9136-9609},}
\author[34,35,36]{{K.~Honscheid}\orcidlink{0000-0002-6550-2023},}
\author[37]{{C.~Howlett}\orcidlink{0000-0002-1081-9410},}
\author[38]{{M.~Ishak}\orcidlink{0000-0002-6024-466X},}
\author[39]{{R.~Joyce}\orcidlink{0000-0003-0201-5241},}
\author[39]{{S.~Juneau}\orcidlink{0000-0002-0000-2394},}
\author[40]{{D.~Kirkby}\orcidlink{0000-0002-8828-5463},}
\author[14]{{A.~Kremin}\orcidlink{0000-0001-6356-7424},}
\author[14]{{M.~Landriau}\orcidlink{0000-0003-1838-8528},}
\author[41]{{L.~Le~Guillou}\orcidlink{0000-0001-7178-8868},}
\author[42,24]{{M.~Manera}\orcidlink{0000-0003-4962-8934},}
\author[39]{{A.~Meisner}\orcidlink{0000-0002-1125-7384},}
\author[23,24]{{R.~Miquel},}
\author[28]{{S.~Nadathur}\orcidlink{0000-0001-9070-3102},}
\author[43,44,45]{{W.~J.~Percival}\orcidlink{0000-0002-0644-5727},}
\author[46]{{F.~Prada}\orcidlink{0000-0001-7145-8674},}
\author[47]{{I.~P\'erez-R\`afols}\orcidlink{0000-0001-6979-0125},}
\author[48]{{G.~Rossi},}
\author[49,50,51]{{L.~Samushia}\orcidlink{0000-0002-1609-5687},}
\author[52]{{E.~Sanchez}\orcidlink{0000-0002-9646-8198},}
\author[14]{{D.~Schlegel},}
\author[11,2]{{M.~Schubnell},}
\author[53]{{H.~Seo}\orcidlink{0000-0002-6588-3508},}
\author[14]{{J.~Silber}\orcidlink{0000-0002-3461-0320},}
\author[2]{{G.~Tarl\'{e}}\orcidlink{0000-0003-1704-0781},}
\author[39]{{B.~A.~Weaver},}
\author[41]{{P.~Zarrouk}\orcidlink{0000-0002-7305-9578},}
\author[14]{{R.~Zhou}\orcidlink{0000-0001-5381-4372},}
\author{(DESI collaboration)}

\abstract{
We present a theoretical model for the power spectrum and bispectrum of galaxy clustering that exploits the complementarity between small-scale power spectrum information and large-scale bispectrum measurements. We extend the \folps\ code by combining its one-loop EFT galaxy power spectrum with a tree-level galaxy bispectrum projected onto the tripolar spherical harmonics (Sugiyama) basis. To access additional small-scale information, we also consider a line-of-sight damping factor in both statistics, mirroring approaches commonly used in studies of redshift-space distortions. We test the model using DESI DR2 galaxy mocks. Even without damping, the joint analysis of the EFT power spectrum and bispectrum significantly improves constraints and reduces parameter degeneracies relative to power spectrum analyses alone. For LRG-like samples, including the damping further extends the range beyond $k\sim0.3 \ihMpc$ in the power spectrum and $k \sim 0.24 \ihMpc$ in the bispectrum without introducing statistically significant parameter biases. This leads to up to $\sim30\%$ tighter constraints on $A_s$ and $\omega_{cdm}$. For low signal-to-noise tracers such as QSOs, however, the damping parameters are weakly constrained and can absorb noise fluctuations, leading to shifts in inferred parameters. Similar limitations may arise in models where cosmological information is encoded in power-spectrum shape features degenerate with the damping, such as scenarios with massive neutrinos. In contrast, for $w_0w_a$CDM we obtain $15\%$ and $21\%$ tighter constraints on $w_0$ and $w_a$, respectively, yielding a deviation from constant dark energy at slightly more than the $1\sigma$ level using full-shape information alone. The code is publicly available at \folpsrepo.
}

\begin{document} 
 \maketitle
\flushbottom

\section{Introduction}\label{sec:intro}

Extracting cosmological information from galaxy clustering data relies on accurate theoretical modeling across a wide range of scales. Full-shape analyses aim to exploit not only baryon acoustic oscillations (BAO), but also the broadband shape of clustering observables, which is sensitive to both cosmology and galaxy bias. In practice, however, the range of scales that can be reliably modeled is limited by nonlinear gravitational evolution and redshift-space effects. Extending the reach of full-shape analyses in a controlled way, while maintaining robust parameter inference, is therefore a central challenge for current and future galaxy surveys.

On sufficiently large scales, the evolution of matter fluctuations can be described using perturbation theory (PT), which provides a systematic expansion in powers of the initial density field \cite{Bernardeau:2001qr}. In modern analyses, this perturbative framework is formulated within the effective field theory (EFT) of large-scale structure
\cite{McDonald:2006mx,McDonald:2009dh,Baumann:2010tm,Carrasco:2012cv,Pajer:2013jj,Assassi:2014fva,Mirbabayi:2014zca,Vlah:2015zda,Aviles:2018thp,Ivanov:2019pdj,DAmico:2019fhj,Chen:2020zjt,Aviles:2020cax,Blas:2015qsi}. In this framework, the impact of unresolved small-scale physics is parameterized through a finite set of counterterms, allowing for a controlled, systematic, and hence improvable description of clustering observables; see, e.g., \cite{Desjacques:2018pfv,Ivanov:2022mrd} for reviews.

In current full-shape analyses, the redshift-space galaxy power spectrum has been the primary observable used to extract cosmological constraints. The power spectrum provides strong sensitivity through its broadband shape, BAO features, and redshift-space distortions (RSD). Within the EFT framework, power spectrum analyses are typically restricted to mildly nonlinear scales, with conservative choices of the maximum wavenumber to avoid modeling systematics. Several studies have explored ways to extend this scale range, in particular by adopting phenomenological descriptions of redshift-space effects \cite{Fisher:1994ks,2011MNRAS.417.1913R,2020MNRAS.492.4189I,Ramirez:2023ads}. In this context, damping models motivated by small-scale velocity dispersion have been shown to capture the suppression of power on small scales and to improve parameter constraints \cite{Taruya:2010mx,BOSS:2016off,Eggemeier:2022anw}. While such approaches go beyond the strict Wilsonian interpretation of EFT, they can provide useful descriptions of galaxy clustering statistics.

An alternative and complementary way to extend the information content of full-shape analyses is to include higher-order statistics, such as the galaxy bispectrum \cite{Scoccimarro:1999ed,Sefusatti:2006pa,Verde:1998zr,Scoccimarro:2000sn,Gil-Marin:2014sta,Ruggeri:2017dda,deBelsunce:2018xtd,Yankelevich:2018uaz,Floerchinger:2019eoj,Bharadwaj:2020wkc,Kamalinejad:2020izi,Eggemeier:2021cam,Ivanov:2021kcd,Philcox:2022frc,DAmico:2022ukl,Chen:2024pyp,Pal:2025hpl,Bakx:2025pop}. The bispectrum probes mode coupling induced by nonlinear gravitational evolution and galaxy bias, and is particularly sensitive to second-order bias parameters that are only weakly constrained by the power spectrum alone \cite{Scoccimarro:1999ed,Sefusatti:2006pa}. In the absence of primordial non-Gaussianity (PNG), the bispectrum is intrinsically nonlinear, and its measurement can therefore help break degeneracies between biases and cosmological parameters and, ultimately, improve overall constraints. Several works have demonstrated the potential of bispectrum analyses using BOSS and eBOSS galaxies \cite{Gil-Marin:2016wya,Philcox:2021kcw,DAmico:2022osl,Lu:2025gki}, as well as more recently with data from the Dark Energy Spectroscopic Instrument (DESI) \cite{NovellMasot:2025fju,Chudaykin:2025aux,Chudaykin:2025lww,Chudaykin:2025vdh}.

The present work is motivated by the prospect of applying full-shape analyses to the DESI Data Release 2 (DR2). During its first three years of operation, from May 2021 to April 2024, DESI collected 13.5 million galaxy and quasar redshifts, along with 1.3 million Lyman-$\alpha$ spectra. DESI, a robotic, fiber-fed, highly multiplexed spectroscopic survey instrument mounted on the Mayall 4-meter telescope at Kitt Peak National Observatory \cite{DESI2022.KP1.Instr}, has been designed to cover approximately $17{,}000$ square degrees over eight years, with a target of 63 million spectroscopically confirmed galaxies and quasars \cite{SurveyOps.Schlafly.2023,Spectro.Pipeline.Guy.2023}. The DESI Data Release 1 (DR1; \cite{DESI2024.I.DR1}), which includes spectra for more than 18 million unique targets, is now publicly available. Early DESI results have already provided new constraints on dark energy \cite{DESI2024.VII.KP7B,DESI.DR2.DR2,Y3.cpe-s1.Lodha.2025,DESI:2025gwf}, modified gravity \cite{DESI2024.VII.KP7B,KP7s1-MG}, neutrino physics \cite{DESI.DR2.DR2,Y3.cpe-s2.Elbers.2025}, and PNG \cite{Chaussidon:2024qni}. In particular, the DR2 release has delivered precise BAO measurements across a wide redshift range, demonstrating that the survey volume and statistical power now justify moving beyond standard two-point analyses.

In this context, we build upon the existing \folps\ code\footnote{\folpsrepo} \cite{Noriega:2022nhf,Aviles:2021que} (Fast Optimized Large-scale structure Perturbation Solver) and develop \texttt{FolpsD}, an EFT-inspired phenomenological extension that incorporates a line-of-sight (LoS) damping factor in the modeling of redshift-space clustering. In the spirit of the Taruya--Nishimichi--Saito (TNS) model \cite{Taruya:2010mx}, this damping allows the galaxy power spectrum to be modeled at higher wavenumbers than typically accessible with the standard one-loop EFT prescription. In addition, we incorporate a tree-level galaxy bispectrum model projected onto the tripolar spherical harmonics basis (hereafter the Sugiyama basis; \cite{Sugiyama:2018yzo}), which can be analyzed both with and without damping factors and provides direct sensitivity to second-order galaxy bias parameters.

In summary, we study the joint use of the galaxy power spectrum and bispectrum for full-shape analyses in the context of DESI. Our goal is to assess how information from different scales and statistics can be combined in a controlled way, and to quantify the complementarity between extending the power spectrum to smaller scales using phenomenological damping models and including bispectrum measurements that constrain galaxy bias on larger scales. We validate this combined modeling using second-generation Abacus mock catalogs \cite{abacus2021}, study its robustness as a function of scale cuts, and quantify its impact on cosmological and bias parameter constraints. Finally, we apply the power-spectrum component of our modeling to DESI DR1 data and discuss the implications for future DESI full-shape analyses.

The paper is organized as follows. In \cref{sec:model} we summarize the theoretical framework for the power spectrum and bispectrum modeling. \Cref{sec:mocks} describes our settings. In \cref{sec:mockresults} we validate the methodology using Abacus mock catalogs, and in \cref{sec:bkmax} we discuss the correlation between different multipoles and its implications for the maximum scale used in the bispectrum analysis. Results from the DESI DR1 analysis are presented in \cref{sec:DR1} and \cref{sec:w0wa}, in the former we assume a $\Lambda$CDM while in the latter, by using the ShapeFit method, we present constraints on the $w_0w_a$CDM model. We conclude in \cref{sec:conclusions}. Additional technical details and complementary analyses are provided in the appendices, including a discussion of neutrino mass constraints in \cref{app:neutrinos}. 

\section{Modeling the power spectrum and bispectrum} \label{sec:model}

\folps\ \cite{Noriega:2022nhf} is an EFT code that computes the one-loop power spectrum for biased tracers, such as galaxies and quasars, 
given a set of cosmological and nuisance parameters. It has the capability to use kernels beyond the Einstein-de Sitter (EdS) approximation, allowing it to handle new scales introduced by massive neutrinos or certain modified gravity theories. The theoretical framework used by \folps\ is described in detail in \cite{Aviles:2020cax,Aviles:2021que,Noriega:2022nhf}; see also \cite{Aviles:2020wme,Rodriguez-Meza:2023rga,KP5s3-Noriega}.

The EFT galaxy power spectrum is computed as 
\begin{align} \label{eq:Pg}
    P_\text{EFT}(k,\mu) &= P_K(k,\mu) + P_\text{1-loop}(k,\mu) + P_\text{ctr}^\text{LO}(k,\mu)   + P_\text{shot}(k,\mu)\; ,  
\end{align}
with $P_K(k,\mu)= (b_1 + f \mu^2)^2 P_L(k)$ the Kaiser power spectrum \cite{Kaiser:1987qv}, $P_L(k)$ the linear power spectrum, $b_1$ the Eulerian linear bias, $\mu = \hat{\vk} \cdot \vhn $ the angle cosine between the wave vector $\vk$ and the LoS direction $\vhn$, and the growth factor is the logarithmic derivative of the growth function $D_+$,
\begin{equation}
 f(t)   = \frac{{\rm d} \log D_+(t)}{{\rm d} \log a(t)}\; .
\end{equation}
In the models where the linear growth function equation is scale-free, as in $\Lambda$CDM and generally dark energy models,  $D_+$ and hence $f$ are only time-dependent functions. In what follows, we assume this to be the case. However, one of the \folps\ framework features is that the introduction of additional scales, such as the neutrino mass, makes the growth function and consequently $f$ scale dependent. We further discuss this in \cref{app:Zkernels}.

The one-loop contribution in PT is given by
\begin{equation}
P_\text{1-loop}(k,\mu) = \sum_{m=0}^{4} \sum_{n=0}^{m} f^{m}\mu^{2n} I_{nm}(k), 
\end{equation}
where $I_{mn}$ are functions of the wavenumber only, computed from the \textit{loop} integrals via Fast Fourier Transform methods \cite{Hamilton:1999uv,Simonovic:2017mhp}. The complete list of $I_{mn}$ integrals can be found in \cite{Aviles:2020wme,Noriega:2022nhf}. 

The leading-order (LO) EFT counterterms are given by 
\begin{equation} \label{PLOctr}
P_\text{ctr}^\text{LO}(k,\mu) = (\alpha_0 + \alpha_2 \mu^2 + \alpha_4 \mu^4 +\alpha_6 \mu^6) k^2 P_L(k),
\end{equation}
while the shot noise is modeled as
\begin{equation} \label{Psn}
P_\text{shot} = \frac{1}{\bar{n}}(\alpha_0^\text{shot} +\alpha_{01}^\text{shot} k^2 + \alpha_2^\text{shot} k^2 \mu^2 ),
\end{equation}
with $\bar{n}$ the number density of tracers. Hence, the parameters $\alpha^\text{shot}$ quantify the deviation of a Poisson noise.  On top of \cref{eq:Pg}, we apply infrared resummations using the method of \cite{Ivanov:2018gjr}, as explained in detail in \cite{Noriega:2022nhf,KP5s3-Noriega}.

As described in the rest of this section, the \folps\ code includes two new key ingredients: damping factors and the bispectrum. In addition, new features such as the treatment of different galaxy bias schemes have been implemented.

\subsection{\texttt{FolpsD}: Line-of-sight damping factor}\label{sec:damping}

In redshift space, there are two characteristic scales that control different expansions of the analytical description of the galaxy distribution \cite{Kaiser:1987qv,Scoccimarro:2004tg}. There is a shorter scale, $k_\text{NL}$, at which nonlinearities in the matter overdensity field become comparable to the linear contribution, and hence the loop-corrections hierarchy in the power spectrum disappears. This scale signs the breaking of the perturbative description and it is around $k_\text{NL}\sim 0.3 \, h\text{Mpc}^{-1}$ for tracers at $z \sim 0.5$. In addition to the PT of fields, one introduces a second expansion to handle the nonlinear map between real and redshift spaces. Such expansion is controlled by a second scale, $k_s$, relevant along the LoS. In particular, it controls the power spectrum suppression due to RSD, leading to the so-called Finger-of-God (FoG) effect. This additional scale is typically smaller than $0.2 \, h\text{Mpc}^{-1}$ for the same tracers as above, hence the description based on \cref{eq:Pg}  actually stops being accurate at $k_\text{s}<k_\text{NL}$. In contrast, in real space one expects the EFT description to break down at $k_\text{NL}$.

Therefore, in order to describe the galaxy field on scales $k_s < k < k_\text{NL}$, one needs to go beyond the prescription used in standard full-shape analyses, e.g., beyond those based on the EFT description of \cref{eq:Pg}. To this end, non-perturbative effects over the LoS may be included: one could try to write down a series on powers of $k_\parallel^2$, with $k_\parallel \equiv \vec{k}\cdot \vhn =k\mu$, by adding next-to-leading order (NLO) counterterms to the LO one in \cref{eq:Pg}, namely
\begin{equation} \label{eq:NLOctr}
P_\text{ctr}^\text{LO} \rightarrow P_\text{ctr}^\text{NLO}  =    P_\text{ctr}^\text{LO} + \sum_n  c_n (k \mu)^{2n} P_K(k,\mu)\;.
\end{equation}
The $c_{n}$ coefficients are formally $\mathcal{O}(P_L^{\,n})$ quantities which, in principle, can be constructed from the correlations of $2n$ small-scale density and velocity fields that have been cut-off from the loop integrals. \revised{Consequently, terms such as $c_n (k\mu)^{2n} P_K$ contribute at the same perturbative order as other $n$-loop corrections.}

Instead of keeping the series in \cref{eq:NLOctr}, the collection of counterterms can be resummed into a single function, $(\mathcal{D}[(k\, \mu)^{2}] - 1) P_K(k,\mu)$. In doing so, we arrive at
\begin{align} \label{eq:PgD}
    P(k,\mu) &=  \mathcal{D}\big(k^{2} \mu^2 \big)  \left[ P_K(k,\mu) + P_\text{1-loop}(k,\mu) \right] 
    + P_\text{ctr}^\text{LO}(k,\mu) + P_\text{shot}(k,\mu)\; , 
\end{align}
which is an expression we use extensively throughout this work. From now on, we will refer to this model as  \texttt{FolpsD}. Since the NLO counterterms are of higher perturbative order, the power spectra of \cref{eq:PgD,eq:Pg} are formally equal up to one-loop.\footnote{Despite cross terms of the form $c_n P_\text{1-loop}$ would sctrictly appear, these are $\mathcal{O}(P_L^{\,n+2})$ contributions.}

In this work, the damping factor is chosen phenomenologically to have a Lorentzian form \cite{1976Ap&SS..45....3P,1994ApJ...431..569P},
\begin{equation}
\mathcal{D}(x^2) = \frac{1}{1+x^2}\; ,
\end{equation}
with 
\begin{equation}
    x = X_{p}f \sigma_v k \mu\; ,
\end{equation}
where $\sigma_v^2 = \int_0^\infty dk\, P_L(k)/(6\pi^2)$ is the velocity dispersion that induces random displacements of typical size $\Delta s \sim \sigma_v$, leading to an apparent suppression of structure along the LoS. As a result, modes with $f k_\parallel \gtrsim 1/\sigma_v$, that are sensitive to short-scale physics, present strong FoG effect. Hence, perturbative control of the redshift-space expansion requires $k_s \ll 1/f\sigma_v$.  Consequently, the free parameter $X_{p}$ and the functional form of the damping factor control the NLO counterterms in powers of $k_\parallel^2/ (f \sigma_v)^{-2}$. Numerically, one finds $\sigma_v \sim 6\,D_+  (z)\,h^{-1}\text{Mpc}$.

Upon expanding the damping function $\mathcal{D}$, the linear coefficient can be absorbed by the LO counterterm, but beyond that, the terms are new to the theory.  We note that this expansion neglects terms proportional to $k^{2m} k_\parallel^{2n}$ with $m>0$, which can also be expected due to the short-range nonlocality induced by nonlinear processes in galaxy formation \cite{McDonald:2009dh}.  That is, we assume that the real-space galaxy power spectrum can be modeled with high accuracy using only the one-loop EFT up to some $k_\text{max}$, and the goal of the NLO counterterms in \cref{eq:NLOctr} is to achieve a similar level of accuracy for the redshift-space power spectrum. However, by resumming the expansion we effectively impose a specific functional relation among the different NLO counterterms, as dictated by the damping form, and describe all of them in terms of a single parameter $X_p$.\footnote{The combination $X_p \sigma_v$ therefore plays a role analogous to the FoG parameter $\sigma_\text{FoG}$ widely used in earlier literature, e.g. \cite{2012MNRAS.426.2719R}.} In this sense, the damping acts as a strong prior on the relative size and scale dependence of these contributions, reducing the freedom present in the standard EFT counterterm expansion. This effectively “Dirac-delta–like” prior imposed by the damping function introduces limitations in the inference of parameters, as we discuss later in \cref{subsec:limitations}. 

Some works use one term of the NLO expansion, e.g  $\tilde{c}\,(k\mu)^4 P_L(k)$,  with $\tilde{c}$ a borrowed counterterm from two-loop theory \cite{Ivanov:2019pdj,Chudaykin:2025aux}.  However, it is unclear whether this allows them to push to higher values of $k_\text{max}$ (see, e.g., \cite{2026arXiv260120826E}). One may expect that, once the one-loop EFT description breaks down, going to higher perturbative orders would not lead to substantial improvements, and that a non-perturbative prescription would ultimately be required to achieve them.\footnote{See, however, the recent work \cite{Chen:2026usz}, showing how $k \sim 1\ihMpc$ can be reached for the 2-loop matter real-space power spectrum at $z\sim 1$.} Furthermore, the NLO counterterm $\tilde{c}$ is highly degenerate with the shot-noise parameter $\alpha_2^{\text{shot}}$, which significantly reduces its independent constraining power within the \texttt{Folps} framework.

\subsubsection{Relation to the TNS and VDG models}

Damping prescriptions have been a recurring feature of RSD modeling \cite{1983ApJ...267..465D,1992MNRAS.258..581P,1994ApJ...431..569P,1995MNRAS.275..515C,1998MNRAS.296...10H,Scoccimarro:1999ed,Scoccimarro:2004tg}, even encompassing approaches such as Gaussian-Streaming models \cite{Fisher:1994ks,Vlah:2016bcl} and related frameworks. In particular, the velocity-difference generator (VDG) models \cite{BOSS:2016off,Eggemeier:2022anw,Eggemeier:2025xwi} has demonstrated that damping factors can extend the modeling of redshift-space clustering to smaller scales than those typically accessible with the standard one-loop EFT prescription \cite{Eggemeier:2022anw,BOSS:2016off,2026arXiv260120826E}. The formulation of \texttt{FolpsD} is partly inspired by this approach. VDG originates from the treatment of the real-to-redshift-space mapping in the TNS model \cite{Taruya:2010mx}, differing from it only in the functional form of the damping factor, and from the earlier work of reference \cite{Scoccimarro:2004tg}, where the statistics of pairwise galaxy velocities play a central role.

Rather than performing a full nonlinear expansion of the real-to-redshift-space transformation, the TNS framework isolates the dominant nonlinear contributions and encapsulates them into a single object: the generating function of galaxy pair velocity differences. This velocity difference generator effectively captures the impact of small-scale random motions, producing a suppression of power on quasi-linear and nonlinear scales, as expected from the FoG effect.
In TNS models, the velocity generator is substituted by a damping function, which historically has been chosen phenomenologically as exponential and Lorentzian.

In contrast to \texttt{FolpsD}, in the TNS prescription the correlators $\langle vv\rangle\langle\delta\delta\rangle$, $\langle vv\rangle\langle\delta v\rangle$ and
$\langle vv\rangle\langle vv\rangle$ are not included in the perturbative expansion, since the exponential term that generates them is replaced by the phenomenological damping function. These missing correlators lead to the terms collected in
\begin{equation} \label{defDeltaP}
\Delta P(k,\mu) \equiv C(k,\mu)-(k\mu f)^2 P_K(k,\mu) \in P_\text{1-loop}(k,\mu),
\end{equation}
where
\begin{equation}
C(k,\mu)=(k\mu f)^2 \int \frac{{\rm d}^3 p }{(2\pi)^3} \frac{(\vp\cdot \vhn)^2}{p^4} P_{\theta\theta}(p) P_K(|\vk-\vp|,\mu_{\vk-\vp}),
\end{equation}
with $\mu_{\vk-\vp}$ the cosine of the angle between the vector $\vk-\vp$ and the LoS direction $\vhn$.

However, the function $C(k,\mu)$  is necessary to preserve the large-scale constraint imposed by momentum conservation. In its absence, the mode coupling term $B(k,\mu)$ (defined in \cite{Taruya:2010mx}) generates an ultraviolet contribution to the power spectrum scaling as $P_\text{UV} \propto k^2$ in the limit $k \to 0$, as shown for example in Sect.~4.4 of Ref.~\cite{Aviles:2020wme}, whereas purely short-scale nonlinear dynamics must produce density fluctuations scaling as $P_\text{UV} \propto k^4$ as required by momentum conservation \cite{Mercolli:2013bsa,Blas:2014hya}.

The second term in \cref{defDeltaP}, $-(k\mu f)^2P_K(k,\mu)$, ensures that the coupling between long wavelength displacement fields and short scale modes has the Galilean invariant structure $k^2P_L(k)$ \cite{Peloso:2013zw,Kehagias:2013yd,Creminelli:2013mca}. In practice, the TNS and VDG models absorb these missing contributions into the phenomenological damping factor. As a consequence, the large scale limit of these models scales as $k^2$ instead of satisfying the constraints imposed by momentum conservation and Galilean invariance. Restoring the full $\Delta P$ term removes this ultraviolet sensitivity and guarantees the correct large scale behavior.

Operationally, TNS theories can be written as a minimal modification to EFT, \cref{eq:PgD}, given by
\begin{align} \label{eq:PgTNS}
P_\text{TNS}(k,\mu) &= \mathcal{D}(k_\parallel^{2}) \big[ P_K(k,\mu) + P_\text{1-loop}(k,\mu) - \Delta P (k,\mu) \big] + P_\text{ctr}^\text{LO}(k,\mu) + P_\text{shot}(k,\mu).
\end{align}

The main difference with \texttt{FolpsD} therefore arises because the contributions that generate the term $\Delta P$ are not included explicitly in the description and are instead replaced by a damping function. By borrowing NLO counterterms from higher loops and resumming them in a damping prefactor, \texttt{FolpsD} is also a damped model, but maintains the full structure imposed by the systematic treatment of all correlators at one-loop order and therefore complies with the constraints imposed by momentum conservation and Galilean invariance. 

While the largest deviations of one-loop TNS/VDG and EFT predictions appear at very large scales, where the ultraviolet cancellations become relevant, these effects are not completely negligible at intermediate scales. The incorrect  scaling implies that short modes influence large-scale fluctuations more strongly than allowed by the symmetry constraints discussed above. In practice, this leads to deviations from the EFT prediction at the level of a few percent in the range of scales relevant for galaxy surveys, which should be interpreted as a systematic error of the model, although they could be partially mitigated by the damping factor. It is beyond the scope of this work to quantify precisely the impact of this effect on the estimation of cosmological parameters.

On the other hand, the main similarity between our modeling and the original TNS model \cite{Taruya:2010mx} is that both use a Lorentzian damping factor. In the case of VDG, \cref{eq:PgTNS} is used as in TNS, but the damping factor is fixed to the asymptotic form of the velocity generator at infinite pairwise separation \cite{Eggemeier:2025xwi}.

\subsection{\texttt{FolpsD}: bispectrum}

Besides the power spectrum and its damping, the new \folps\ code implementation includes the tree-level bispectrum. We construct the galaxy bispectrum, $B(\vk_1,\vk_2,\vk_3)$, defined by
\begin{equation}
\langle \delta_g(\vk_1) \delta_g(\vk_2)\delta_g(\vk_3)\rangle   =  (2\pi)^3 \dD(\vk_1+\vk_2+\vk_3) B(\vk_1,\vk_2,\vk_3)\; ,
\end{equation}
with $\delta_g$ the galaxy density fluctuation,
using the PT prescription of \cite{Ivanov:2021kcd} as
\begin{align}\label{Bk}
B(\vk_1,\vk_2,\vk_3) =& 2 Z_2(\vk_1,\vk_2) Z_1^\text{FoG}(\vk_1)P^\text{IR-res}_L(\vk_1)    Z_1^\text{FoG}(\vk_2)  P^\text{IR-res}_L(\vk_2) \nonumber\\ &+ \text{cyclic permutations}\; , 
\end{align}
with the first-order kernel
\begin{equation}\label{eq:Z1}
    Z_1^\text{FoG}(\vk) = Z_1(\vk) - c_1 k^2\mu^2 - c_2 k^2\mu^4\; .
\end{equation}
The NLO counterterms $c_1$ and $c_2$ have been included to model FoG into $Z_1(\vk)=b_1+f\mu^2$.  The second-order kernel takes the form
 \begin{align}
 Z_2(\vk_1,\vk_2) =& \frac{b_2}{2} + b_{s} \left( \frac{\vk_1\cdot\vk_2}{k_1^2k_2^2} - \frac{1}{3}\right) + b_1 F_2(\vk_1,\vk_2) + f \mu_{12}^2 G_2(\vk_1,\vk_2)   \nonumber\\
 &+ \frac{k_{12} \mu_{12}}{2} f \left[ \frac{\mu_1}{k_1} \big(1+f \mu_2^2 \big)+ \frac{\mu_2}{k_2} \big(1+f \mu_1^2\big) \right], 
\end{align}
with $\mu_i$ the cosine of the angle between the wave vector $\vk_i$ and the LoS direction, $k_{12} \equiv |\vk_1 + \vk_2| $ and $k_{12}\mu_{12} = k_1 \mu_1 + k_2 \mu_2$. The parameters $b_2$ and $b_s$ are, respectively, the local and tidal second-order biases. In \cref{app:Zkernels}, we derive the $Z_{1,2}$ kernels for scale-dependent theories, as are treated within \folps.
Infrared resummations in the bispectrum are  directly included in the linear power spectra of \cref{Bk} as \cite{Ivanov:2018gjr,Ivanov:2021kcd}
\begin{align}
P_L(k) \quad \rightarrow \quad P^\text{IR-res}_L(k,\mu) \equiv P_{\rm nw}(k) + P_{\rm w}(k)\, {\rm e}^{-k^2 \Sigma^2_\text{tot}(k,\mu)}\; ,
\end{align}
where $P_{\rm nw}$ is the linear real-space power spectrum with the BAO wiggles removed, and $P_{\rm w}$ is the pure wiggle piece, such that $P_L(k)=P_{\rm nw}(k)+P_{\rm w}(k)$.  The function $\Sigma^2_\text{tot}$ is defined by
\begin{equation}\label{Sigma2T}
\Sigma^2_\text{tot}(k,\mu) = \big[1+f \mu^2 \big( 2 + f \big) \big]\Sigma^2 + f^2 \mu^2 (\mu^2-1) \delta\Sigma^2,    
\end{equation}
with
\begin{align}\label{Sigma2}
\Sigma^2 &= \frac{1}{6 \pi^2}\int_0^{k_\text{IR}} {\rm d}p \,P_{\rm nw}(p) \left[ 1 - j_0\left(p \,\ell_\text{BAO}\right) + 2 j_2 \left(p \,\ell_\text{BAO}\right)\right]\; , \\
\delta\Sigma^2 &= \frac{1}{2 \pi^2}\int_0^{k_\text{IR}} {\rm d}p \,P_{\rm nw}(p)  j_2 \left(p \,\ell_\text{BAO}\right)\; .
\label{deltaSigma2}
\end{align}
Here, $j_n$ is the spherical Bessel function of degree $n$, and $\ell_{\text{BAO}} = 105\,h^{-1}\,\text{Mpc}$ is approximately the BAO scale. The wavenumber $k_{\text{IR}}$ defines a transition scale between long- and short-wavelength modes; the final results depend only weakly on its value, as long as $k_{\text{IR}} \gtrsim 0.1  \ihMpc$.

In addition, we use the same damping factor prescription as discussed in section \ref{sec:damping} for the power spectrum but projected on each wave vector of the triangle configuration over the LoS. The result is given by substituting \cite{Scoccimarro:1999ed,Matarrese:1997sk}
\begin{equation}
  B(\vk_1,\vk_2,\vk_3) \rightarrow D_b\Big(X_b \sum_{i=1,2,3} (\sigma_v f\mu_i k_i)^2 \Big) B (\vk_1,\vk_2,\vk_3)
\end{equation}
in \cref{Bk}, where we choose the damping factor, $D_b$, to be Lorentzian as with the power spectrum, namely
\begin{align}
D_b &= \frac{1}{1+\Big((k_1\mu_1)^2 + (k_2\mu_2)^2 + (k_3\mu_3)^2 \Big) (f \sigma_v X_b)^2/2}\; ,
\end{align}
with $X_b$ a FoG free parameter in addition to $X_p$.

The final ingredient, the shot noise, is added through the parameters $P_\text{shot}$ and $B_\text{shot}$, given by
\begin{align} \label{eq:damping.bispectrum}
   B_\text{shot-noise}(\vk_1,\vk_2,\vk_3) =& B_\text{shot} \left(b_1 + 2\frac{P_\text{shot}}{B_\text{shot}}f\mu^2 \right) Z_1(\vk_1)  P^\text{IR-res}_L(\vk_1) \nonumber\\
   &+  (\vk_1\rightarrow \vk_2) + (\vk_1\rightarrow \vk_3) +  P_\text{shot}^2\; .
\end{align}
Poissonian noise corresponds to $P_\text{shot} = B_\text{shot} = 1/\bar{n}$, with $\bar{n}$ the number density of galaxies. Notice that, as in the case of the \texttt{FolpsD} power spectrum, the damping factor is applied before adding the shot noise; otherwise, the noise modeling would be suppressed. Finally, when we refer to the \texttt{EFT} bispectrum we use the same description outlined above but without considering the damping. 


\subsubsection{Coordinates} \label{app:coordinates}

In the presence of statistical homogeneity and isotropy, the bispectrum depends on three scalar variables, which can be chosen as the triangle sides $k_i$ ($i=1,2,3$). Alternatively, one can choose two sides of the triangle, ($k_1, k_2$), and its opening angle $\theta_{12}=\cos^{-1}(\hat{\vk}_1 \cdot \hat{\vk}_2)$. In practice, \folps\ works with the cosine of this angle, $x \equiv x_{12} = \cos \theta_{12}$. 
Redshift space partially breaks this isotropy by introducing two additional variables, which can be identified with the two angles needed to describe the relative orientation between a given triangle and the LoS direction, namely $(\omega, \phi)$.
We choose a coordinate system where the $z$-direction aligns with $\vk_1$, such that the wave vectors read
\begin{align}
    \vk_1 &= \left( 0,0,k_1\right), \\
    \vk_2 &= \left( -k_2 \sqrt{1-x^2}, 0, k_2 x \right), \\
    \vk_3 &=\left(k_2 \sqrt{1-x^2}, 0,-k_1 -k_2 x   \right), 
\end{align}
with corresponding opening angles 
\begin{align}
  x \equiv x_{12} &\equiv \hat{\vk}_1 \cdot \hat{\vk}_2 = \frac{k_3^2 -k_1^2 - k_2^2}{2 k_1 k_2}, \\
  x_{13} &\equiv \hat{\vk}_1 \cdot \hat{\vk}_3 = -\frac{k_1 + k_2 x}{k_3}, \\
  x_{23} &\equiv \hat{\vk}_2 \cdot \hat{\vk}_3 = - \frac{k_2 + k_1 x}{k_3},
\end{align}
where we have used that $\vk_3=-\vk_1-\vk_2$. 
Notice that the minus sign in the $\hat{x}$-axis entry of $\vk_2$ is due to the left-handedness of the oriented plane $\hat{z} \rightarrow \hat{x} $, which we choose to be the plane where the triangle $\vk_1 \rightarrow \vk_2 \rightarrow \vk_3 \rightarrow \vk_1 $ lives. 

The orientation of a triangle with respect to the LoS is characterized by the azimuthal angle $\phi$ about $\hat{z}=\hat{\vk}_1$, and by the angle $\omega$ between $\hat{z}$ and the LoS direction $\vhn$, or by its cosine 
\begin{equation}
     \mu = \hat{z} \cdot \vhn  = \cos \omega\; .
\end{equation}
In these coordinates, the LoS direction becomes 
\begin{align}
    \vhn &= \left( \cos \phi \sqrt{1-\mu^2},\, \sin \phi \sqrt{1-\mu^2}, \, \mu \right),
\end{align}
and as a result, the projected wave vectors over the LoS are given by
\begin{align}
    \mu_1 &= \hat{\vk}_1 \cdot \vhn = \mu ,\\
    \mu_2 &= \hat{\vk}_2 \cdot \vhn =  -\sqrt{1-\mu^2} \sqrt{1-x^2} \cos \phi   +   \mu x, \\
    \mu_3 &=\hat{\vk}_3 \cdot \vhn = - \frac{k_1}{k_3} \mu - \frac{k_2}{k_3} \mu_2.
\end{align}

One can do different decompositions of the bispectrum in the $(k_1,k_2,\theta_{12},\omega,\phi)$ coordinates which lead, among other choices, to the so-called Sugiyama and Scoccimarro bases, as we describe in what follows.

\subsubsection{Sugiyama basis}

Throughout this work we use the decomposition of the bispectrum proposed in \cite{Sugiyama:2018yzo}, that we call Sugiyama basis (see also \cite{sugiyama2023new}). This consists on expanding the full bispectrum $B(\vk_1,\vk_2,\vk_3)=B(\vk_1,\vk_2,\vhn)$ as
\begin{align} \label{sugiexp}
	B(\vk_1,\vk_2,\vhn)
	&=
	\sum_{\ell_1,\ell_2,L} B_{\ell_1\ell_2 L}(k_1,k_2)\, S_{\ell_1\ell_2L}(\hat{\vk}_1,\hat{\vk}_2,\vhn)\; ,
\end{align}
with the tripolar spherical harmonics basis \cite{Varshalovich1988} 
\begin{align}
	S_{\ell_1\ell_2L}(\hat{\vk}_1,\hat{\vk}_2,\vhn) 
   &=
   \frac{1}{H_{\ell_1\ell_2L}} \sum_{m_1m_2M}  \left( \begin{smallmatrix} \ell_1 & \ell_2 & L \\ m_1 & m_2 & M \end{smallmatrix}  \right)  y_{\ell_1}^{m_1}(\hat{\vk}_1) y_{\ell_2}^{m_2}(\hat{\vk}_2) y_L^M(\vhn)\; .
    \label{Eq:Slll}
\end{align} 
The coefficients $H$ are defined in terms of Wigner 3-$j$ symbols as $H_{\ell_1\ell_2L}=\left( \begin{smallmatrix} \ell_1 & \ell_2 & L \\ 0 & 0 & 0 \end{smallmatrix}  \right)$. Since these are zero when the total angular momentum sum is an odd integer, the sum in \cref{sugiexp} is performed over all combinations with the sum $\ell_1+\ell_2+L$ an even integer. This even integer restriction is a consequence of imposing parity symmetry in three dimensions \cite{Sugiyama:2018yzo}. The reduced spherical harmonics $y_\ell^m = \sqrt{\frac{4\pi }{2 \ell+1}} Y_\ell^m$ form an orthogonal basis for functions on the sphere ---in the convention adopted here, the orthonormal basis is given by spherical harmonics. 

Inverting the decomposition above leads to the bispectrum multipoles, given by
\begin{align}
B_{\ell_1\ell_2 L} (k_1,k_2) =& N_{\ell_1\ell_2 L} H_{\ell_1\ell_2 L} \sum_{m_1,m_2,M} \left( \begin{smallmatrix} \ell_1 & \ell_2 & L \\ m_1 & m_2 & M \end{smallmatrix}  \right) \nonumber\\
&\times \int\frac{{\rm d}\Omega_{\vk_1}  {\rm d}\Omega_{\vk_2} {\rm d}\Omega_{\vhn}}{(4\pi)^3} 
y_{\ell_1}^{m_1 *}(\hat{\vk}_1) y_{\ell_2}^{m_2 *}(\hat{\vk}_2)  y_L^{M *} (\vhn) B(\vk_1,\vk_2,\vhn),
\end{align}
with $N_{\ell_1\ell_2 L} = (2\ell_1+1)(2\ell_2+1)(2L+1)$. 

In our choice of coordinates from the previous subsection,  $(k_1,k_2,\theta_{12},\omega,\phi)$, the Sugiyama multipoles read \cite{Sugiyama:2018yzo}
\begin{align}
B_{\ell_1\ell_2 L} (k_1,k_2) = & N_{\ell_1\ell_2 L} H^2_{\ell_1\ell_2 L} \nonumber\\
&\times 
\int_{-1}^{1} \frac{{\rm d} (\cos \omega)}{2} \int_0^{2\pi} \frac{{\rm d}\phi}{2 \pi} \int_{-1}^{1} \frac{{\rm d} (\cos \theta_{12})}{2} S^*_{\ell_1\ell_2L}(\theta_{12},\omega,\phi) B(k_1,k_2,\theta_{12}, \omega,\phi)\; , 
\end{align}
where the tripolar bases now takes the form 
\begin{equation}
    S^*_{\ell_1\ell_2L}(\theta_{12},\omega,\phi) = \frac{1}{H_{\ell_1\ell_2 L}}\sum_{M=-L}^L \left( \begin{smallmatrix} \ell_1 & \ell_2 & L \\ 0 & -M & M \end{smallmatrix}  \right) y_{\ell_2}^{-M *} (\theta_{12},0) y_L^{M *} (\omega,\phi)\; . 
\end{equation}
The spherical harmonic with angular momentum $\ell_2$ corresponds to the direction of the wave vector $\vk_2$ with respect to the $\hat{z}$ axis, and the one with the total angular momentum $L$ to the LoS direction. The spherical harmonic in the $\vk_1$ direction became $\delta_{0m_1}$ since it corresponds to the $\hat{z}$-axis direction. 

A final comment is in place. For consistency with previous work, \cite{Sugiyama:2018yzo,Sugiyama:2020uil,Wang:2023zkv,Wang:2024qyx}, we rescale the multipoles coefficients in the following way
\begin{equation} \label{redefinemult}
   B_{\ell_1\ell_2 L} (k_1,k_2) \rightarrow H_{\ell_1\ell_2 L} B_{\ell_1\ell_2 L} (k_1,k_2)\; . 
\end{equation}
For example, after applying this rescaling, the multipoles 000 and 202 become
\begin{align}
B_{000} (k_1,k_2) &= \frac{1}{8\pi} \int_{-1}^{1} d\mu \int_0^{2\pi} {\rm d}\phi \int_{-1}^{1} {\rm d}x  B(k_1,k_2,\mu,x,\phi)\; , \\
B_{202} (k_1,k_2) &= \frac{5}{8\pi} \int_{-1}^{1} {\rm d}\mu \,\mathcal{L}_2(\mu)\int_0^{2\pi} {\rm d}\phi \int_{-1}^{1} {\rm d}x  \,B(k_1,k_2,\mu,x,\phi)\; ,
\end{align}
where $\mathcal{L}_2$ is the second degree Legendre polynomial. $B_{000}$ and $B_{202}$ are the dominant monopole ($L=0$) and quadrupole ($L=2$) moments, respectively. According to ref. \cite{Sugiyama:2018yzo}, these multipoles are expected to yield the largest signal-to-noise ratio.

\subsubsection{Scoccimarro basis}

\begin{figure*}
	\begin{center}
	\includegraphics[width=2.93 in]{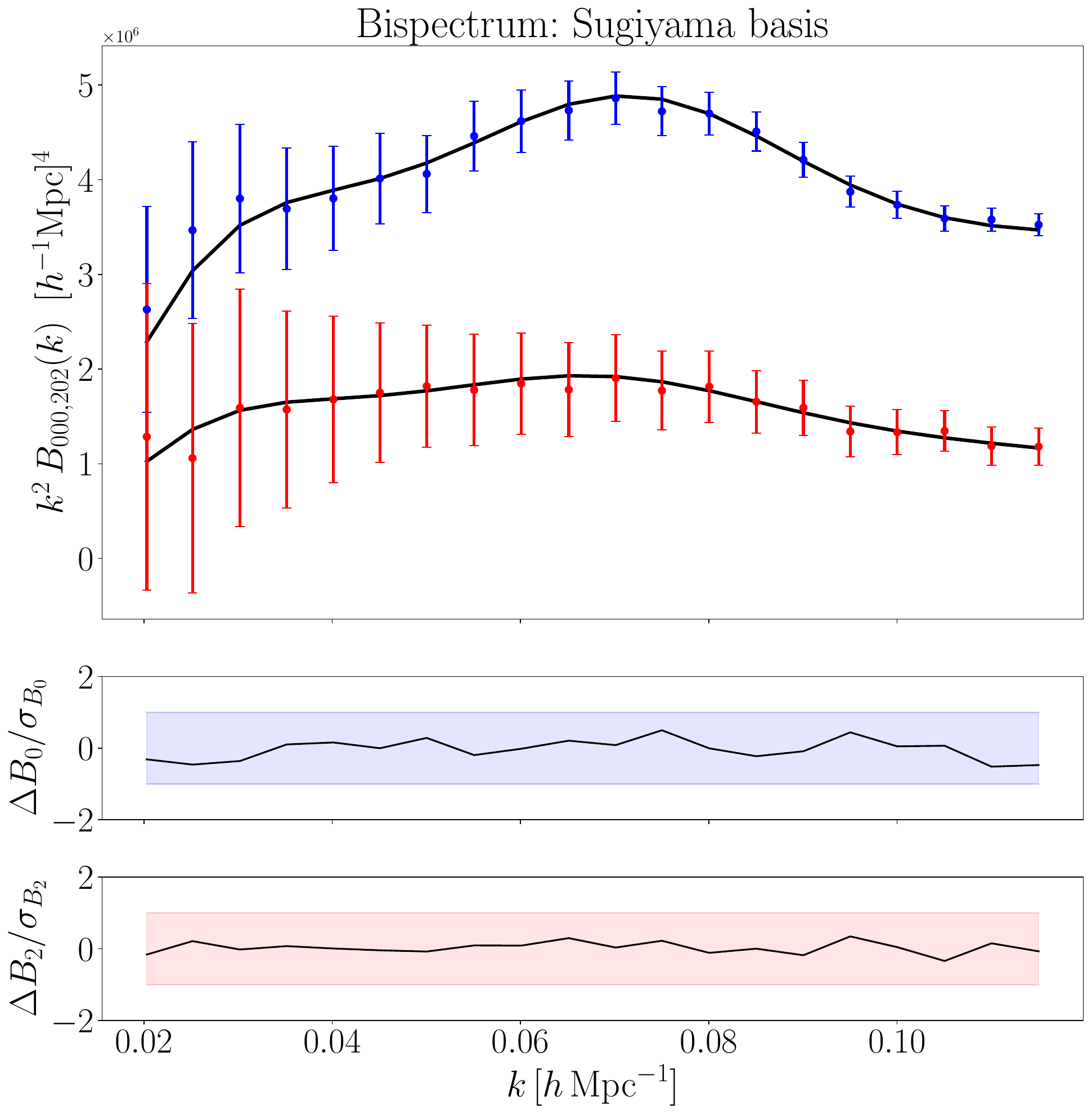}
	\includegraphics[width=3.0 in]{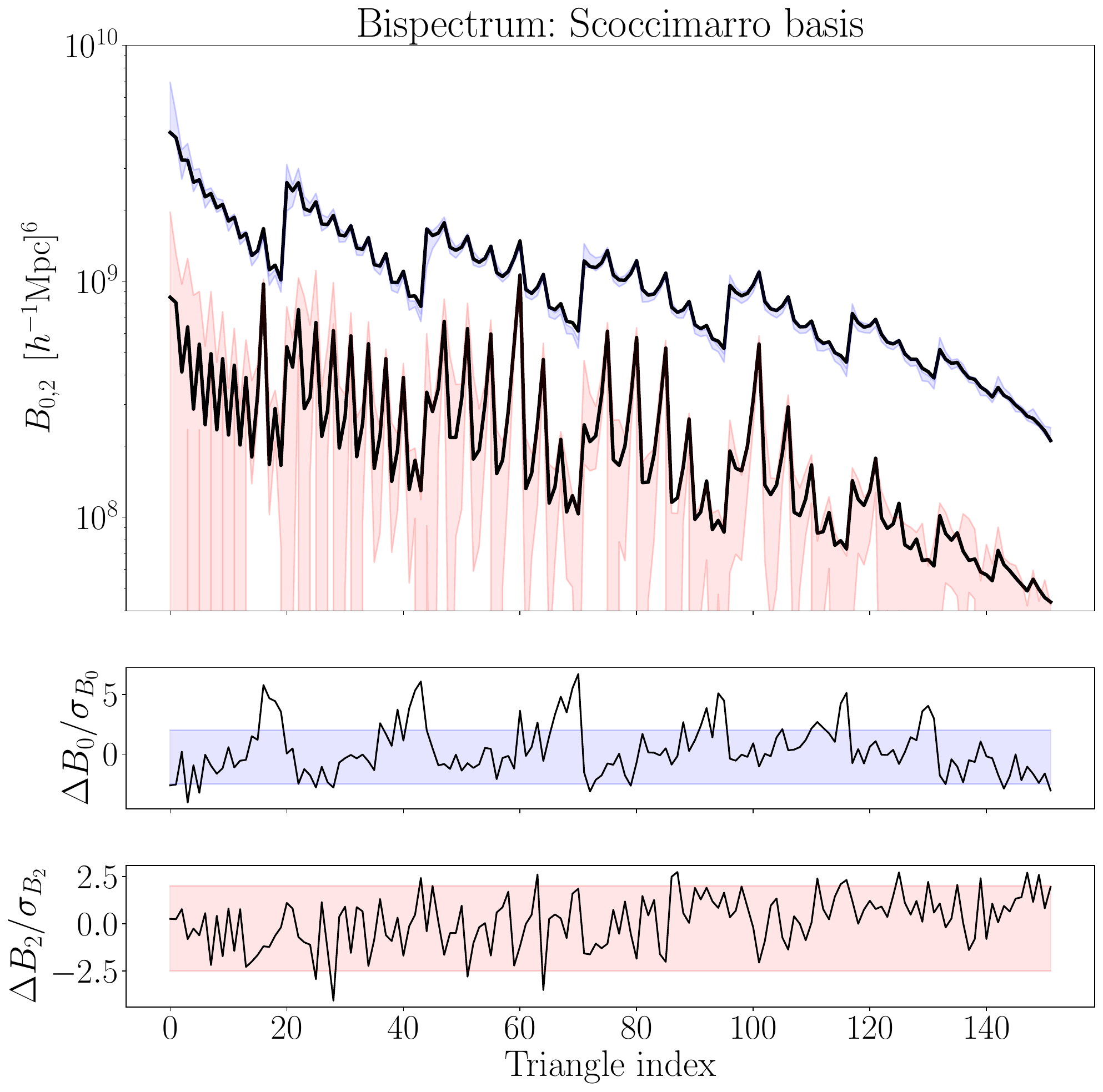}
    \caption{Monopole and quadrupole bispectrum multipoles in the Sugiyama basis (left panel) and the Scoccimarro basis (right panel). Black curves in both panels show the best-fit models obtained from a joint fit using the power spectrum and the bispectrum in the Sugiyama basis over the mean of 25 LRG2 mocks. Error bars are derived from the same covariance matrices used in the fits and corresponding to a single volume box ($=8 \,(h^{-1}\text{Gpc})^3$). Further details about the fits are given in \cref{sec:mocks,sec:mockresults}.
    } \label{fig:SugiyamaVsScoccimarro}
	\end{center}
\end{figure*}

An alternative decomposition of the bispectrum is obtained by integrating over the azimuthal angle $\phi$ and taking Legendre multipoles of the angle perpendicular to the plane where a triangle lives and the LoS direction, $\omega$. We call the coefficients of such an expansion Scoccimarro basis bispectrum multipoles~\cite{Scoccimarro:1999ed}.

Following a bi-step method to obtain the Scoccimarro multipoles, one can perform the following harmonic expansion
\begin{align}\label{Bscocck}
B(\vk_1,\vk_2,\vk_3) =& \sum_{L} \sum_{M=-L}^{L} B_{LM}(k_1,k_2,k_3) Y_L^M(\omega,\phi)\; .
\end{align}
However, for simplicity, the bispectrum is often averaged over the angle $\phi$. That is, usually only the $M=0$ moments of the previous decomposition are considered, leading to the Scoccimarro multipoles
\begin{equation}
    B_L(k_1,k_2,k_3) = \frac{2 L + 1}{2} \int_{0}^{2\pi} \frac{{\rm d}\phi}{2\pi} \int_{-1}^1 {\rm d}(\cos \omega) \mathcal{L}_L(\cos \omega) B(k_1,k_2,k_3,\omega,\phi)\; .
\end{equation}

Although the subsequent sections focus the analysis primarily using the Sugiyama decomposition, a new capability of \folps\ is that it can compute the bispectrum in either the Sugiyama or Scoccimarro bases, with compatible results.

\Cref{fig:SugiyamaVsScoccimarro} shows an example comparing the modeling in the two bases. For this comparison, we use Luminous Red Galaxies mocks over the interval $0.6<z<0.8$ (LRG2), and measurements obtained with the Sugiyama and Scoccimarro estimators, which are described in the next section. The black lines in both panels of the figure are computed using the best-fit parameters obtained by fitting the mean signal of 25 LRG2 mock realizations in the Sugiyama basis only. The data points shown in the left panel correspond to this mean signal, with error bars scaled to the volume of a single simulation box. Details of this fitting procedure, referred to as \texttt{FolpsD-Pk+Bk}, are given in the next section. Overall, this comparison shows good consistency between the two bispectrum bases modelings and measurements within the uncertainty of a single mock realization. A detailed quantitative comparison is left for future work.

\subsection{Alcock–Paczynski effect}

In addition to the anisotropies introduced by RSD, an anisotropy also arises because the fiducial cosmology `fid' used to compute distances from observed redshifts and angles, and to construct the observed power spectrum and bispectrum, does not necessarily coincide with the true cosmology, giving rise to the Alcock–Paczynski (AP) effect \cite{Alcock:1979mp}. That is, given an observed wave vector $\vk$ with an associated fiducial cosmology, the true wave vector $\vk_\text{t}$  is given by
\begin{equation} \label{kobsktrue}
   \vk_\text{t}(\vk,\vhn) = \frac{1}{q_\perp} \vk + \left( \frac{1}{q_\perp} - \frac{1}{q_\parallel} \right) \mu \, \vhn
\end{equation}
with $\mu=\hat{\vk} \cdot \vhn$, and the AP parameters
\begin{align}
    q_\parallel = \frac{D_H(z)}{D_H^\text{fid}(z)} \frac{H^\text{fid}_0}{H_0}  \quad \text{and} \quad
    q_\perp= \frac{D_A(z)}{D_A^\text{fid}(z)} \frac{H^\text{fid}_0}{H_0}\; ,
\end{align}
where $D_A$ is the comoving angular diameter distance and $D_H = 1/H$ the Hubble distance. 
From \cref{kobsktrue}, we obtain the standard relations for the magnitude and orientation of the wave vector $\vk$,
\begin{align}
    k_\text{t} &= k \frac{1}{q_{\perp}} \left[  1 + \mu^2 \left( \left(\frac{q_{\perp}}{q_\parallel} \right)^2 -1 \right) \right]^{1/2},  \\
    \mu_\text{t} &= \frac{k \mu}{k_\text{t} q_\parallel} = \mu \frac{q_{\perp}}{q_\parallel} \left[  1 + \mu^2 \left( \left(\frac{q_{\perp}}{q_\parallel} \right)^2 -1 \right) \right]^{-1/2}\;,
\end{align}
with $\mu_\text{t}=\hat{\vk}_\text{t} \cdot \vhn$. 
The power spectrum multipoles become,
\begin{equation}
   P_\ell(k) = \frac{2\ell+1}{2 q_\perp^2 q_\parallel} \int_{-1}^1 {\rm d}\mu\,\mathcal{L}_\ell(\mu) P_g(k_t,\mu_t)\;. 
\end{equation}
Analogously, the Sugiyama basis bispectrum is
\begin{equation}
    B_{\ell_1,\ell_2L}(k_1,k_2) = \frac{N_{\ell_1\ell_2 L} H^2_{\ell_1\ell_2 L}}{(q_\perp^2 q_\parallel)^2} \int_{-1}^1 \frac{{\rm d}\mu}{2}\int_0^{2\pi} \frac{{\rm d}\phi}{2 \pi} \int_{-1}^{1} \frac{{\rm d}x}{2}  S^*_{\ell_1,\ell_2L}(\mu,x,\phi) B_g(k_{1t},k_{2t},\mu_t,x_t,\phi_t)\;,
\end{equation}
with $x_t = (k_{3t}^2-k_{2t}^2-k_{1t}^2)/(2 k_{1t} k_{2t})$ and $\phi_t=\phi$. 
Equivalently, the Scoccimarro basis bispectrum is
\begin{equation}
    B_L(k_1,k_2,k_3) = \frac{2 L + 1}{2(q_\perp^2 q_\parallel)^2} \int_{0}^{2\pi} \frac{{\rm d}\phi}{2\pi} \int_{-1}^1 {\rm d}(\cos \omega) \mathcal{L}_L(\cos \omega) B(k_{1t},k_{2t},k_{3t},\omega_t,\phi_t)\;,
\end{equation}
with $\omega_t=\cos^{-1}\mu_t$.

\subsection{Galaxy biasing}

The new \folps\ has three different galaxy bias implementations. One of them is based on the following bias expansion \cite{Desjacques:2016bnm}
\begin{align} \label{biasexp}
    \delta_g =& b_1\delta + \frac{1}{2} b_2\delta^2 + b_{s} s_{ij}s_{ij}  + b_{\rm td} O_{\rm td}  + \epsilon_k,
\end{align}
with $\delta$ and $\delta_g$ being the matter and galaxy overdensities, and $\epsilon_k$ is the stochastic field, uncorrelated with large-scale wavelength modes. Here, $s_{ij}$ is  the tidal field
\begin{align}
   s_{ij} &= \left( \frac{k_i k_j}{k^2} - \frac{1}{3}\delta_{ij}\right) \delta,
\end{align}
and $O_{\rm td}$ is the third order nonlinear operator, 
\begin{align}
   O_{\rm td} &= \frac{8}{21} s_{ij} \left( \frac{k_i k_j}{k^2} - \frac{1}{3}\delta_{ij}\right) \left( \delta^2 - \frac{3}{2} s_{kl} s_{kl} \right). 
\end{align}
In writing the expansion of \cref{biasexp} we have omitted the third order biases $b_3$, $b_{K^3}$ and $b_{\delta K^2}$, which do not enter our final expressions for the power spectrum and tree-level bispectrum. We notice that the notation of  \cite{Desjacques:2016bnm} is $K_{ij}=s_{ij}$ and $b_{K^2}=b_s $. 

The second bias scheme, which is set by default in \folps\ and used in previous works \cite{Aviles:2020cax,Noriega:2022nhf,KP5s3-Noriega}, utilizes the bias expansion of \cite{McDonald:2009dh} (labeled below as \texttt{McDonald}), and it is related to the basis of \cref{biasexp} used in this work, by the following map
\begin{align} \label{biasrelations}
b_1^\texttt{McDonald} = b_1, \qquad& b_2^\texttt{McDonald} = b_2, \nonumber\\
b_{s^2}^\texttt{McDonald} = 2 b_s, \quad&
b_{3nl}^\texttt{McDonald}= -\frac{32}{21}\left(b_{s} + \frac{2}{5} b_{td} \right)\;.
\end{align}
Finally and for completeness, \folps\ also has the capability of working with a third bias prescription, introduced in \cite{Assassi:2014fva} and adopted by the \textsc{class-pt} code \cite{Chudaykin:2020aoj}. This prescription is related to the \folps\ default \texttt{McDonald} basis by
\begin{align} \label{biasrelationsC}
b_1^\texttt{McDonald} = b_1^\textsc{class-pt}, \qquad& b_2^\texttt{McDonald} = b_2^\textsc{class-pt} -\frac{4}{3} b_{\mathcal{G}^2}^\textsc{class-pt}, \nonumber\\
b_{s^2}^\texttt{McDonald} = 2 b_{\mathcal{G}^2}^\textsc{class-pt}, \quad&
b_{3nl}^\texttt{McDonald}= -\frac{32}{21}\left(b_{\mathcal{G}^2}^\textsc{class-pt} + \frac{2}{5} b_{\Gamma_3}^\textsc{class-pt} \right)\;.
\end{align}

For the choice of EdS kernels in \folps, the one-loop power spectrum level matches exactly that of \textsc{class-pt}.


\section{Data and settings}\label{sec:mocks}

\begin{figure}
	\begin{center}
	\includegraphics[width=6 in]{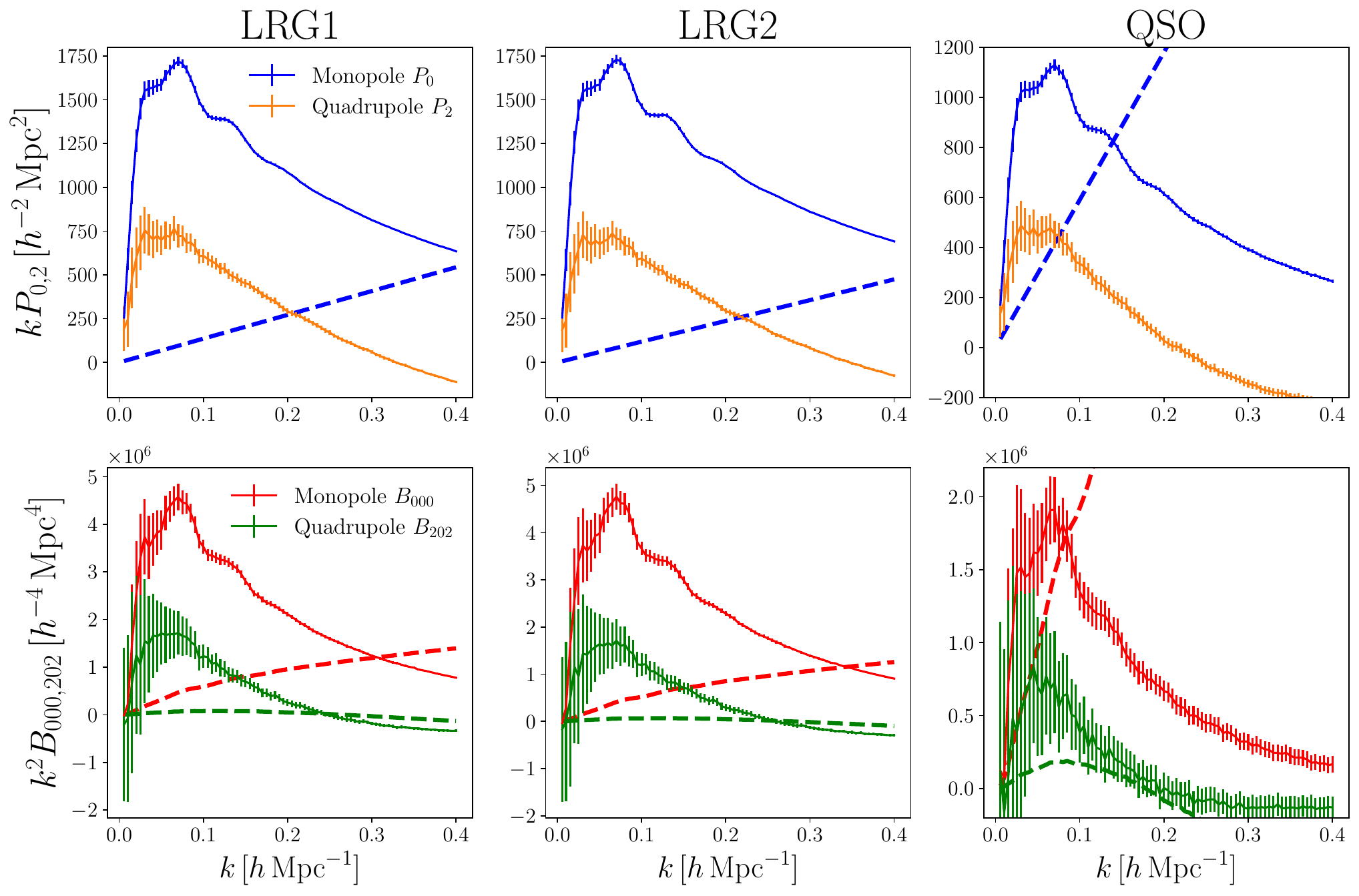} 
    \caption{ Power spectrum (top row) and bispectrum (bottom row) multipoles measured from second-generation Abacus mocks. Each panel shows the monopole and quadrupole of the respective spectra. The three columns correspond to LRGs in two redshift bins and QSOs. Solid lines show the spectra with the noise subtracted, while the Poissonian shot noise is shown with dashed lines. The bispectrum multipoles $B_{\ell_1\ell_2L}(k_1,k_2)$ are evaluated along the diagonal $k_1 = k_2$.
    }\label{fig:tracers}
	\end{center}
\end{figure}

In this section we present the data and settings we use to test our pipeline for the power spectrum and bispectrum using simulated data. To do so, we use 25 independent realisations of the baseline $\Lambda$CDM cosmology of the AbacusSummit $N$-body simulation suite \cite{abacus2021},\footnote{\href{https://abacussummit.readthedocs.io}{https://abacussummit.readthedocs.io}} each consisting of $3\times 10^{11}$ particles in a box of volume $(2\,h^{-1}\text{Gpc})^3$. The cosmological parameters are set to the mean of the full temperature, polarisation, and lensing (TT,TE,EE+lowE+lensing) data obtained from the Planck 2018 \texttt{Plik} results \cite{2020A&A...641A...6P}, with baryonic matter abundance $\omega_b = 0.02237$, cold dark matter abundance $\omega_{cdm} = 0.12$, reduced Hubble constant $h = 0.6736$, primordial amplitude $A_s = 2.0830 \times 10^{-9}$ and spectral index $n_s = 0.9649$. We consider a single species of massive neutrinos with $m_\nu=0.06\,\text{eV}$, such that $N_{\rm ur} = 2.0328$, $N_\mathrm{ncdm} = 1$ and $\omega_{\rm ncdm} = 0.0006442$. 
    
    For the galaxy catalogues, we use the so-called second-generation (2nd Gen) Abacus mock catalogues, in which halo occupation distribution (HOD) parameters, following \cite{zheng2007}, have been fitted to match the two-point correlation functions of the DESI Early Data Release \cite{adame2024validation}.  
    The resulting catalogues reproduce the clustering of several DESI tracers, including the Luminous Red Galaxies (LRGs) in three redshift bins. $ 0.4 < z_1 < 0.6$ (LRG1), $0.6 < z_2 < 0.8$ (LRG2), and $0.8 < z_3 < 1.1$ (LRG3) and quasars (QSO) between $0.8 < z < 2.1$. The HOD accounts for an incompleteness factor $0 < f_{\rm ic} \leq 1$, and velocity-bias parameters for central and satellite galaxies are considered when constructing the redshift-space catalogues. 
    
    In practice, the mocks analysed here correspond to simulation snapshots at $z=0.5$ for LRG1, $z=0.8$ for LRG2, and $z=1.4$ for the QSO sample, which are used as effective representations of the corresponding redshift bins. The number densities, in units of $h^3\,{\rm Mpc}^{-3}$, are $\bar{n}_{\rm LRG1} = 0.00070$, $\bar{n}_{\rm LRG2} = 0.00085$, $\bar{n}_{\rm LRG3} = 0.00081$, and $\bar{n}_{\rm QSO} = 0.00017$. We refer the interested reader to \cite{2024MNRAS.530..947Y} for details on the HOD modelling and its parameters.

    The multipoles of the power spectrum and bispectrum are measured from the mocks with the \textsc{Triumvirate} code \cite{Wang:2023zkv},\footnote{\href{https://github.com/MikeSWang/Triumvirate}{https://github.com/MikeSWang/Triumvirate}} an FFT-based Python/C++ implementation of two- and three-point clustering estimators that compute bispectrum multipoles in the Sugiyama basis. Since we only work with periodic boxes, the estimator uses the global plane-parallel limit along the $\hat{z}$-direction. The density fields were assigned to a regular Cartesian mesh using the Triangular-Shaped Cloud mass-assignment scheme, following a particle normalization convention for the bispectra, and mesh-mixed normalization for the power spectra. Aliasing effects were automatically corrected with the code. 
    For both summary statistics, the measurements were performed between $0.005  < k < 0.4 \ihMpc$  in bins of $\Delta k = 0.005  \ihMpc$, but we only consider measurements within $0.02  < k < k_{\rm max}$ for the fits.

The sample covariance matrices are estimated using the Effective Zel'dovich mocks (EZmocks). These consist of $N_\text{mocks}=2000$  independent synthetic realizations for LRGs, ELGs, and QSOs, generated with fast approximate methods based on the Zel'dovich approximation \cite{1970A&A.....5...84Z,2015MNRAS.446.2621C,2021MNRAS.503.1149Z}. A key advantage of EZmocks is their high computational efficiency relative to full  $N$-body simulations, while still accurately reproducing the two- and three-point statistics up to mildly non-linear scales. Each EZmock cubic box has a volume of $(6\,h^{-1}\text{Gpc})^3$, hence we rescale the covariance matrix by a factor of 27 $\,= (6/2)^3$ to match the volume of a single Abacus cubic box mock. We correct the bias in the inverse of the sample covariance matrix by applying the Hartlap correction factor  \cite{Hartlap:2006kj}.
\begin{equation}
    C^{-1}  \rightarrow \frac{N_\text{mocks} - N_\text{bins}- 2}{N_\text{mocks} - 1}  \, C^{-1},
\end{equation}
where $N_\text{bins}$ is the dimensionality of the data vector. 
In our analysis, the size of the data vector ranges from 74 to 146 elements, corresponding to power spectrum-only fits and to joint power spectrum plus bispectrum fits, respectively. For these values, the Hartlap factor modifies the covariance amplitude by approximately $3.7\%$ to $7.5\%$.
Additionally, we include the Percival factor \cite{2014MNRAS.439.2531P,Dodelson:2013uaa} to propagate the error of the inverse of the covariance matrix into the standard deviation of the estimated parameters.

\Cref{fig:tracers} shows the multipoles of the mock tracers used in this section: LRG1, LRG2, and QSO. Solid lines correspond to the mean of the 25 Abacus mock realizations for each tracer. The Poissonian shot noise has been subtracted from the monopole in these curves to highlight the signal, while the dashed lines show the shot noise contribution. Error bars are computed from the EZmocks sample covariance.

\subsection{Settings}

\begin{table}[t]
\centering
\begingroup
\begin{spacing}{1.3}
\begin{tabular}{l c l}
\toprule
Parameter & Prior & Comments \\
\midrule\midrule

\multicolumn{3}{c}{\textbf{Cosmological parameters}} \\
\midrule

$h$ & $\mathcal{U}[0.5,0.9]$ &  \\

$\omega_{cdm}$ & $\mathcal{U}[0.05,0.2]$ &  \\

$\omega_b$ & $\mathcal{N}(0.02218,0.00055^2)$ & BBN prior \\
$\omega_b$ (mocks) & $\mathcal{N}(0.02237,0.00037^2)$ &  \\

$\log(10^{10}A_s)$ & $\mathcal{U}[2,4]$ &  \\

$n_s$ & $\mathcal{N}(0.9649,0.042^2)$ & Planck $n_{s10}$ prior \\
$n_s$ (mocks) & Fixed to fiducial &  \\

\midrule\midrule
\multicolumn{3}{c}{\textbf{Galaxy bias parameters}} \\
\midrule

$\tilde{b}_1=b _1 \sigma_8 \sqrt{A_\text{AP}}$& $\mathcal{U}[0.1,4]$ &  \\

$\tilde{b}_2 = b_2 \sigma_8^2 \sqrt{A_\text{AP}}$& $\mathcal{N}(0,5^2)$ &  \\

$\tilde{b}_s = b_s \sigma_8^2 \sqrt{A_\text{AP}}$& $\mathcal{N}\!\left(-\frac{2}{7}(b_{1,\rm fid}-1)\sigma_{8,\rm fid}^2,\,5^2\right)$ &  \\

$\tilde{b}_3 = b_3 \sigma_8^3 \sqrt{A_\text{AP}}$& $\mathcal{N}\!\left(\frac{23}{42}(b_{1,\rm fid}-1)\sigma_{8,\rm fid}^3,\,
(1 \sigma_{8,\rm fid}^3)^2\right)$&  \\

\midrule\midrule
\multicolumn{3}{c}{\textbf{Power spectrum only parameters ($P_k$)}} \\
\midrule

$A_{\rm AP}\,\sigma_8^2\,\tilde{\alpha}_0$ & $\mathcal{N}(0,12.5^2)$ &  \\

$A_{\rm AP}\,\sigma_8^2\,\tilde{\alpha}_2$ & $\mathcal{N}(0,12.5^2)$ &  \\

$A_{\rm AP}\,\sigma_8^2\,\tilde{\alpha}_4$ & $\mathcal{N}(0,12.5^2)$ &  \\

$A_{\rm AP}\,\alpha_0^{\rm shot}$ 
& $\mathcal{N}(0,2^2)\times \overline{n}_g^{-1}$ &  \\

$A_{\rm AP}\,\alpha_2^{\rm shot}$ 
& $\mathcal{N}(0,5^2)\times \dfrac{f_{\rm sat}\sigma^2_{1,\rm eff}}{\overline{n}_g}$ &  \\

\midrule\midrule
\multicolumn{3}{c}{\textbf{Bispectrum only parameters ($B_k$)}} \\
\midrule

$A_{\rm AP}\,\sigma_8^2 c_1$ & $\mathcal{N}(0,5^2)$ &  \\

$A_{\rm AP}\,\sigma_8^2 c_2$ & $\mathcal{N}(0,0.1^2)$ &  \\

$A_{\rm AP}\,P_{\rm shot}$ 
& $\mathcal{N}(0,1^2)\times \overline{n}_g^{-1}$ &  \\

$A_{\rm AP}\,B_{\rm shot}$ 
& $\mathcal{N}(0,1^2)\times \overline{n}_g^{-1}$ &  \\

\bottomrule
\end{tabular}
\end{spacing}
\endgroup
\caption{Model parameters and priors. Rows labelled ``mocks'' indicate priors used in chains run on mocks. The ``tilde'' counterterms ($\tilde{\alpha}_0,\, \tilde{\alpha}_2,\,\tilde{\alpha}_4$) are related to the folpsD counterterms (${\alpha}_0,\, {\alpha}_2,\,{\alpha}_4$) via the relations $\alpha_0= b_1^2\tilde{\alpha}_0,\, \alpha_2=fb_1(\tilde{\alpha}_0+\tilde{\alpha}_2),\, \alpha_4=f^2\tilde{\alpha}_2+fb_1\tilde{\alpha}_4 $.
}
\label{table:parameters_priors}
\end{table}

We use the re-parametrization of nuisance parameters proposed in \cite{Maus:2024dzi,2025arXiv250909562T}, which absorbs the amplitude parameters that most strongly control the observed power spectrum signal, specifically powers of $\sigma_8$ and the AP amplitude $A_\text{AP}$,
\begin{equation}
  A_\text{AP} = \left(\frac{H_0^\text{fid}}{H_0} \right)^3 \frac{H(z)}{H^\text{fid}(z)} \left( \frac{D_A^\text{fid}(z)}{D_A(z)} \right)^2,
\end{equation}
with $H_0$ being the Hubble constant and $D_A$ the comoving angular diameter distance. The AP amplitude can be written as $A_\text{AP} =  (q_\parallel q_\perp^2)^{-1}$, which accounts for the power spectrum amplitude change due to the AP effect.

Hence we consider the re-parametrized biases basis,
\begin{align}
  \tilde{b}_1 &= b_1 \sigma_8 \sqrt{A_\text{AP}},   \quad&
  \tilde{b}_2 &= b_2 \sigma_8^2 \sqrt{A_\text{AP}},  \nonumber\\
  \tilde{b}_s &= b_s \sigma_8^2 \sqrt{A_\text{AP}},   \quad&
  \tilde{b}_3 &= b_{td} \sigma_8^3 \sqrt{A_\text{AP}}.  
\end{align}

Such re-parametrizations redefine the nuisance terms so that they scale with the overall amplitude of the power spectrum ($\sim b^2 \sigma_8^2 A_\text{AP}$) rather than acting as independent offsets. By doing so, the nuisance and cosmological parameters become less degenerate in the likelihood, which reduces the prior-volume or projection effects \cite{Carrilho:2022mon,Simon:2022lde,2023OJAp....6E..23H,MBonici2025}. We, as well, reparametrize the EFT counterterms, such that we use the ``tilde'' parameters in \cref{table:parameters_priors}, that are related to the counterterms of \cref{PLOctr} via  $\alpha_0= b_1^2\tilde{\alpha}_0,\, \alpha_2=fb_1(\tilde{\alpha}_0+\tilde{\alpha}_2),\, \alpha_4=f^2\tilde{\alpha}_2+fb_1\tilde{\alpha}_4 $ \cite{KP5s2-Maus}.\footnote{\href{https://github.com/sfschen/velocileptors/blob/master/notebooks/LPT_RSD_Optimal_Settings.ipynb}{https://github.com/sfschen/velocileptors/blob/master/notebooks/LPT\_RSD\_Optimal\_Settings.ipynb}}

We adopt similar reparametrizations that include powers of $\sigma_8$ and $A_\text{AP}$ for the rest of the nuisance parameters.  In \cref{table:parameters_priors} we show all the parameters varied and their priors. 

We infer parameters by integrating \folps\ into the \textsc{desilike}\footnote{\href{https://github.com/cosmodesi/desilike}{https://github.com/cosmodesi/desilike}} package. Specifically, we use the Metropolis-Hasting algorithm to sample the space of parameters with the \textsc{Cobaya} Bayesian sampler \cite{2021JCAP...05..057T}.\footnote{\href{https://cobaya.readthedocs.io/}{https://cobaya.readthedocs.io/}} We run Monte Carlo Markov Chains (MCMC), which are stopped once the Gelman–Rubin statistic \cite{1992StaSc...7..457G} reaches $R-1 = 0.01$, for fits to the power spectrum only, and $R-1 = 0.05$ when we also include the bispectrum, as well as for fits to the DESI DR1 data in \cref{sec:DR1,sec:w0wa}, which are performed using only the galaxy power spectrum.

In the inference, we analytically marginalize over parameters that enter linearly in the power spectrum as explained in detail in \cite{KP5s3-Noriega}, namely the power-spectrum shot noise and counterterms.





\begin{center}
\begin{table*}[t]
\centering
\begingroup
\begin{spacing}{1.3}
       \begin{tabular} { l  c c c}
\toprule       
\noalign{\vskip 3pt}\\[-37pt]


  &  \phantom{abcdefghabcdef} &  \phantom{abcdefghabcdef} &  \phantom{abcdefghabcdef}\\
  &  $P_{0,2}$ &   $B_{000}$ &  $B_{202}$\\[2pt]

\noalign{\vskip 3pt}\\[-20pt]

&\multicolumn{3}{c}{\Cref{sec:mockresults} baseline scale-cuts in $\,k_\text{max}/(h\text{Mpc}^{-1})\,$ units. } \\[2pt]
  
\hline\\[-10pt]
{\phantom{a}\texttt{EFT-Pk}\phantom{abcdefgh}} & $0.201         $ &   --- & ---\\[4pt]

{\phantom{a}\texttt{EFT-Pk+Bk}          } & $0.201         $ &  $0.12$ & $0.03$\\[4pt]

{\phantom{a}\texttt{FolpsD-Pk}}  &  $0.301         $ &   --- & ---\\[4pt]

{\phantom{a}\texttt{FolpsD-Pk+Bk}}  &  $0.301         $ &  $0.12$ & $0.08$\\[4pt]

\bottomrule
\end{tabular}
\end{spacing}
\endgroup
\caption{Summary of the main PT models used for the mock data analysis in \cref{sec:mockresults}. 
The minimum wavenumber is $k_\text{min} = 0.02\ihMpc$ for all models. In \cref{sec:bkmax} we show the Sugiyama bispectrum can be extended to higher wavenumbers. }
\label{table:PTmodels}
\end{table*}
\end{center}

\revised{For the power spectrum, we use the same binning scheme adopted in \cite{KP5s3-Noriega}. For the Sugiyama bispectrum, the model is evaluated by interpolating the AP-corrected diagonal at the effective $k$ of each bin. We tested different binning prescriptions and found no observable differences in the inferred parameters. We therefore adopt this interpolation scheme since it is computationally the most efficient option.}

\section{Results on mocks} \label{sec:mockresults}

In this section, we first compare the performance of the \texttt{Folps-EFT} model (hereafter \texttt{EFT}), defined in \cref{eq:Pg}, with that of \texttt{FolpsD}, given in \cref{eq:PgD}, when fitting the second-generation Abacus mock power spectra. After that, we extend the analysis to include fits that also incorporate the galaxy bispectrum.

Throughout this section, we adopt conservative scale cuts listed in \cref{table:BaselineAnalysis}, which define our baseline analyses. These $k$-ranges will be extended in  \cref{sec:bkmax}, yielding tighter and more accurate constraints for LRG mocks.

In \cref{subsec:LRG1QSO}, we fit QSO mocks and examine the impact of noise on damping-based models. Extensions beyond the $\Lambda$CDM model, including massive neutrinos, are presented in \cref{app:neutrinos}, where we study how phenomenological damping interacts with neutrino-induced modifications of the power spectrum shape.

\subsection{Constraints from the power spectrum}

\begin{figure*}
	\begin{center}
    \includegraphics[width=6 in]{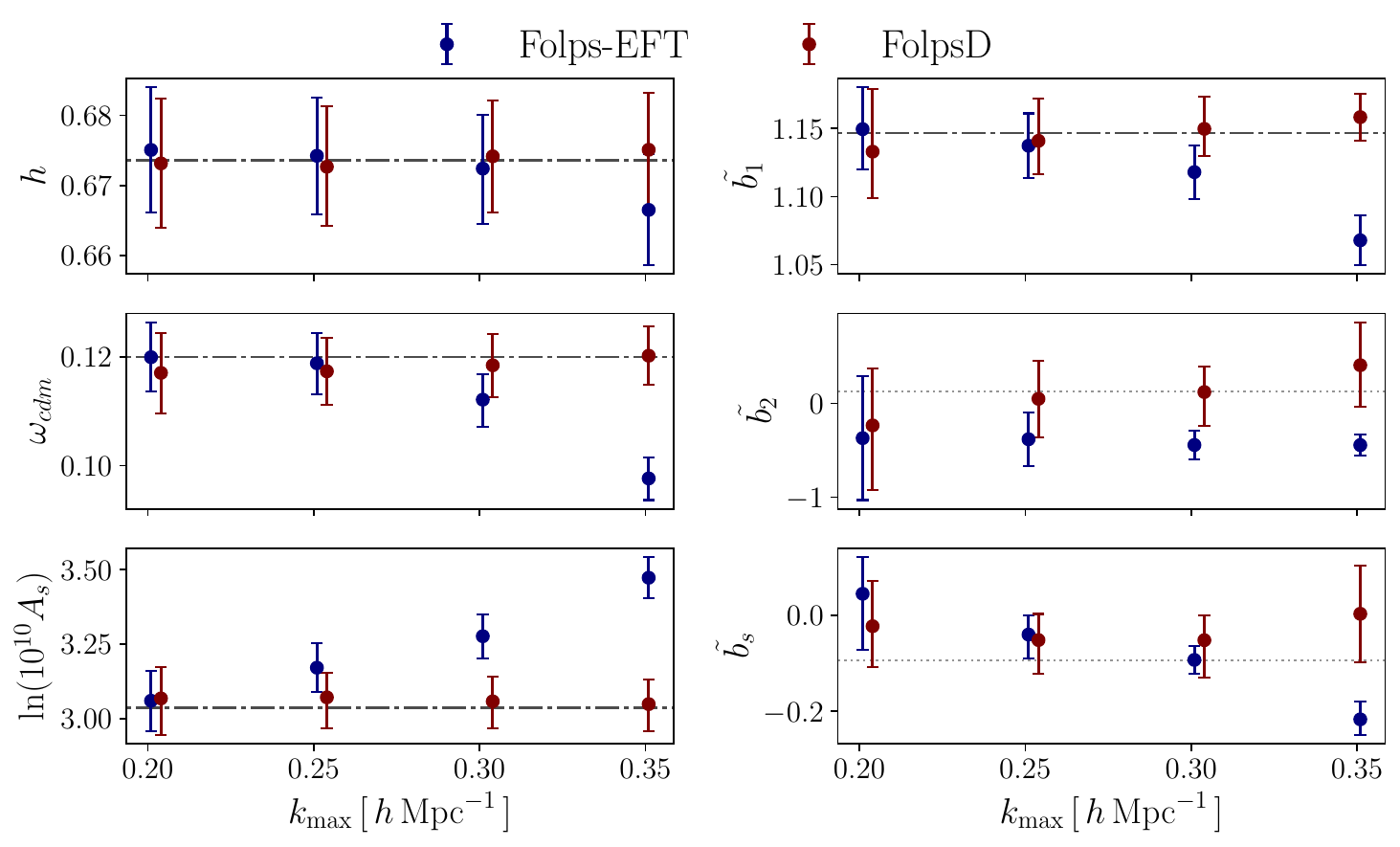}
	\caption{Mean values and $1\sigma$ confidence intervals for cosmological parameters ($h$, $\omega_{cdm}$, $\log(10^{10}A_s)$) and rescaled biases $\tilde{b}_1$, $\tilde{b}_2$, and $\tilde{b}_s$, obtained from fits to the LRG2 Abacus mocks for different values of $k_\text{max}$. The blue points correspond to \texttt{EFT} while the red points represent \texttt{folpsD}. The long dashed lines for the parameters $h$, $\omega_{cdm}$, $\log(10^{10}A_s)$ and $\tilde{b}_1$ represent their true values known from the simulations. The dotted lines for parameters $\tilde{b}_2$, and $\tilde{b}_s$ depict their approximate values expected from coevolution. The fits are performed over the interval $0.02\,h\,\mathrm{Mpc}^{-1} < k < k_\text{max}$. 
    }\label{fig:Pk_kmax}
	\end{center}
\end{figure*}

We begin with a comparison between the \texttt{EFT} and \texttt{FolpsD} models for the LRG2 Abacus second-generation mocks using only the power spectrum. \Cref{fig:Pk_kmax} shows the mean posterior values and $68\%$ confidence intervals obtained from fits to the power spectrum over the range $0.02 \ihMpc < k < k_{\max}$, for different choices of $k_{\max}$. Blue whiskers correspond to the \texttt{EFT} results, while red whiskers correspond to \texttt{FolpsD}. In the left panel, the dot-dashed horizontal lines indicate the simulation values of the cosmological parameters, whereas in the right panel they show the values of the re-parametrized bias $\tilde{b}_1$ obtained from previous analyses that give $b_1^\text{LRG2} \approx 2.1$. For the second-order biases, the dashed lines 
correspond to the co-evolution predictions \cite{Baldauf:2012hs,Chan:2012jj,Saito:2014qha}, $b_2^\text{coev}=\frac{8}{21} (b_1^\text{fid}-1)$ and $b_s^\text{coev}=-\frac{2}{7} (b_1^\text{fid}-1)$, which galaxy tracers are expected to follow only approximately.

From \cref{fig:Pk_kmax}, we observe that fitting the power spectrum to higher $k_\text{max}$ values reduces the error bars of the estimated cosmological parameters, particularly $A_s$ and $\omega_{cdm}$ for \texttt{EFT}. However, the measurements become noticeably biased. This mismatch arises from biased constraints on the bias parameters, primarily through $\tilde{b}_1$, as shown in the right panel of \cref{fig:Pk_kmax}, but also from poor fits to $\tilde{b}_2$ and $\tilde{b}_s$. 
On the other hand, when fitting the data using \texttt{FolpsD} the bias parameters $\tilde{b}_2$ and $\tilde{b}_s$ lie very close to the co-evolution values and they become tightly constrained, compared to the \texttt{EFT} case with $k_\text{max}=0.201 \ihMpc$. This behavior suggests that the additional constraining power provided by the damped model comes mainly from an improved fitting to these second-order bias parameters, which is obtained from the small scales. We explore this in more detail in \cref{app:b2bsinfo}. 

Although \cref{fig:Pk_kmax} shows unbiased cosmological parameters up to $k\sim 0.3 \ihMpc$, in the following, we adopt the power spectrum \texttt{FolpsD} model with a conservative scale cut of $k_\text{max} = 0.301 \ihMpc$, and refer to it as \texttt{FolpsD-Pk}.\footnote{Although \cref{fig:Pk_kmax} shows that the model can reach up to $k_\text{max} = 0.351 \ihMpc$, we maintain a conservative choice of $k_\text{max} = 0.301 \ihMpc$.} This choice is also supported by recent studies showing that the TNS model with a Lorentzian damping recovers unbiased constraints up to $k_\text{max}\simeq0.3\,h\text{Mpc}^{-1}$ \cite{Eggemeier:2025xwi}. On the other hand, for the EFT model, we choose $k_\text{max} = 0.201 \, h\text{Mpc}^{-1}$, which is the maximum wave number adopted for the first-generation Abacus mock analysis in \cite{KP5s1-Maus,KP5s2-Maus,KP5s3-Noriega,KP5s4-Lai}, and for the DESI-DR1 analysis in \cite{DESI2024.V.KP5,DESI2024.VII.KP7B}. A summary of the models used in this section is given in \cref{table:PTmodels}.

\begin{figure*}
	\begin{center}
	\includegraphics[width=3.2 in]{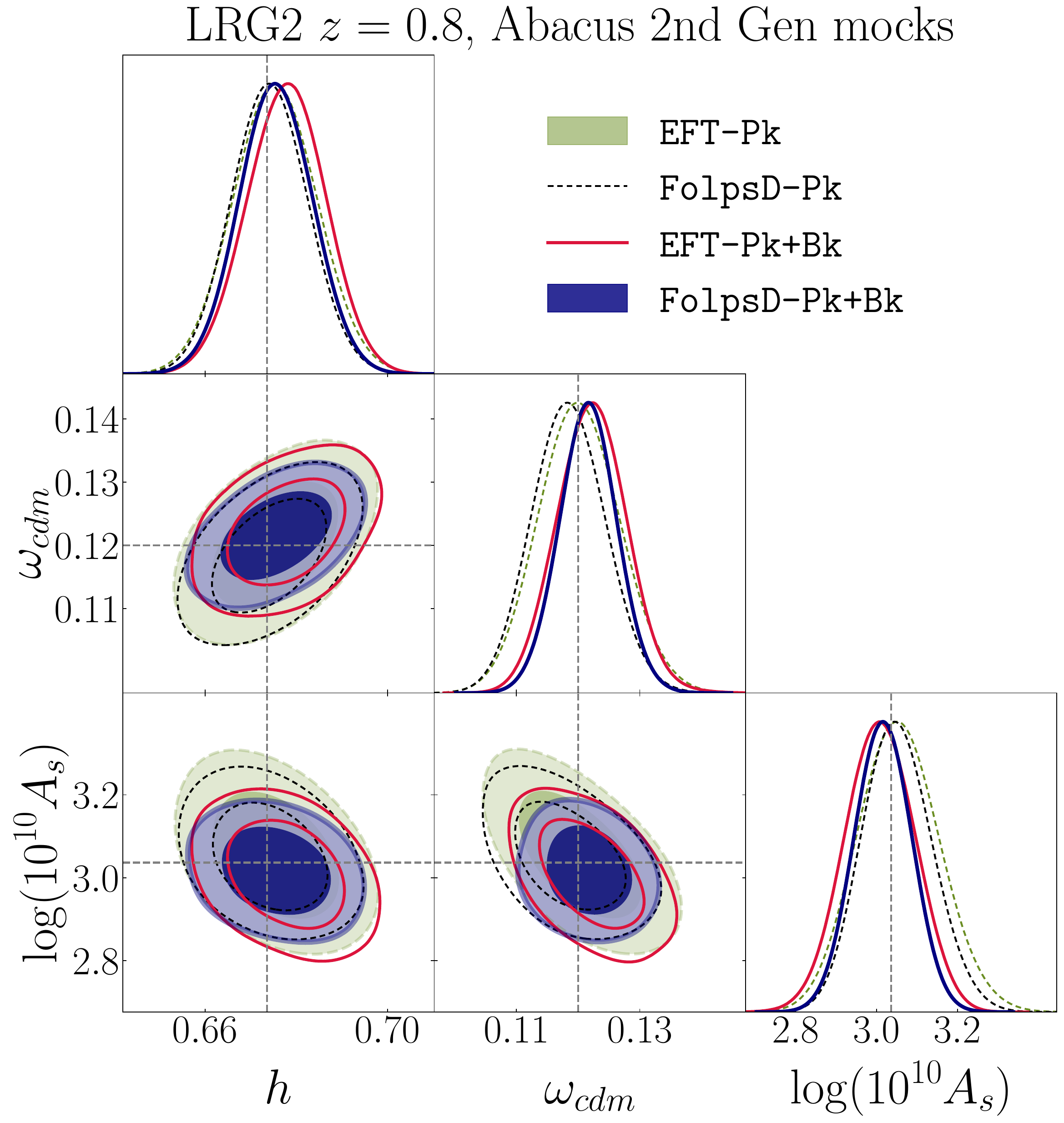}
	\includegraphics[width=2.7 in]{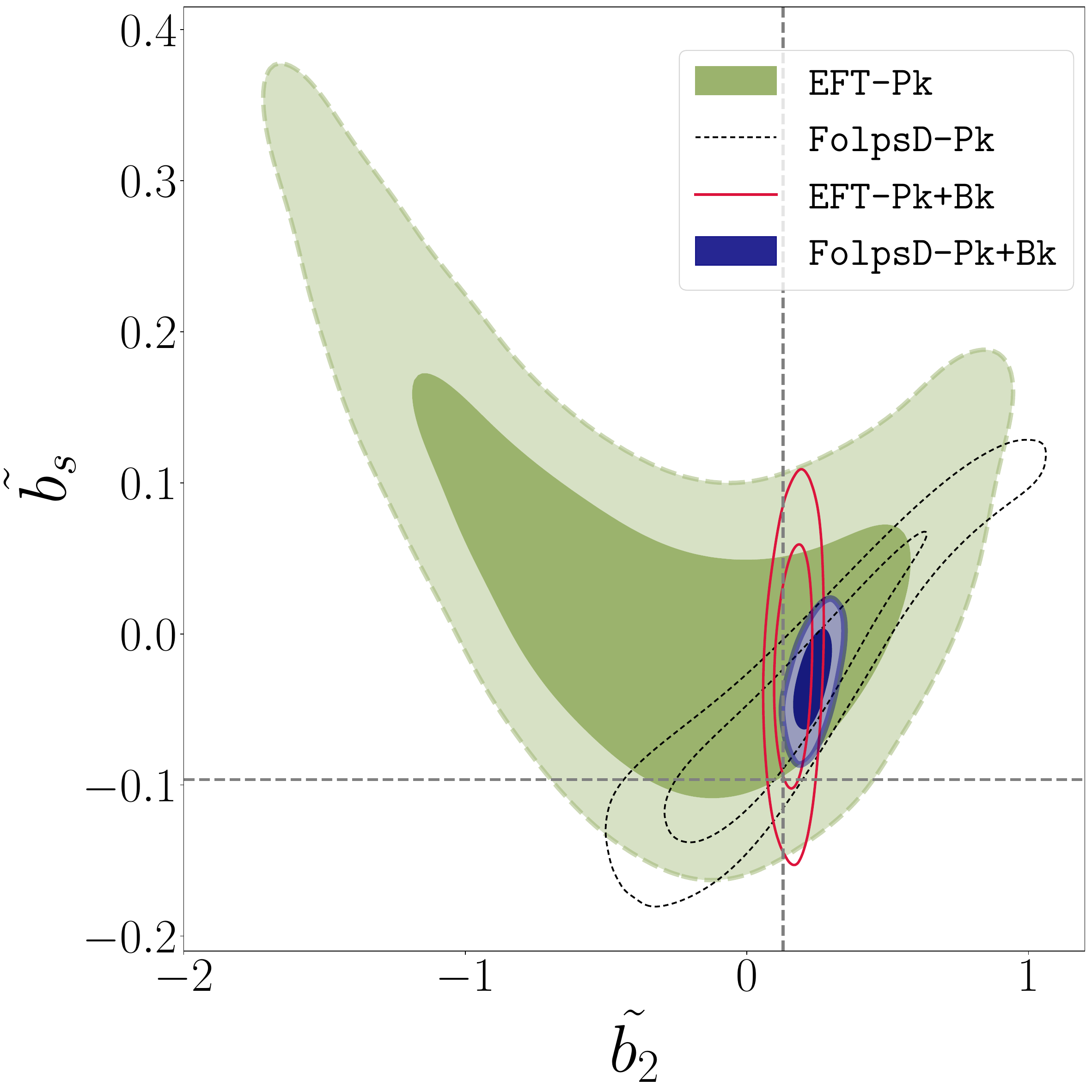}
    \caption{Fitting to LRG2 mocks sample using the four main modelings: \texttt{EFT} considering the power spectrum alone and adding the bispectrum, and the same for \texttt{FolpsD}. \textit{Left panel:} Triangle plot for the cosmological parameters $h$, $\omega_{cdm}$ and $\log(10^{10}A_s)$. \textit{Right panel:} Contours in the $b_2$-$b_s$ plane, where the dot-dashed lines denote the coevolution values with $b_1=2.1$. }\label{fig:LRG2}
	\end{center}
\end{figure*}

\begin{figure*}
	\begin{center}
	\includegraphics[width=3.01 in]{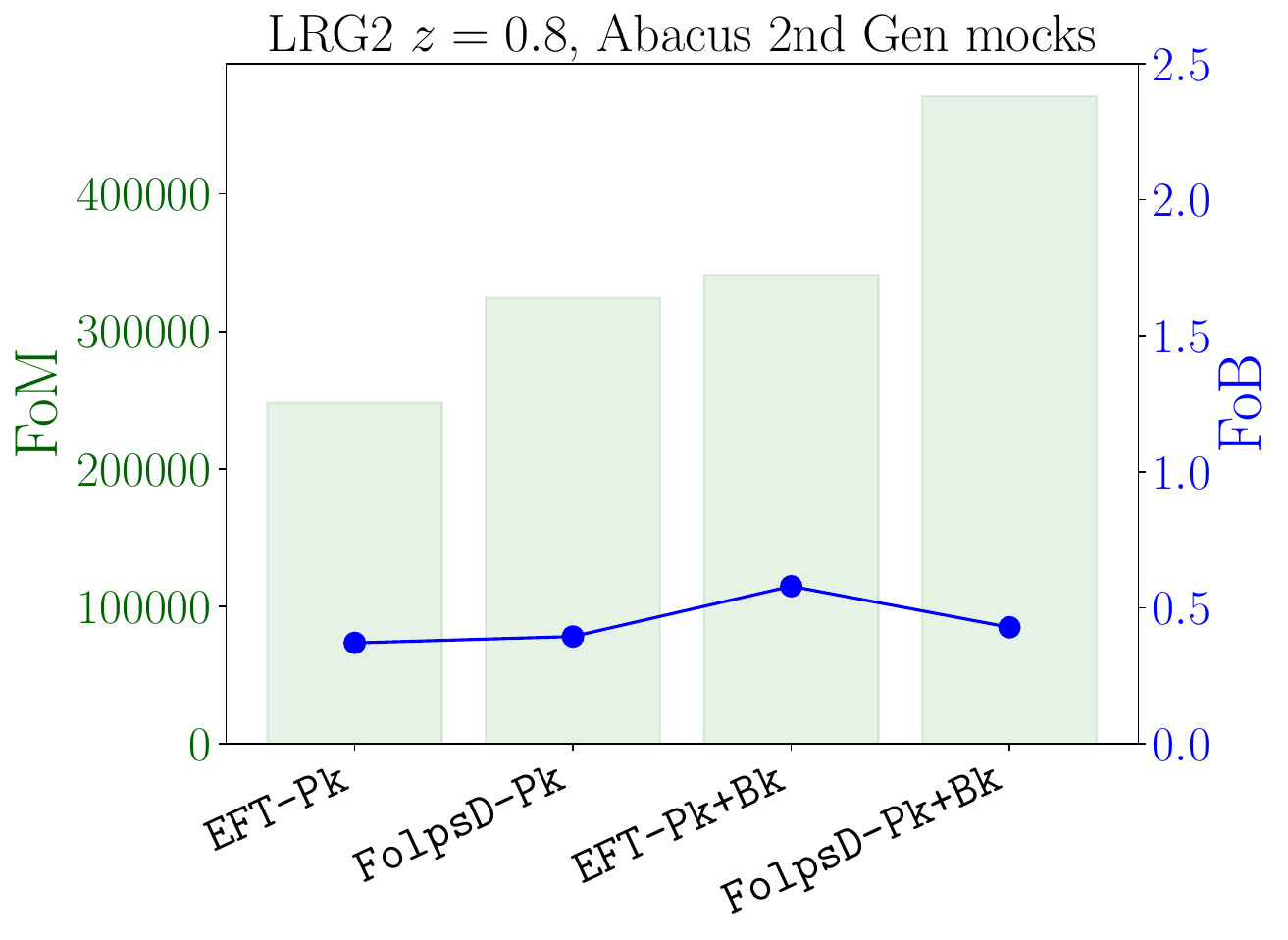}
    \includegraphics[width=2.94 in]{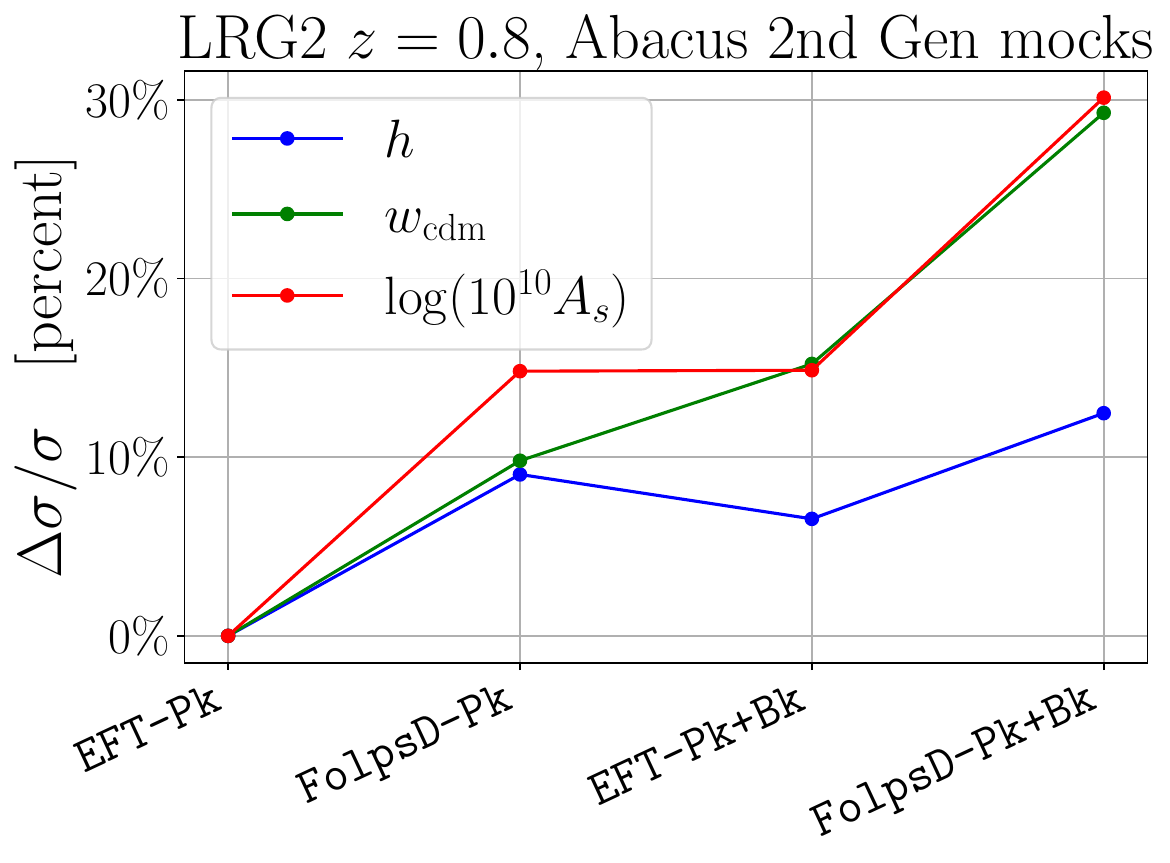}
    \caption{\textit{Left panel:} Figure of Merit (FoM) (green bars) and Figure of Bias (FoB) (blue dots connected by lines) for the four main modelings fitted to the LRG2 mock data. \textit{Right panel:} percentage improvement in the standard deviation of the fitted cosmological parameters $h$, $w_{cdm}$, and $A_s$, with the latter two showing up to a $30\%$ reduction when comparing \texttt{FolpsD-Pk+Bk} to \texttt{EFT-Pk}. }\label{fig:fom_fob}
	\end{center}
\end{figure*}

\begin{figure}
	\begin{center}
	\includegraphics[width=5.9 in]{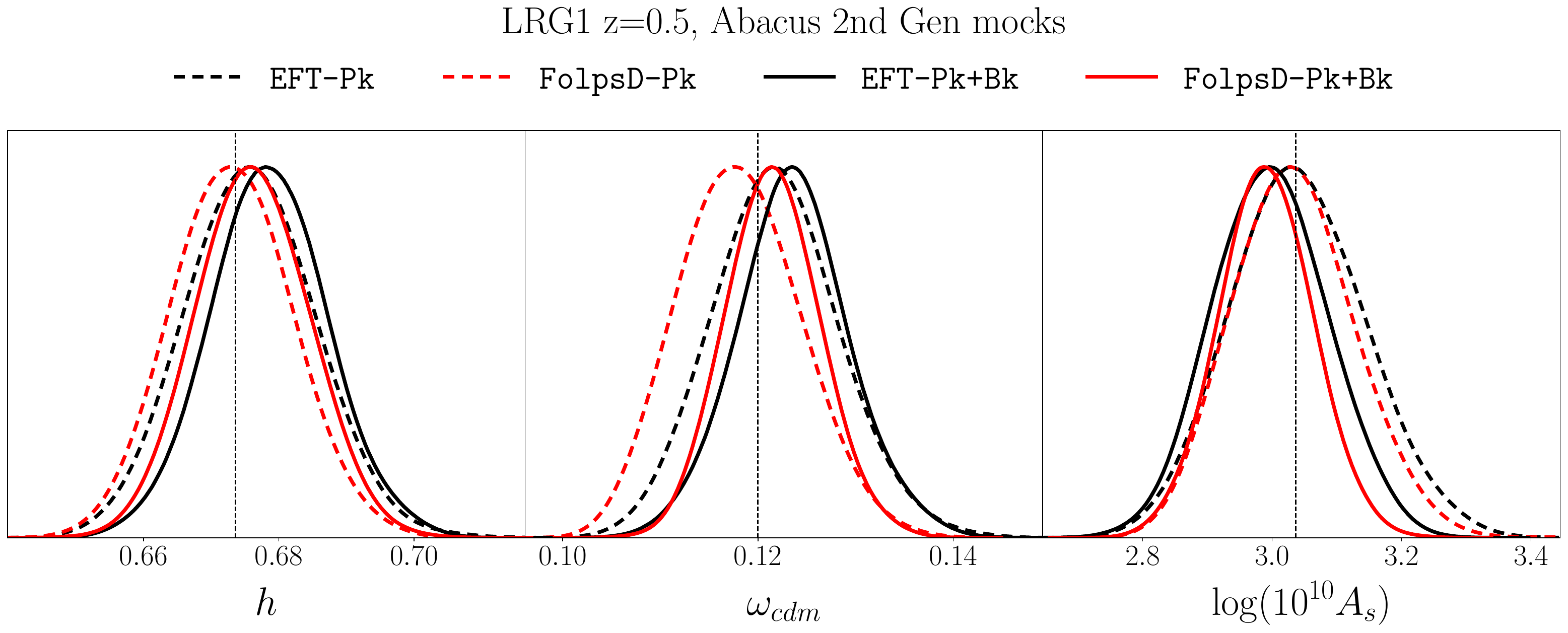} \\[10pt]
	\includegraphics[width=6.5 in]{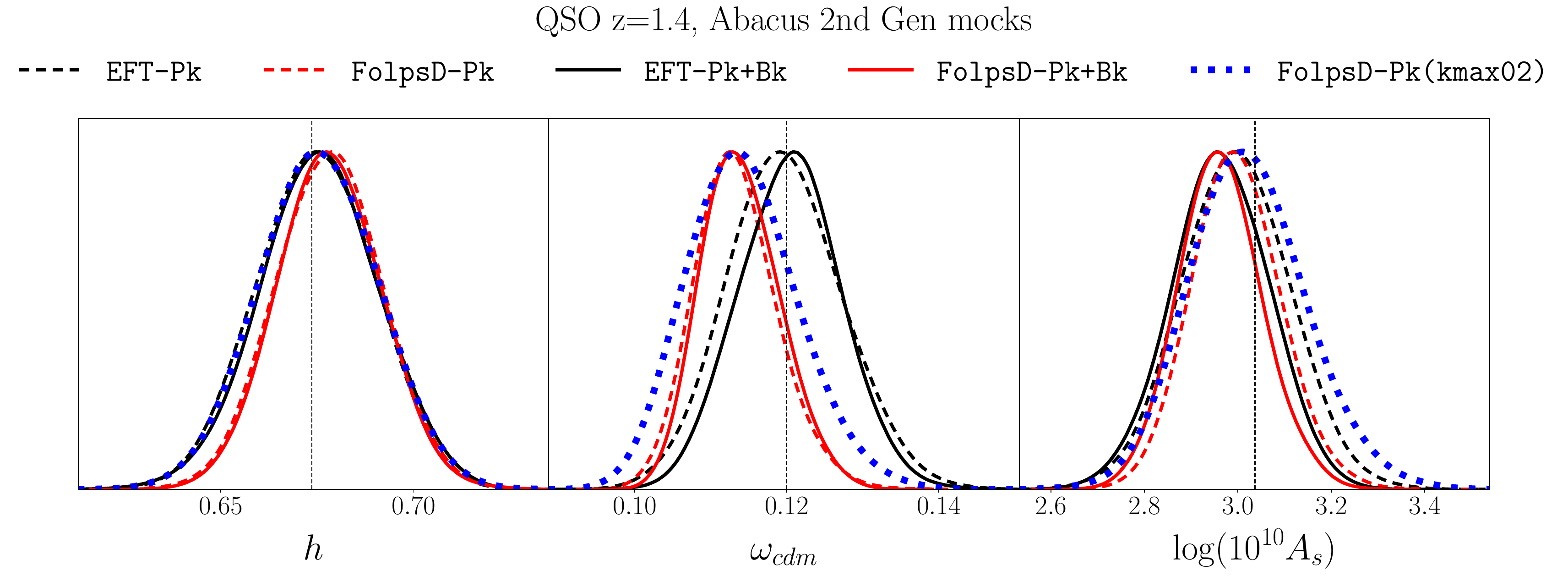}
    \caption{One-dimensional posteriors for the cosmological parameters $h$, $\omega_{cdm}$ and $\log(10^{10}A_s)$ from the fits to LRG1 (top panel) and QSO (bottom panel) mocks. We utilize the baseline models of \cref{table:BaselineAnalysis}. In the case of QSOs, we further fit with the model \texttt{FolpsD-Pk} but over a more conservative range with $k_\text{max}=0.201\ihMpc$.}\label{fig:LRG1-QSO_1Dim}
	\end{center}
\end{figure}

\begin{center}
\begin{table*}[t]
\scriptsize
\renewcommand{\arraystretch}{0.6}
\begingroup
\begin{spacing}{1.3}
{\setlength{\tabcolsep}{2pt}%
\begin{tabular*}{\textwidth}{@{\extracolsep{\fill}}l ccc ccc ccc@{}}
\toprule
& \multicolumn{3}{c}{LRG1} & \multicolumn{3}{c}{LRG2} & \multicolumn{3}{c}{QSO} \\
\cmidrule(lr){2-4} \cmidrule(lr){5-7} \cmidrule(lr){8-10}
& \small{68\% c.i.} & \small{$\sigma_\Omega$} & \small{$\Delta \Omega / \sigma_\Omega$}
& \small{68\% c.i.} & \small{$\sigma_\Omega$} & \small{$\Delta \Omega / \sigma_\Omega$}
& \small{68\% c.i.} & \small{$\sigma_\Omega$} & \small{$\Delta \Omega / \sigma_\Omega$} \\
\midrule \\[-3pt]

\texttt{EFT-Pk}:\\[-2pt]
\cmidrule{1-1}\\[-4pt]
 
$h$
& $ 0.6758^{+0.0095}_{-0.0096} $  &  $0.0097$  &  $0.23$
& $ 0.6754^{+0.0089}_{-0.009} $  &  $0.009$  &  $0.2$
& $ 0.6747^{+0.0156}_{-0.0157} $  &  $0.0153$  &  $0.07$
\\[6pt]
 
$\omega_{cdm}$
& $ 0.1218^{+0.0064}_{-0.0065} $  &  $0.0066$  &  $0.28$
& $ 0.1202^{+0.0062}_{-0.0067} $  &  $0.0065$  &  $0.03$
& $ 0.1197^{+0.0071}_{-0.008} $  &  $0.0076$  &  $-0.04$
\\[6pt]
 
$\log(10^{10}A_s)$
& $ 3.04^{+0.105}_{-0.105} $  &  $0.105$  &  $0.03$
& $ 3.054^{+0.099}_{-0.099} $  &  $0.099$  &  $0.18$
& $ 2.996^{+0.116}_{-0.116} $  &  $0.117$  &  $-0.35$
\\[6pt]
\phantom{a}\\

\texttt{FolpsD-Pk}:\\[-2pt]
\cmidrule{1-1}\\[-4pt]
 
$h$
& $ 0.6731^{+0.009}_{-0.009} $  &  $0.009$  &  $-0.05$
& $ 0.6742^{+0.0081}_{-0.0082} $  &  $0.0082$  &  $0.07$
& $ 0.6781^{+0.0138}_{-0.0139} $  &  $0.0138$  &  $0.32$
\\[6pt]
 
$\omega_{cdm}$
& $ 0.1181^{+0.006}_{-0.0067} $  &  $0.0063$  &  $-0.3$
& $ 0.1185^{+0.0058}_{-0.0058} $  &  $0.0058$  &  $-0.26$
& $ 0.1133^{+0.0048}_{-0.0059} $  &  $0.0055$  &  $-1.2$
\\[6pt]
 
$\log(10^{10}A_s)$
& $ 3.033^{+0.088}_{-0.101} $  &  $0.095$  &  $-0.03$
& $ 3.051^{+0.078}_{-0.09} $  &  $0.084$  &  $0.17$
& $ 2.997^{+0.09}_{-0.099} $  &  $0.096$  &  $-0.41$
\\[6pt]
\phantom{a}\\

\texttt{EFT-Pk+Bk}:\\[-2pt]
\cmidrule{1-1}\\[-4pt]
 
$h$
& $ 0.6786^{+0.0085}_{-0.0086} $  &  $0.0087$  &  $0.57$
& $ 0.6778^{+0.0083}_{-0.0085} $  &  $0.0084$  &  $0.51$
& $ 0.6757^{+0.0156}_{-0.0151} $  &  $0.0155$  &  $0.13$
\\[6pt]
 
$\omega_{cdm}$
& $ 0.1236^{+0.0055}_{-0.0056} $  &  $0.0058$  &  $0.62$
& $ 0.1222^{+0.0054}_{-0.0057} $  &  $0.0055$  &  $0.4$
& $ 0.1204^{+0.0066}_{-0.0069} $  &  $0.0067$  &  $0.06$
\\[6pt]
 
$\log(10^{10}A_s)$
& $ 2.997^{+0.086}_{-0.093} $  &  $0.089$  &  $-0.44$
& $ 3.008^{+0.085}_{-0.083} $  &  $0.084$  &  $-0.33$
& $ 2.963^{+0.107}_{-0.103} $  &  $0.105$  &  $-0.7$
\\[6pt]
\phantom{a}\\

\texttt{FolpsD-Pk+Bk}:\\[-2pt]
\cmidrule{1-1}\\[-4pt]
 
$h$
& $ 0.6762^{+0.0084}_{-0.0084} $  &  $0.0084$  &  $0.31$
& $ 0.6756^{+0.0078}_{-0.0078} $  &  $0.0078$  &  $0.25$
& $ 0.6781^{+0.0136}_{-0.0134} $  &  $0.0134$  &  $0.34$
\\[6pt]
 
$\omega_{cdm}$
& $ 0.1215^{+0.0049}_{-0.0049} $  &  $0.0049$  &  $0.32$
& $ 0.1216^{+0.0045}_{-0.0045} $  &  $0.0046$  &  $0.35$
& $ 0.114^{+0.0047}_{-0.0058} $  &  $0.0052$  &  $-1.15$
\\[6pt]
 
$\log(10^{10}A_s)$
& $ 2.991^{+0.073}_{-0.073} $  &  $0.073$  &  $-0.62$
& $ 3.016^{+0.07}_{-0.07} $  &  $0.069$  &  $-0.3$
& $ 2.962^{+0.085}_{-0.093} $  &  $0.09$  &  $-0.82$
\\[6pt]
\phantom{a}\\

\bottomrule
\end{tabular*}
}%
\end{spacing}
\endgroup
\caption{One-dimensional posterior means and 0.68 confidence intervals (c.i.) for the cosmological parameters $\Omega=\{h$, $\omega_{cdm}$, $\log(10^{10}A_s)$\}. We also show the errors $\sigma_\Omega$ given by the standard deviation, and the quantity $(\Omega^{\text{sims}} - \Omega^{\text{best fit}})/\sigma_\Omega$, which indicates how many sigmas away the recovered values are from the simulation truth.}
\label{table:BaselineAnalysis}
\end{table*}
\end{center}

\subsection{Adding the bispectrum}\label{subsec:addingBk}

Now, we add the Sugiyama bispectrum to the data. Throughout this work, we fit only to the diagonal of the multipoles; that is, we fit $B_{\ell_1\ell_2L}(k_1,k_2)$ with $k_1 = k_2$, since this choice has shown to give the largest signal-to-noise ratio \cite{Sugiyama:2018yzo}.

We first consider pure EFT modeling, fitting the power spectrum up to $k = 0.201\, h\text{Mpc}^{-1}$ and the EFT bispectrum up to $k = 0.12\, h\text{Mpc}^{-1}$ for the monopole $B_{000}(k,k)$, and up to only $k = 0.03\, h\text{Mpc}^{-1}$ for the quadrupole $B_{202}(k,k)$.
We call this combination \texttt{EFT-Pk+Bk}; see \cref{table:PTmodels}. We also fit \texttt{FolspD-Pk+Bk} which, as before, considers the power spectrum up to $k = 0.301\, h\text{Mpc}^{-1}$, while the bispectrum is used up to $k = 0.12\, h\text{Mpc}^{-1}$ for the monopole and up to $k = 0.08\, h\text{Mpc}^{-1}$ for the quadrupole.

In \cref{fig:LRG2}, we show contour plots of the posterior distributions. The strongest constraining power is obtained from \texttt{FolpsD} using both the power spectrum and the bispectrum, while the weakest constraints come from \texttt{EFT} model using only the power spectrum. We have tested the EFT bispectrum using the quadrupole up to $k=0.08 \ihMpc$, and although we observe a slight improvement on constraining power, some of the best-fit parameter values become biased from the simulation values, by up to  $1\sigma$. For this reason, we have fixed $ k_{\max} = 0.03 \ihMpc$ for this case.

From the right panel of \cref{fig:LRG2}, we further notice that the degeneracy between $\tilde{b}_2$ and $\tilde{b}_s$ differs between the \texttt{FolpsD-Pk} and \texttt{EFT-Pk+Bk} cases. This is likely because, in the latter case, these two parameters contribute primarily at large scales (both increasing the bispectrum amplitude, hence the negative degeneracy), whereas in the former case the additional constraining power necessarily arises from small scales. In contrast, the \texttt{FolpsD-Pk+Bk} analysis extracts additional information from both small and large scales.

In \cref{fig:fom_fob}, we summarize the constraining performance of the four main model configurations applied to the LRG2 mock sample. The top panel shows the Figure of Merit (FoM; green bars) and the Figure of Bias (FoB; blue line with dots) defined as
\begin{align}
   \text{FoM}&= (\det \ve C)^{-1}, \\ 
   \text{FoB}&= \Delta^\text{T} \,\ve C^{-1} \,\Delta . 
\end{align}
Here, $\ve C$ is the posterior covariance matrix of the parameters $\log(10^{10}A_s)$, $\omega_{cdm}$, and $h$, while $\Delta$ is the vector of differences between these parameters posterior means and the values of the simulations.

The inclusion of the bispectrum, both in the \texttt{EFT} and \texttt{FolpsD} models, enhances the constraining power, as indicated by the increasing FoM values shown in the left panel of \cref{fig:fom_fob}. 
Among all cases, \texttt{FolpsD-Pk+Bk} achieves the highest FoM, while the FoBs are similar for all cases, demonstrating that the combination of the FolpsD framework with bispectrum information provides the most robust and still reasonably unbiased parameter estimation, with specific numbers quantifying this in \cref{table:BaselineAnalysis}.

The right panel of \cref{fig:fom_fob} quantifies the relative improvement in the marginalized uncertainties of the cosmological parameters $h$, $w_{cdm}$, and $A_s$ with respect to the baseline \texttt{EFT-Pk} case. We find that the addition of the bispectrum significantly tightens the constraints, particularly on the amplitude $A_s$ and the matter abundance $w_{cdm}$ whose uncertainties are reduced by up to $30\%$. For parameter $h$ the improvement on the constraining power is around $13\%$. Hence the bispectrum information, especially when modeled with damping in the power spectrum, substantially increases the effective constraining power of the LRG2 analysis.

In \cref{table:BaselineAnalysis}, we show the $1\sigma$ confidence intervals for the cosmological parameters ($h$, $\omega_{cdm}$, $\log(10^{10}A_s)$) for each of the methods used so far in this section. We also show the errors given by the sample variance, $\sigma_\Omega$, and the number of $\sigma$ away from the simulations (sims) value, as given by the weighted distance between the best-fit estimated values and the true simulation values, $\Delta\Omega/\sigma_\Omega = (\Omega^{\text{sims}} - \Omega^{\text{best fit}})/\sigma_\Omega$. This table shows a systematic improvement in both accuracy and precision when enlarging the data vector, either by adding the bispectrum or by extending the power spectrum to higher wavelengths.  In \cref{sec:bkmax} we will show that for \texttt{folpsD} we can extend the fits to the LRGs bispectrum monopole to smaller scales. %

\subsection{Constraints from LRG1 and QSO} \label{subsec:LRG1QSO}

In \cref{fig:LRG1-QSO_1Dim}, we show the one-dimensional posterior distributions for the LRG1 (top panel) and QSO (bottom panel) Abacus second-generation mocks. Although it is not clear from the plots whether one method outperforms the others,\footnote{Degeneracies among bias parameters differ from one tracer to another. Therefore, stronger constraints on $b_2$ or $b_s$ do not necessarily translate into tighter constraints on $b_1$, nor on the cosmological parameters.
} an inspection of \cref{table:BaselineAnalysis} shows consistent improvements in constraining power relative to \texttt{EFT-Pk}, either when switching to damped models or when including bispectrum information. 

For LRG1, we observe reductions in parameter uncertainties of up to nearly 30\%, while the best-fit values lie within $0.6\sigma$ of the true values. We notice that the fitting patterns for the LRG1 targets are very similar to those of LRG2.

From the last column of \cref{table:BaselineAnalysis}, we observe that the largest shifts in the recovered parameters occur for QSOs when we consider the \texttt{FolpsD} methods, for both the power spectrum only analysis and when including the bispectrum. These biases in the recovered parameters when fitting with damped models is also shown in \cref{fig:LRG1-QSO_1Dim}, particularly in $\omega_{cdm}$ where they go beyond the $1\sigma$. Meanwhile, \texttt{EFT} do not exhibit any bias.  This behavior may be related to the large shot-noise contribution for QSOs. As shown in \cref{fig:tracers}, for the QSO monopole the Poisson shot noise becomes comparable to the measured power spectrum at $k \simeq 0.14 \ihMpc$, and exceeds it by roughly a factor of five at $k \simeq 0.3 \ihMpc$. 
Motivated by this behavior, we repeat the \texttt{FolpsD-Pk} fit using a reduced range, $0.02 < k < 0.201 \ihMpc$, matching the scale cut adopted for the \texttt{EFT-Pk} analysis. The corresponding one-dimensional posteriors are shown in the bottom panel of \cref{fig:LRG1-QSO_1Dim} (blue dotted lines). Even within this restricted range, the best-fit parameters remain shifted relative to the simulated values by the same amount as with the baseline $k$-range, in contrast to the unbiased results obtained with \texttt{EFT-Pk}. Therefore, the observed bias in the estimated parameter $\omega_{cdm}$ in \texttt{FolpsD} models is not driven by the choice of scale cuts.
Rather, the results seem to indicate that in regimes where the power spectrum becomes noise dominated, the phenomenological damping can bias the inferred cosmological parameters, as we discuss below.

\subsection{Limitations of \texttt{FolpsD} damping factor} \label{subsec:limitations}

As shown in the previous sections, damping models can lead to tighter constraints in $\Lambda$CDM scenarios with tracers that have a sufficiently high signal-to-noise, such as LRGs. The same ocurrs if the background evolution is $w_0w_a$CDM, as discussed in \cref{sec:w0wa}. However, this is not always the case in realistic situations. For example, for the QSO sample, which has a significantly noise-dominated power spectrum at relatively large scales, we find that the inclusion of damping functions can lead to biases even when conservative scale cuts are adopted ($k_\text{max}=0.201 \ihMpc$ for the power spectrum). We interpret this behavior as arising from the fact that, when the signal-to-noise ratio is low, the damping parameters are weakly constrained and add extra shape freedom to the model. In this regime, the damping can fit noise fluctuations instead of the true clustering signal, leading to shifts in the inferred cosmological parameters. In contrast, the EFT model remains more constrained because its counterterms enter in a controlled way. They do not fit the noise because they do not have that freedom.

A related limitation appears when inferring the sum of neutrino masses from mock catalogs, as explored in detail in \cref{app:neutrinos}. In this case, we find that including the damping in the power spectrum alone can degrade the constraints, although this effect is mitigated when the bispectrum is included in the joint analysis. We attribute this result to the fact that, in full-shape analyses, the neutrino mass is primarily constrained through the suppression of the power spectrum, rather than through the background expansion history. Since the damping factor acts in a similar manner by smoothing the BAO wiggles and suppressing the broadband, it reduces the sensitivity of the analysis to the neutrino mass.

Is expected that similar degradation of parameter estimations will be present generally in models that contain features that degenerate with the damping.  This is the case of models that introduce new physical scales in late-time structure formation, such as several modified gravity theories that enhance the power spectrum above certain wavenumber. Finally, one should notice that these limitations of the damping factor is not as marked when including the bispectrum, as we discuss in the following sections.

\section{Bispectrum beyond conservative scale cuts} \label{sec:bkmax}

\begin{figure*}
	\begin{center}
	\includegraphics[width=3.34 in]{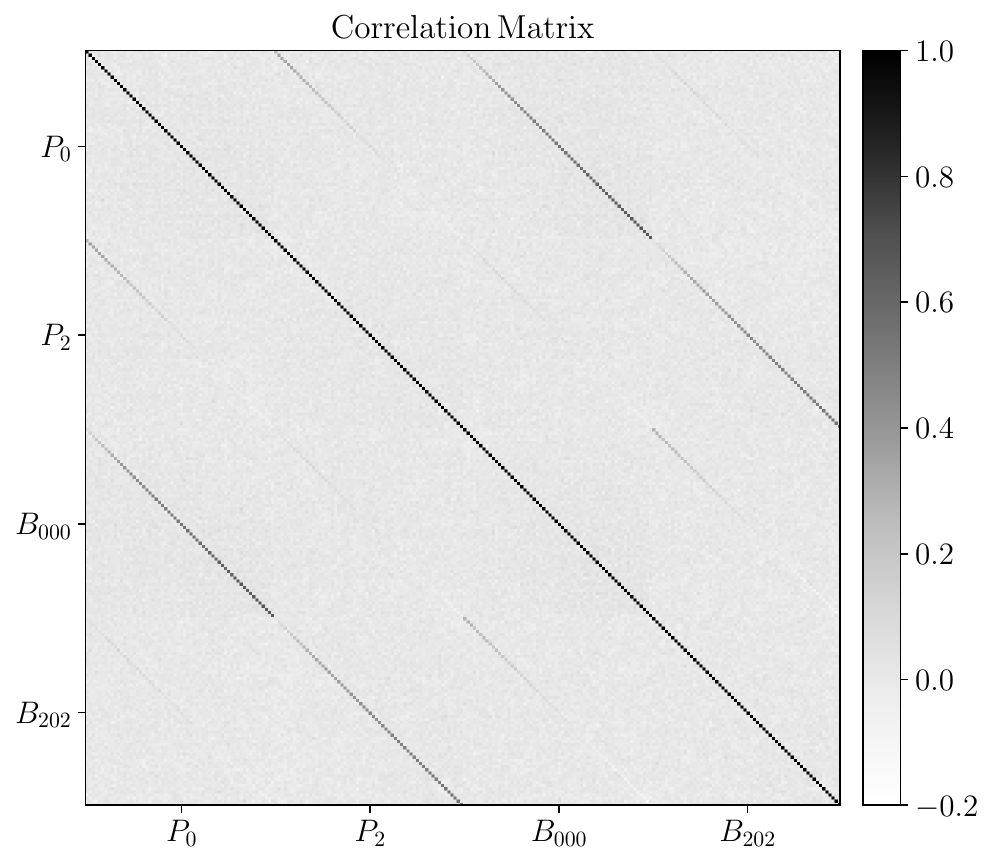}
	\includegraphics[width=2.54 in]{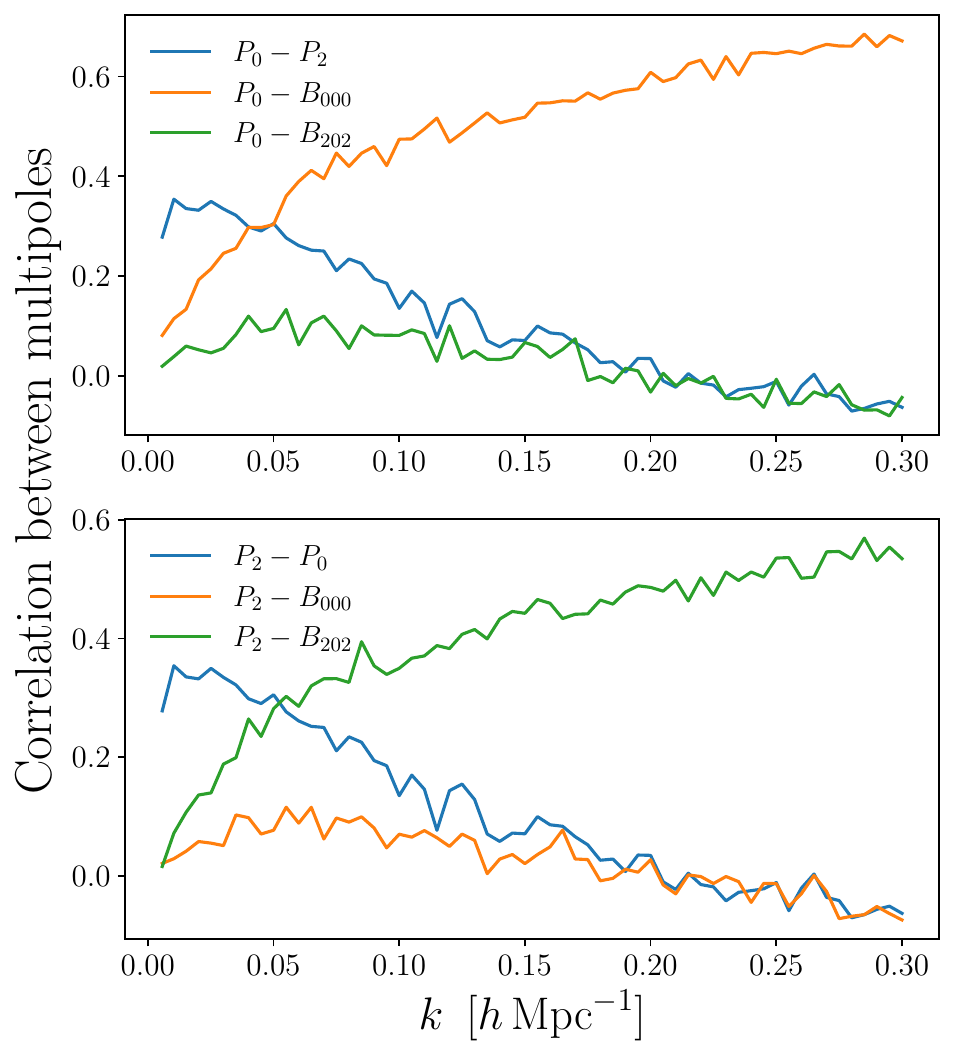}
    \caption{
    \textit{Left panel:} Correlation matrix obtained from 2000 EZmocks we use to construct the covariance matrix. \textit{Right panel:} correlation between different multipoles at the same $k$. This figure shows that the correlation between $P_0$ ($P_2$) is higher with $B_{000}$ ($B_{202}$) than with the rest of the multipoles.}\label{fig:correlationmatrix}
	\end{center}
\end{figure*}

\begin{figure*}
	\begin{center}
	\includegraphics[width=4.5 in]{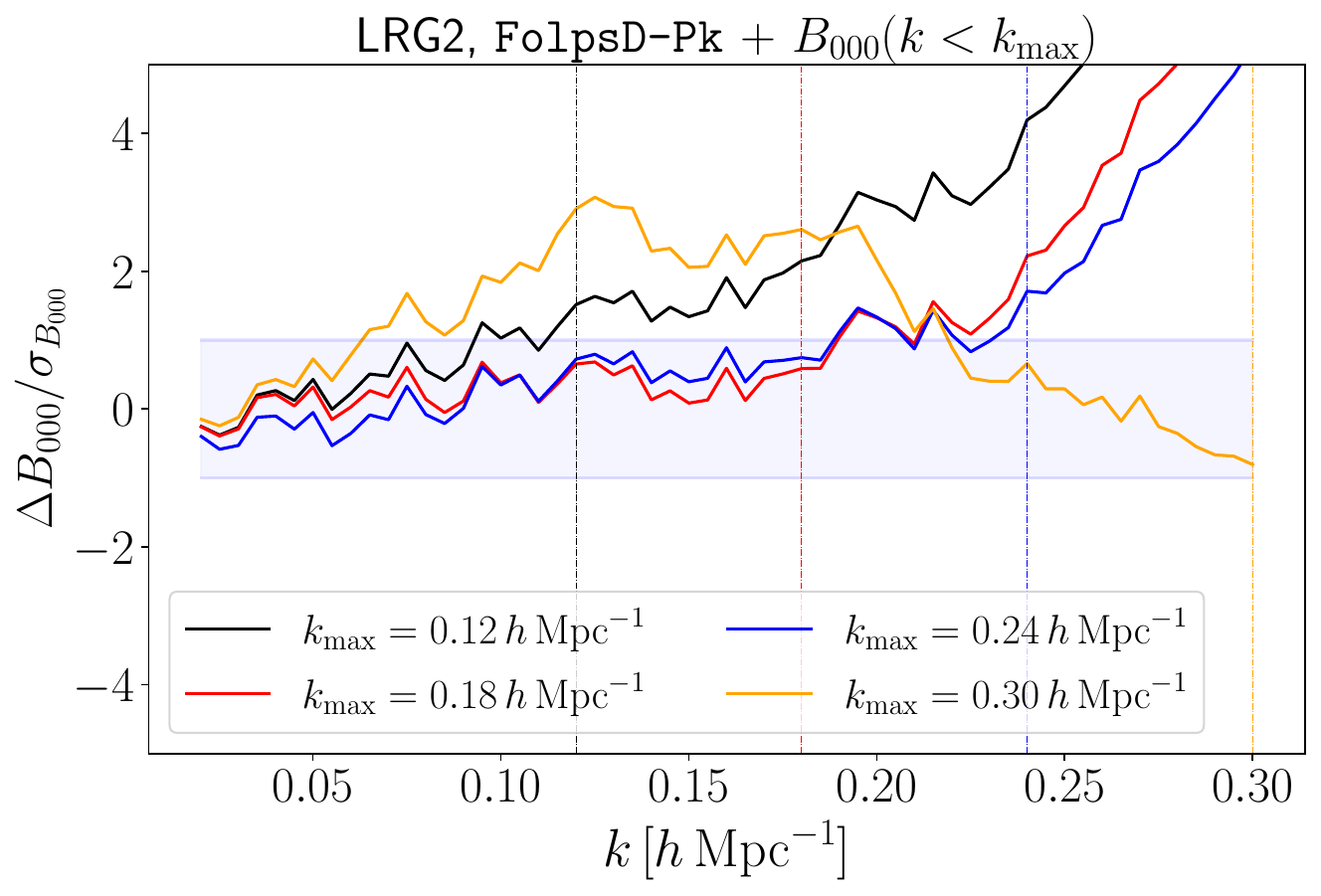}
    \caption{Bispectrum monopole $B_{000}$ evaluated at the best-fit parameters obtained from fits with different maximum wavenumbers $k_{\max}$. The fits are performed to the LRG2 Abacus mocks using the power spectrum monopole and quadrupole modeled with \texttt{FolpsD}, together with the bispectrum multipoles. For the bispectrum, $B_{000}$ is included over the range $0.02 < k < k_{\max}$, while $B_{202}$ is included over $0.02 < k < 0.03\,h\,\mathrm{Mpc}^{-1}$. The colored curves show the normalized difference $\Delta B_{000}/\sigma_{B_{000}}$ between the best-fit model and the data for fits with $k_{\max} = 0.12,\,0.18,\,0.24,$ and $0.30\,h\,\mathrm{Mpc}^{-1}$. Vertical dashed lines indicate the corresponding values of $k_{\max}$.}

    \label{fig:B000kmax}
	\end{center}
\end{figure*}

\begin{figure*}
	\begin{center}
	\includegraphics[width=5.2 in]{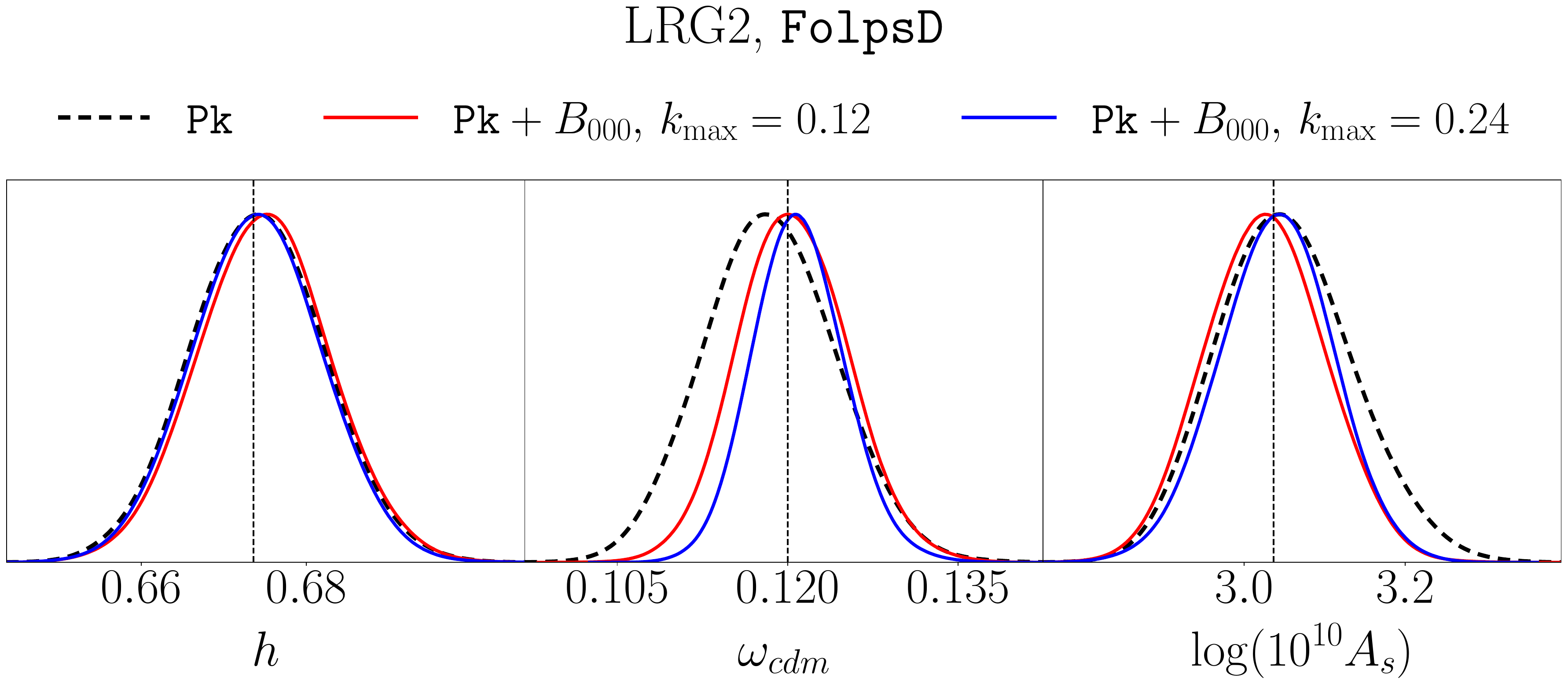} \\[10pt]
	\includegraphics[width=5.2 in]{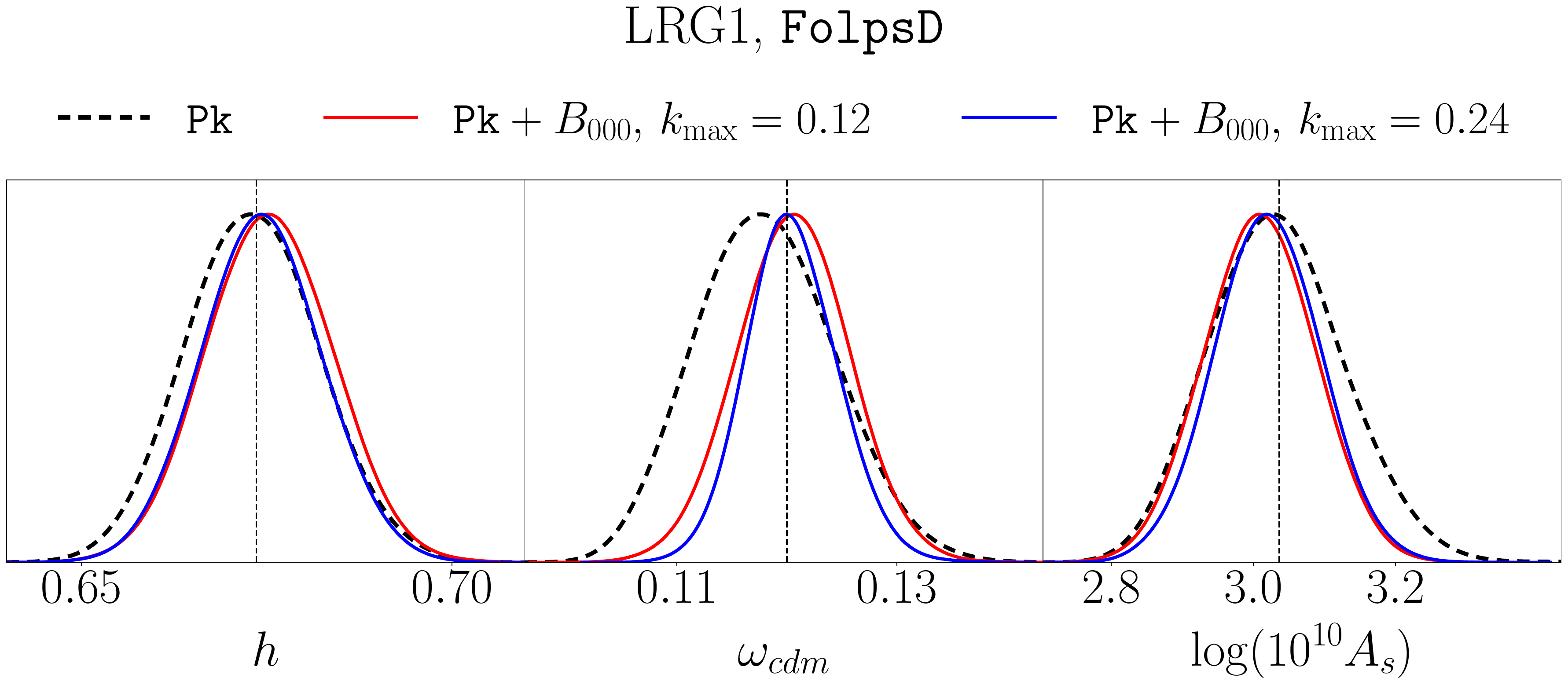} 
    \caption{
One-dimensional posterior distributions for $h$, $\omega_{\rm cdm}$, and $\log(10^{10}A_s)$ for the LRG samples. Top and bottom panels show results for LRG2 and LRG1 using the \texttt{FolpsD} model, respectively. In each case, constraints from the power spectrum alone (\texttt{Pk}; dashed black) are compared to those obtained by including the bispectrum monopole $B_{000}$ up to $k_{\max}=0.12\,h\,\mathrm{Mpc}^{-1}$ (red) and higher $k_{\max}$ values (blue). Vertical dashed lines indicate the fiducial values.
}
    \label{fig:B000kmax_1D}
	\end{center}
\end{figure*}

\begin{figure*}
	\begin{center}
	\includegraphics[width=5.2 in]{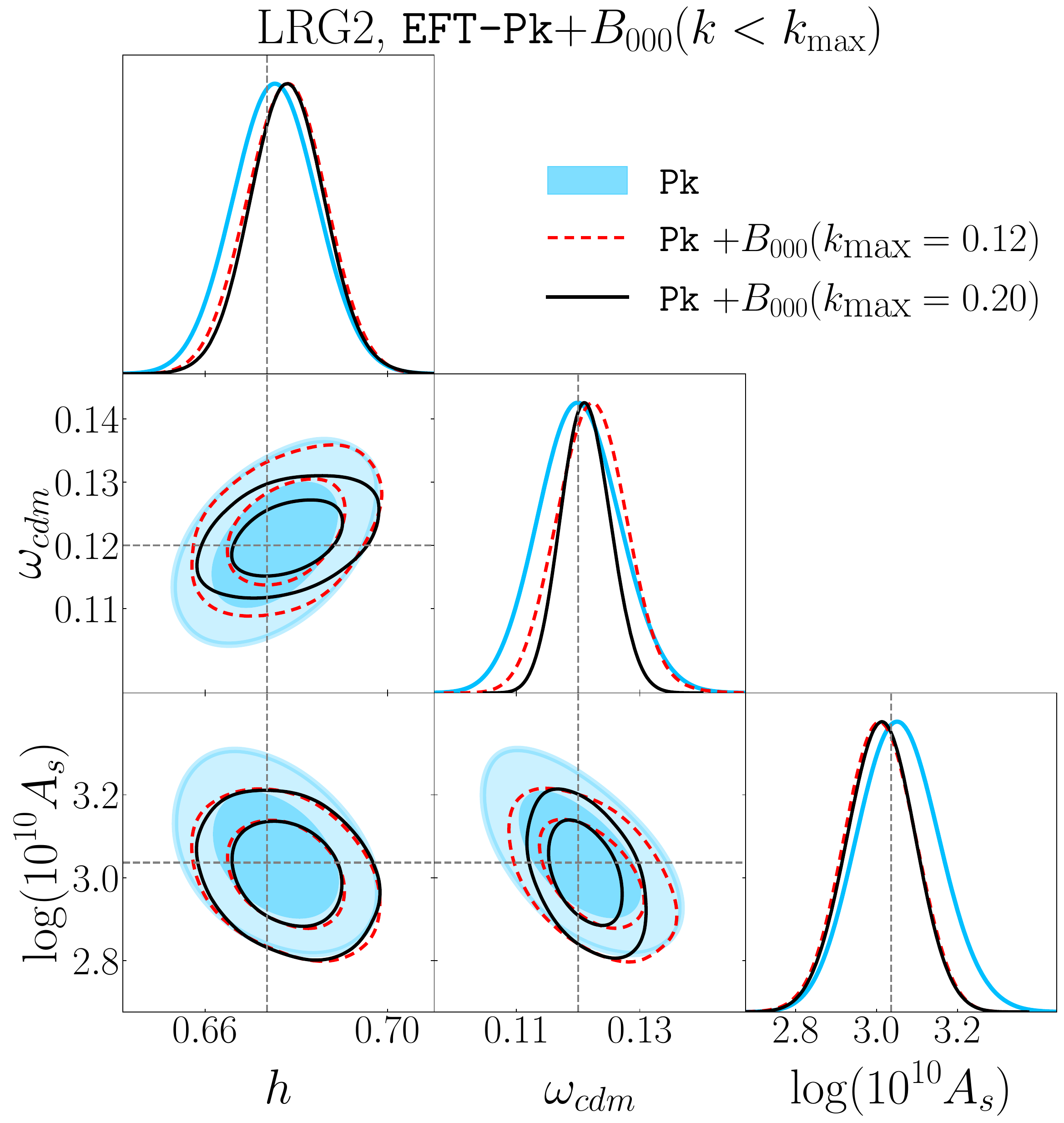} 
    \caption{
Contour plots for $h$, $\omega_{\rm cdm}$, and $\log(10^{10}A_s)$ for the LRG2 sample using \texttt{EFT} modeling, showing constraints from the power spectrum alone (\texttt{Pk}; blue) and including the bispectrum monopole $B_{000}$ up to $k_{\max}=0.12\,h\,\mathrm{Mpc}^{-1}$ (dashed red) and higher $k_{\max}$ values (solid black). Vertical dashed lines indicate the fiducial values.
}
    \label{fig:B000kmax_tri}
	\end{center}
\end{figure*}

In this section we move beyond the conservative $k$-ranges adopted in \cref{sec:mockresults}, quantifying both the biases and the constraining power obtained when extending the bispectrum analysis to higher $k_\text{max}$ in fits to LRG mocks.

In \cref{fig:correlationmatrix} we show the correlation matrix of the full data vector, including the monopole and quadrupole of both the power spectrum and bispectrum, evaluated over the range $k \in [0.02,0.301]\ihMpc$ for all multipoles. The left panel displays the full matrix, while the right panel shows the correlations between different multipoles evaluated at the same wavenumber. The largest correlations are found between $P_0$ and $B_{000}$ and between $P_2$ and $B_{202}$, reaching values above $0.5$ already at moderate wavenumbers. These correlations are more than twice as large as those between the power spectrum monopole and quadrupole. While the correlation between $P_0$ and $P_2$ decreases with increasing $k$, as expected from nonlinear redshift-space effects and random velocities, the correlation between $P_0$ and $B_{000}$ (and similarly between $P_2$ and $B_{202}$) instead grows at small scales. Since the cross-covariance between the power spectrum and bispectrum  vanishes in the Gaussian limit, and therefore at small $k$, this behavior reflects the increasing importance of nonlinear mode coupling, which correlates small-scale power with higher-order statistics.

This trend, in which the bispectrum becomes increasingly correlated with the power spectrum at small scales, may indicate a reduced amount of independent information gained by extending the analysis to higher $k$. 
To further explore this, we study the impact of varying the maximum wavenumber $k_\text{max}$ included in the bispectrum fit. Our findings are summarized in \cref{fig:B000kmax}. We fit \texttt{FolpsD} using the power spectrum monopole and quadrupole together with the bispectrum monopole $B_{000}$ and quadrupole $B_{202}$. For the bispectrum monopole, we use the range $0.02 < k/(h\text{Mpc}^{-1}) < k_\text{max}$, while for the quadrupole we restrict the range to $0.02 < k/(h\text{Mpc}^{-1}) < 0.03$, picking only two points of $B_{202}$. We make this choice because, as shown in \cref{app:B202}, the contribution of the quadrupole bispectrum to the LRG2 constraining power is small, and this choice allows us to isolate the effect of extending the monopole alone. We consider $k_\text{max} = 0.12, 0.18, 0.24,$ and $0.30 \ihMpc$.

In \cref{fig:B000kmax} we show the theoretical bispectrum monopole evaluated at the best-fit parameters for each choice of $k_\text{max}$. We subtract the LRG2 Abacus mock measurements and normalize the residuals by the errors derived from the EZmock covariance. The fits remain accurate up to $k_\text{max} = 0.24\ihMpc$, while a clear breakdown is observed for $k_\text{max} = 0.30\ihMpc$. The corresponding one-dimensional posteriors are shown in the top panel in \cref{fig:B000kmax_1D}. For $k_\text{max}=0.24\ihMpc$ we obtain the tightest constraints while remaining unbiased, yielding up to a $10\%$ improvement in $\omega_{cdm}$ relative to the $k_\text{max}=0.12\ihMpc$ case. We further show the case of \texttt{FolpsD-Pk} obtained in \cref{sec:mockresults}, which does not contain bispectrum information.  Extending the fit to $k_\text{max}=0.30\,h\mathrm{Mpc}^{-1}$ produces clear parameter shifts from the true values (not shown in the plot). The $1\sigma$ constraints obtained with \texttt{FolpsD}, including $P_0$ and $P_2$ over $0.02 < k < 0.301\,h\mathrm{Mpc}^{-1}$ and $B_{000}$ over $0.02 < k < 0.24\ihMpc$, are

\begin{equation}
\begin{aligned}
\text{parameter } (\Omega) \qquad & 68\%~\text{interval} \qquad (\sigma_\Omega) \qquad (\Delta\Omega/\sigma_\Omega) \\
h \;=\;\,& 0.6741^{+0.0079}_{-0.0078} \qquad 0.0078 \qquad 0.0628 \\
\omega_{cdm} \;=\;\, & 0.1210^{+0.0042}_{-0.0038} \qquad 0.0040 \qquad 0.2423 \\
\log(10^{10}A_s) \;=\;\,& 3.040^{+0.071}_{-0.070} \,\,\,\,\,\qquad 0.070 \,\,\,\qquad 0.0447
\end{aligned}
\end{equation}

Comparing with \cref{table:BaselineAnalysis} (block LRG2 --- \texttt{FolpsD-Pk+Bk}), we find that the constraints on $h$ and $\log(10^{10}A_s)$ remain essentially unchanged, while the uncertainty on $\omega_{cdm}$ improves by $13\%$. More significantly, the biases relative to the true simulation values are reduced. For $h$, the bias decreases from $0.25\sigma$ to $0.06\sigma$, for $\omega_{cdm}$ from $0.35\sigma$ to $0.24\sigma$, and for $\log(10^{10}A_s)$ from $0.30\sigma$ to $0.05\sigma$.
Considering the three parameters jointly, we obtain
\begin{align}
\text{LRG2}: \,\,\,\,\,\, &\text{\texttt{FolpsD-Pk}} \,\, + \,B_{000}(k_\text{max}=0.24\ihMpc):\nonumber\\[8pt]
&\mathrm{FoM} = 537944, \\[4pt] 
&\mathrm{FoB} =  0.2985, 
\end{align}
which correspond to a $14\%$ increase in FoM and a $30\%$ reduction in FoB relative to the \texttt{FolpsD-Pk+Bk} case in \cref{subsec:addingBk}.

Similar improvements are found for LRG1 mocks. The one-dimensional posteriors are shown in the middle panel of \cref{fig:B000kmax_1D}. As in the LRG2 case, the most constraining and least biased results are obtained for $k_\text{max}=0.24\ihMpc$, while extending to smaller scales leads to biasing of the recovered parameters. For LRG1 mocks we obtain
\begin{align}
\text{LRG1}: \,\,\,\,\,\, &\text{\texttt{FolpsD-Pk}} \,\, + \,B_{000}(k_\text{max}=0.24\ihMpc):\nonumber\\[8pt]
&\mathrm{FoM} = 540727, \\[4pt]
&\mathrm{FoB} =  0.1935.
\end{align}

The fact that \texttt{FolpsD} remains reliable up to $k_\text{max} \simeq 0.24\ihMpc$ for the LRG mocks is somewhat unexpected. One possible explanation for this extended reach is the inclusion of the damped power spectrum in the joint fit, which tightly constrains the second-order bias parameters $b_2$ and $b_s$. This could reduce degeneracies in the bispectrum likelihood, allowing the bispectrum monopole to be included to smaller scales without introducing statistically significant biases.
To test this hypothesis, we perform the same analysis using \texttt{EFT} modeling to compare with the damping-based case. In this case we find improvements up to $k_\text{max}=0.20\ihMpc$, while extending to $k_\text{max}=0.24\ihMpc$ leads to noticeable biases.  The results for this analysis are shown in \cref{fig:B000kmax_tri}. The corresponding figures of merit and bias are
\begin{align}
\text{LRG2}: \,\,\,\,\,\, &\text{\texttt{EFT-Pk}} \,\, + \,B_{000}(k_\text{max}=0.20\ihMpc):\nonumber\\[8pt]
&\mathrm{FoM} = 483111, \\[4pt]
&\mathrm{FoB} =  0.5987.
\end{align}

In the \texttt{EFT} case, the FoB remains comparable to that of the \texttt{EFT-Pk+Bk} analysis in \cref{sec:mockresults}, while the FoM increases by $41\%$. The fact that we can extend the valid range of scales of the theoretical tree-level (with added NLO counterterms)  Sugiyama bispectrum with no damping terms, may be a consequence of the strong correlations observed between the power spectrum and bispectrum monopoles. These correlations may help explain why extending the bispectrum to moderately smaller scales does not immediately introduce significant biases. That is, while damping factors can help extend the reached $k$-range, it is not the main driver of the improvement.

A similar analysis for QSO using \texttt{EFT} shows that the least biased constraints with significant constraining power are obtained for $k_\text{max}=0.16 \ihMpc$. The corresponding figures of merit and bias are

\begin{align}
\text{QSO}: \,\,\,\,\,\, &\text{\texttt{EFT-Pk}} \,\, + \,B_{000}(k_\text{max}=0.16\ihMpc):\nonumber\\[8pt]
&\mathrm{FoM} = 118926, \\[4pt]
&\mathrm{FoB} =  0.6757.
\end{align}
Extending the fit to higher $k_\text{max}$ increases the FoM but leads to a growth in FoB, approaching unity for $k_\text{max}=0.20 \ihMpc$. The gain in constraining power at smaller scales is accompanied by parameter shifts, signaling a breakdown of the model in the noise-dominated regime of the QSO sample.

\section{Constraints from the DESI DR1 full-shape power spectrum}\label{sec:DR1}
\begin{figure*}
	\begin{center}
	\includegraphics[width=5 in]{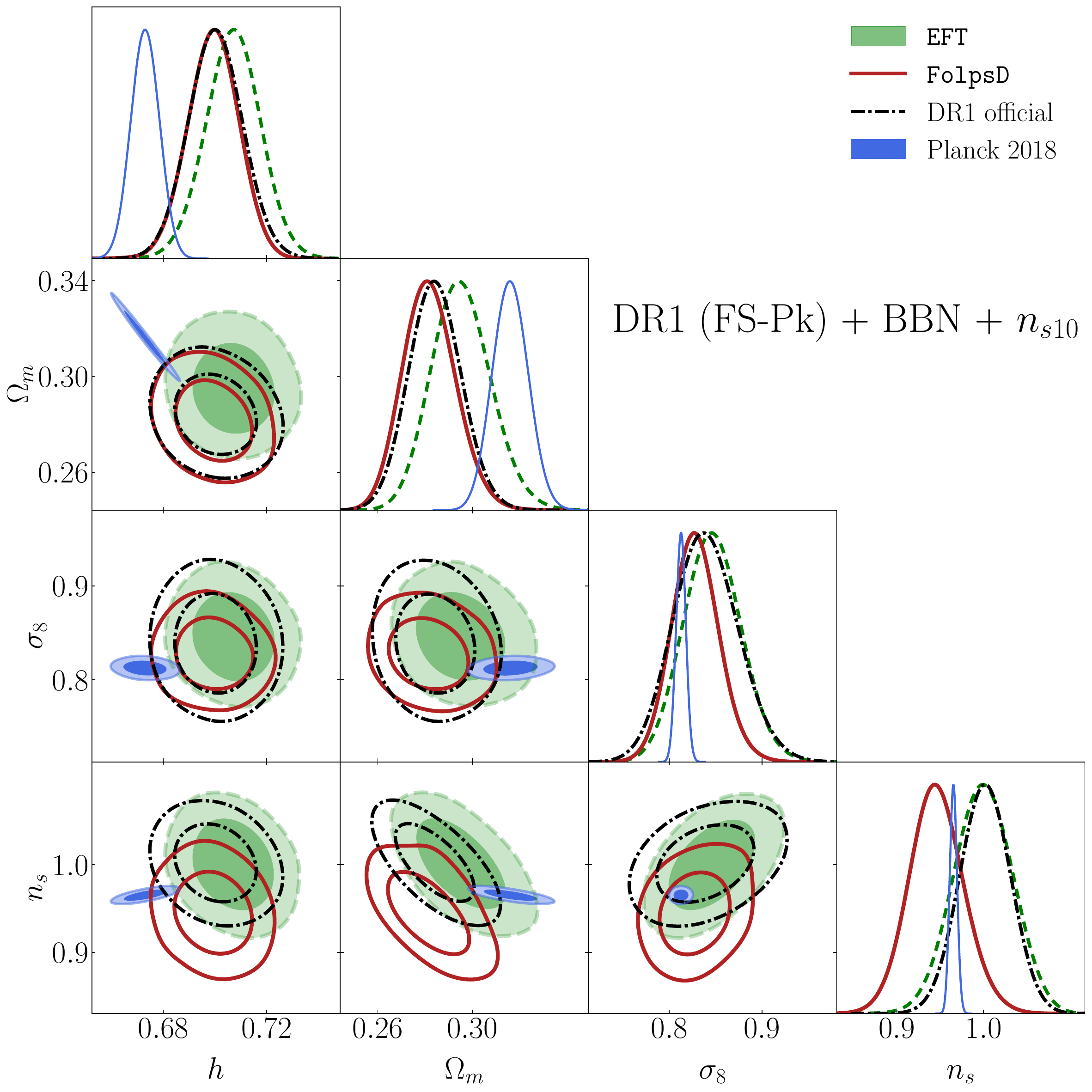}
    \caption{Posterior contour plots at $1$ and $2\sigma$ for  parameters $h$, $\Omega_{m}$, $\sigma_8$ and $n_s$. The data combination adopted is are DESI DR1 full shape data +  BBN prior on $\omega_b$ + Planck $n_{s10}$ prior on the scalar spectral index of primordial density perturbations $n_s$. We test the models \texttt{EFT} (with power spectrum up to $k_\text{max}=0.2\ihMpc$) and  \texttt{FolpsD} (with power spectrum up to $k_\text{max}=0.3\ihMpc$). We add the DR1 official chain and the Planck 2018 results. 
    }\label{fig:DR1_tri_priordoc}
	\end{center}
\end{figure*}

\begin{center}
\begin{table*}[t]
\scriptsize
\begingroup
\begin{spacing}{1.3}
       \begin{tabular} { l  c c c}
\toprule       
\noalign{\vskip 3pt}\\[-45pt]

  &  \phantom{abcdefghabcdefghabcdefgh} &  \phantom{abcdefghabcdefghabcdefgh} &  \phantom{abcdefghabcdefghabcdefgh}\\

  &  \texttt{EFT-Pk} &  \texttt{FolpsD-Pk} (improvement over \texttt{EFT-Pk})&  DR1 Official\\[2pt]
\hline\\[-10pt]
{\phantom{a}$h$\phantom{abcdefgh}            } & $0.707\pm 0.010          $ & $0.6994\pm 0.0096     \quad (7.5\%)       $ & $0.700\pm 0.010            $\\[4pt]

{\phantom{a}$n_s$          } & $0.9998\pm 0.033           $ & $0.946\pm 0.031     \quad (7.6\%)            $ & $1.002\pm 0.029            $\\[4pt]

{\phantom{a}$\omega_{cdm}$ } & $0.1248\pm 0.0071          $ & $0.1151^{+0.0053}_{-0.0060}\quad (19.4\%)            $ & $0.1167\pm 0.0060          $\\[4pt]

{\phantom{a}$\log(10^{10}A_s)$} & $3.028\pm 0.098            $ & $3.136\pm 0.088  \quad (9.8\%)$ & $3.10\pm 0.10              $\\[4pt] 

\hline\\[-10pt]


{\phantom{a}$\Omega_m$     } & $0.295^{+0.011}_{-0.013}   $ & $0.282^{+0.010}_{-0.012}     \quad (10.7\%)   $ & $0.284\pm 0.011            $\\[4pt]

{\phantom{a}$\sigma_8$     } & $0.846\pm 0.031            $ & $0.828\pm 0.025     \quad (18.5\%)           $ & $0.839^{+0.032}_{-0.036}   $\\[2pt]

\bottomrule
\end{tabular}
\end{spacing}
\endgroup
\caption{DESI DR1 one-dimensional posterior means and 68\% confidence intervals for  the varied parameters $h$, $\omega_{cdm}$, $\log(10^{10}A_s)$ and $n_s$, and for the derived parameters $\Omega_m$ and $\sigma_8$. The fits are performed using the DESI DR1 full-shape data, a BBN prior on the baryon abundance, and a Planck 2018 $n_s$ prior on the spectral index.  The percentages in parentheses denote the improvement when moving from \texttt{EFT} to \texttt{Folps-D}. 
}
\label{table:DR1Analysis-priorsdoc}
\end{table*}
\end{center}

\begin{figure*}
	\begin{center}
	\includegraphics[width=6 in]{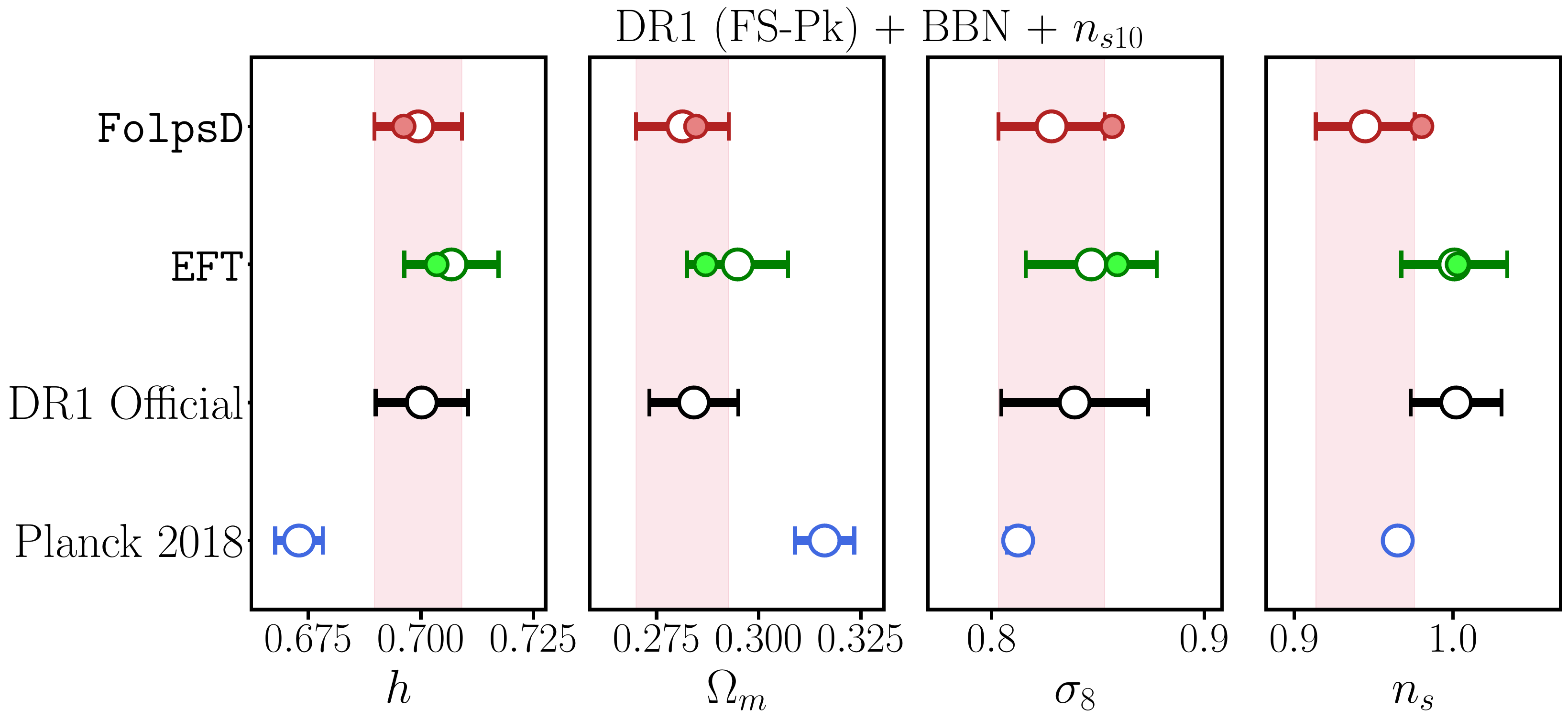}
    \caption{Whisker plots showing the marginalised mean (open circles) and maximum a posteriori (MAP) estimate (filled circles) together with its $1\sigma$ uncertainty for parameters $h$, $\Omega_{m}$, $\sigma_8$ and $n_s$. The datasets are DESI DR1 full-shape data +  BBN prior on $\omega_b$ + Planck $n_{s10}$ prior on the spectral index. We test the models \texttt{EFT} (power spectrum up to $k_\text{max}=0.2 \ihMpc$) and  \texttt{FolpsD} (with power spectrum up to $k_\text{max}=0.3 \ihMpc$). We add the DR1 official and the Planck 2018 results. For reference, we extend the \texttt{FolpsD} whisker with a vertical shadow in red.}\label{fig:DR1_whiskers_priordoc}
	\end{center}
\end{figure*}
In this section, we fit the DESI DR1 LSS catalogs presented in \cite{DESI2024.II.KP3}. We use the same samples as in the full-shape analysis of \cite{DESI2024.V.KP5}, which consist of 3,854,488 galaxies and 856,652 quasars covering a total effective volume of $V_\text{eff} = 17.5\,\text{Gpc}^3$. The data are split into six redshift bins: one bin corresponding to the Bright Galaxy Survey (BGS) sample at an effective redshift $z_\text{eff} = 0.295$; three bins of LRGs at $z_\text{eff} = 0.510$, $z_\text{eff} = 0.706$, and $z_\text{eff} = 0.919$; one bin of Emission Line Galaxies (ELGs) at $z_\text{eff} = 1.317$; and one bin of QSOs at $z_\text{eff} = 1.491$.\footnote{\revised{The DESI DR1 Full Shape and BAO clustering products, including data vectors, window functions and covariances are publicly at \url{https://data.desi.lbl.gov/doc/releases/dr1/vac/full-shape-bao-clustering/}.}}

In addition to the DESI data, we include a prior on the baryon density $\omega_b$, obtained by comparing the relative abundances of light elements with the predictions of Big Bang Nucleosynthesis (BBN) theory  \cite{Schoneberg:2019wmt}.
We also include a Gaussian prior on the spectral index, centered on the best-fit Planck 2018 value \cite{2020A&A...641A...6P} and with a standard deviation ten times larger than its quoted $1\sigma$ uncertainty. These and the remaining parameters and priors are listed in \cref{table:parameters_priors}.

We infer parameters using the \texttt{EFT} and \texttt{FolpsD} models based on the power spectrum alone, adopting the scale ranges $0.02 < k/(h\text{Mpc}^{-1}) < 0.201$ and $0.02 < k/(h\text{Mpc}^{-1}) < 0.301$, respectively. We use the same power spectrum measurements, window functions, and covariance matrices as in the official DR1 analysis \cite{DESI2024.V.KP5}. 

In \cref{fig:DR1_tri_priordoc}, we show contour plots for the varied parameters $h$ and $n_s$, as well as for the derived parameters $\Omega_m$ and $\sigma_8$, chosen to facilitate comparison with previous studies, in particular earlier DESI DR1 results \cite{DESI2024.V.KP5}. The top panel of \cref{table:DR1Analysis-priorsdoc} presents the posterior means and 68\% confidence intervals for all varied parameters, including $\log(10^{10}A_s)$ and $\omega_{cdm}$, and for the derived parameters in the bottom panel. The percentage shown in parentheses is the relative improvement on constraining power, given by $(1 - \sigma^{\texttt{FolpsD}}_\Omega / \sigma^{\texttt{EFT}}_\Omega)$, where $\sigma_\Omega$ denotes the standard deviation of the posterior of the parameter $\Omega$.

\Cref{fig:DR1_whiskers_priordoc} shows whisker plots for the \texttt{EFT} and \texttt{FolpsD} models. The whiskers indicate the $1\sigma$ uncertainties of the posterior distributions centered on their means, which are denoted by empty circles. The red vertical band follow the results obtained with \texttt{FolpsD-Pk}. This represents the most constraining model considered here, yielding improvements of up to $20\%$ relative to \texttt{EFT-Pk}. When compared to the DESI DR1 ``Official'' chains \cite{DESI2024.V.KP5}, the improvement reaches up to $26\%$ in the case of $\sigma_8$. We notice that our analysis adopts different modeling choices and priors than those used in \cite{DESI2024.V.KP5}. Since full-shape results are sensitive to the choice of priors,  a direct comparison between \texttt{FolpsD} and the official DESI-DR1 results should be interpreted with caution.

We further compute Maximum A posteriori (MAP) estimates, using the  \code{iminuit}\footnote{\href{https://github.com/scikit-hep/iminuit}{https://github.com/scikit-hep/iminuit}} code \cite{James:1975dr}. Projection effects are manifested as discrepancies between the MAP values (shown as filled circles in \cref{fig:DR1_whiskers_priordoc}) and the posterior maximum distributions (empty circles). \revised{In particular, for $\sigma_8$ we observe a $\sim 1\sigma$ shift between the posterior mean and the MAP value, which may suggest a residual prior volume effect.} In our analysis, all differences lie within the 1$\sigma$ intervals. However, projection effects are expected to be larger in models beyond $\Lambda$CDM, which we do not consider in this section. As an addendum, in \cref{sec:w0wa} we implement ShapeFit analyses to study the $w_0w_a$CDM model.

\section{DESI DR1 ShapeFit constraints on $w_0w_a$CDM}\label{sec:w0wa}

\begin{figure}
    \centering
    \includegraphics[width=0.49\textwidth]{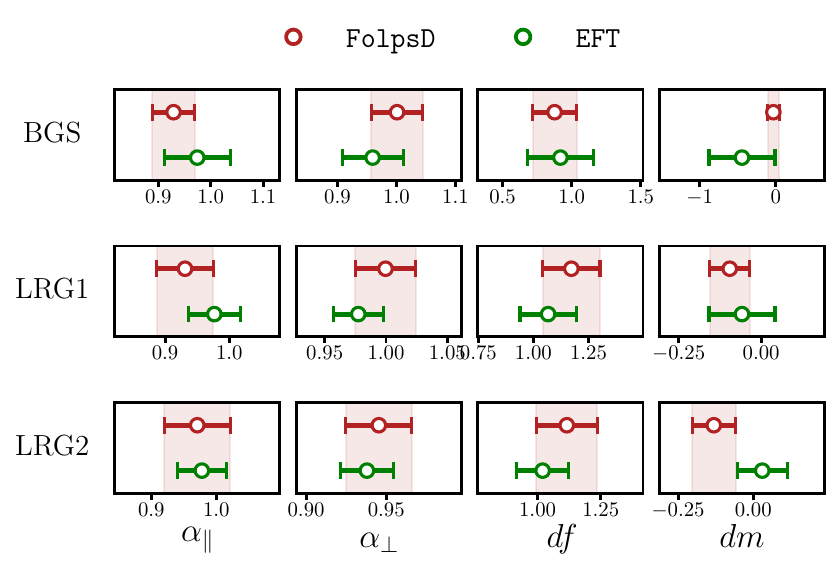}\hfill
    \includegraphics[width=0.49\textwidth]{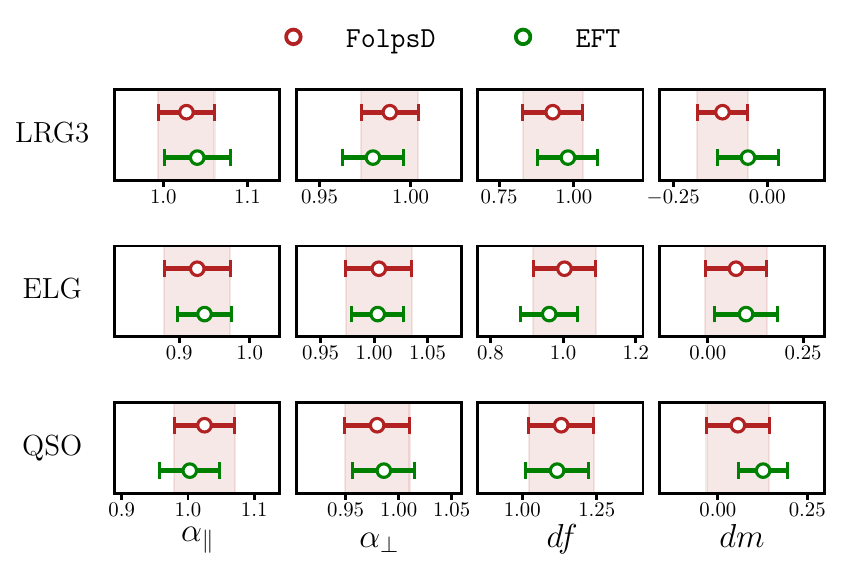}
    \caption{Whisker plots for the ShapeFit analysis, showing marginalised means (open circles) and $1\sigma$ uncertainties for $\alpha_{\parallel,\perp}$, $df \equiv f/f_{\rm fid}$, and $dm \equiv m - m_{\rm fid}$ across the six DESI DR1 redshift bins. The data and model setup follow \cref{fig:DR1_whiskers_priordoc}; \texttt{FolpsD} whiskers are highlighted with red vertical shading.}
    \label{fig:SF_compressed}
\end{figure}

\begin{figure}
  \centering
  \includegraphics[height=7cm]{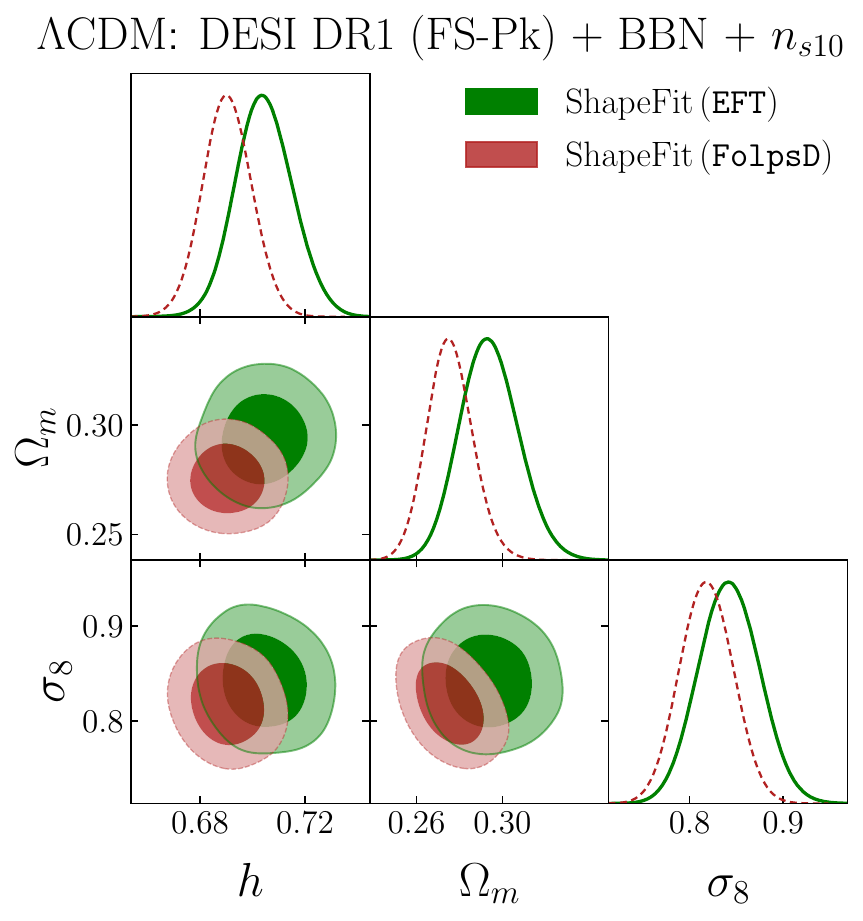}\hspace{0.6cm}%
  \includegraphics[height=7cm]{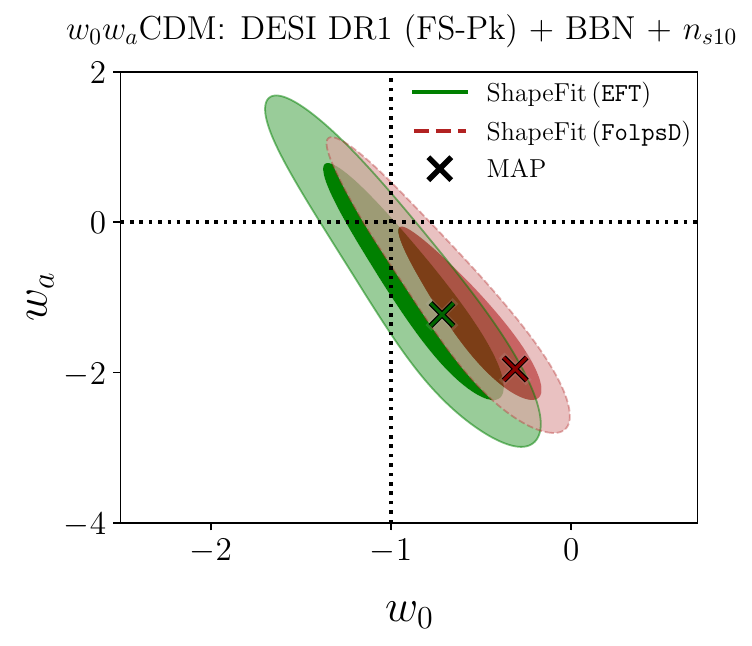}
  \caption{Two-dimensional posterior contours at 68\% and 95\% c.l. from the ShapeFit analysis on DESI DR1 full-shape data. The data also include a BBN prior on $\omega_b$ and a Planck $n_{s10}$ prior. We compare the standard EFT approach, using power spectrum up to $k_\text{max}=0.201\ihMpc$, 
  with \texttt{FolpsD} (up to $k_\text{max}=0.301\ihMpc$) for $\Lambda$CDM (left panel) and $w_0w_a$CDM (right panel). 
  }
  \label{fig:SF_cosmo}
\end{figure}

So far we have considered the so-called full-modeling approach to full-shape analyses, which in \cite{DESI2024.V.KP5,DESI2024.VII.KP7B} refers to the use of EFT-based models. A different approach to full shape is given by ShapeFit \cite{Brieden:2021edu, Brieden:2022lsd}, which compresses the clustering information into a set of physically motivated summary parameters: the AP scalings $\alpha_\parallel$ and $\alpha_\perp$, the growth amplitude combination $f\sigma_{s8}$, and the shape parameters $m$ and $n$, which are designed to capture information related to matter–radiation equality and the primordial spectral index, respectively. Here $\sigma_{s8}$ denotes the ShapeFit amplitude parameter, corresponding to a rescaled version of $\sigma_8$ defined relative to the fiducial power spectrum template so that variations in the shape parameter $m$ do not absorb part of the growth signal. In this framework the linear power spectrum is modified as
\begin{equation}\label{eq:shapefit_transform}
   P'_{\rm lin}(k)  =   P_{\rm lin}(k)\exp \left\{\frac{m}{a} \tanh\left[ a \log\left( \frac{k}{k_p}\right)\right] 
    +
    n \log\left( \frac{k}{k_p}\right)
    \right\},
\end{equation}
where $a$ and $k_p$ are fixed to $a = 0.6$ and $k_p = 0.03\, h^\text{fid}\, \text{Mpc}^{-1} \approx \pi/r^\text{fid}_d$ \cite{Brieden:2021edu}. Because $m$ and $n$ are strongly anti-correlated (e.g., \cite{KP5s3-Noriega}), we vary only $m$ and fix $n=0$, following \cite{DForero2025, NovellMasot:2025fju}.

ShapeFit operates on top of an underlying theoretical model, which may be TNS (as in the original ShapeFit analyses) or EFT. In this section, we compare the constraining power of the ShapeFit formalism using both \texttt{EFT-Pk} and \texttt{FolpsD-Pk} models, adopting power spectrum scale cuts $0.02<k/(\ihMpc)<0.201$ and $0.02<k/(\ihMpc)<0.301$, respectively, when fitting the DESI DR1 power spectrum. 

\Cref{fig:SF_compressed} shows the whisker plots of the ShapeFit parameters for the six DESI DR1 tracer samples, inferred using both \texttt{EFT-Pk} and \texttt{FolpsD-Pk} modeling. The constraints are consistent at the $1\sigma$ level across all tracers and parameters. Notably, \texttt{FolpsD} yields tighter constraints on the shape parameter $m$ than \texttt{EFT}, with the largest improvements observed for the BGS and LRG samples. The QSO results are also consistent between the two approaches, despite the lower signal-to-noise of this tracer.

We then map the ShapeFit compressed parameters onto cosmological parameters. First, we perform this map to the $\Lambda$CDM parameters. The left panel of \cref{fig:SF_cosmo} shows the obtained two-dimensional posteriors, where we observe improved constraints, mainly on $\sigma_8$ and $\Omega_m$ when using \texttt{FolpsD} relative to \texttt{EFT}.

An appealing feature of the ShapeFit formalism is that it shows indications of being less sensitive to prior-volume effects, as discussed in \cite{DForero2025}. Consequently, nuisance reparametrization, as in \cite{KP5s2-Maus,2025arXiv250909562T}, may not be mandatory. We therefore map the inferred ShapeFit parameters onto the $w_0w_a$CDM model for both \texttt{FolpsD} and \texttt{EFT}. The resulting two-dimensional contours in the $w_0$---$w_a$ space are shown in the right panel of \cref{fig:SF_cosmo}, where we find a noticeable reduction in the widths of $w_0$ and $w_a$ for \texttt{FolpsD} compared to \texttt{EFT}. Specifically, the 68\% confidence intervals are
\begin{align}
\text{\texttt{EFT}:}    & \quad w_0 = -0.87^{+0.36}_{-0.28}\;, \qquad w_a = -0.86^{+0.88}_{-1.20}, \\
\text{\texttt{FolpsD}:} & \quad w_0 = -0.57^{+0.31}_{-0.20}\;, \qquad w_a = -1.21^{+0.58}_{-0.92}.
\end{align}
This corresponds to 15\% and 21\% tighter constraints on $w_0$ and $w_a$, respectively. Jointly, the FoM increases from $\mathrm{FoM}^{\texttt{EFT}} = 68$ to $\mathrm{FoM}^{\texttt{FolpsD}} = 161$.


We further compute the MAP estimates, which lie within their respective $1\sigma$ contours, indicating that prior-volume effects are small, if any. 

If the analysis presented in this section is confirmed within a more detailed analysis, possibly involving BAO data, they would indicate the first deviation from $\Lambda$CDM at the level of slightly above $1\sigma$ using LSS data alone.

\section{Conclusions}\label{sec:conclusions}

In this work we study the joint use of the galaxy power spectrum and bispectrum for full-shape analyses in the context of DESI, with the goal of understanding how information from different scales and statistics can be combined in a controlled way. To this end, we extend the model used in the original \folps\ code and test it on DESI-like LRG and QSO mock catalogs. The new model, called \texttt{FolpsD}, also includes a phenomenological damping factor along the LoS to account for the apparent suppression of small-scale structure when galaxies are observed in redshift space. This prescription goes beyond the Wilsonian EFT framework and depends on the assumed functional form, which we take to be Lorentzian. To evaluate its impact on real data, we apply both the \texttt{EFT} and \texttt{FolpsD} power spectrum models to DESI DR1. Assuming a $\Lambda$CDM cosmology, we find that the damping improves the constraints relative to the standard EFT analysis, with an improvement of about $20\%$ in $\sigma_8$.

We also study the limitations of the damping factor. Although Lorentzian damping is widely used in the literature, to our knowledge this is the first detailed assessment of its limitations using realistic spectroscopic data. We find that for noise-dominated targets, such as DESI QSOs, the damping can degrade parameter estimation. Similar effects arise in scenarios that introduce new physical scales in late-time large-scale structure, such as massive neutrinos or some modified gravity models. We therefore recommend caution when applying this approach in such cases.

Higher-order statistics contain additional information on the non-Gaussian galaxy field beyond that encoded in the power spectrum. We therefore implement in the new \folps\ model the galaxy bispectrum in the Sugiyama tripolar spherical harmonic basis, which we analyze with and without the LoS damping factor.

We validate the performance of the combined two- and three-point statistics using second-generation Abacus mock catalogs. For LRG mocks, the damped power spectrum can be reliably used up to $k \sim 0.35 \ihMpc$, improving the constraints relative to the standard \texttt{EFT} analysis limited to $k \sim 0.2 \ihMpc$. This improvement appears both in constraining power and in the reduction of parameter offsets from the true values, quantified through the FoM and FoB, respectively. In contrast, although damping tightens constraints for QSO-like mocks, the posterior means shift from the true simulation values. This occurs both for conservative scale cuts and when extending the power spectrum to $0.301 \ihMpc$, indicating that the shifts are caused by the damping model rather than by the adopted scale cuts.

For the bispectrum, we begin with conservative cuts that include the monopole up to $k_{\max}=0.12 \ihMpc$ for all DESI-like targets, and the quadrupole up to $k=0.08 \ihMpc$ for models with damping (\texttt{FolpsD}), while restricting it to $k=0.03 \ihMpc$ in the standard \texttt{EFT} case. Motivated by the strong correlation between the monopoles of the power spectrum and bispectrum, which increases from negligible values at very large scales to $\gtrsim0.5$ at $k\sim0.10 \ihMpc$, we explore extending the bispectrum range. For LRG2 mocks with damping, the bispectrum monopole can be robustly included up to $k_{\max}=0.24 \ihMpc$, leading to a $14\%$ increase in FoM and a $30\%$ reduction in FoB relative to the conservative $k_{\max}=0.12 \ihMpc$ case. Similar results hold for LRG1. Importantly, even without damping, the EFT model allows the bispectrum monopole to reach $k_{\max}=0.20 \ihMpc$ for LRGs and $k_{\max}=0.16 \ihMpc$ for QSOs. This shows that the improvement is not driven solely by the damping factor but also by the strong cross-correlation between power spectrum and Sugiyama bispectrum multipoles. The damping can nevertheless help by constraining second-order bias parameters through small-scale power spectrum information.

In summary, our results highlight the importance of joint power spectrum and bispectrum analyses for extracting cosmological information from DESI-like data. Even without phenomenological damping, the bispectrum significantly increases the constraining power of full-shape analyses and reduces parameter degeneracies. The damping prescription can further improve performance in high signal-to-noise regimes and for models in which the shape of the power spectrum does not degenerate with the damping, such as dark energy scenarios. As shown in \cref{sec:w0wa}, for the $w_0w_a$CDM model using the ShapeFit procedure we obtain $15\%$ and $21\%$ tighter constraints on $w_0$ and $w_a$, respectively, compared to the \texttt{EFT} full-shape power spectrum analysis. Furthermore, for the first time in a pure full-shape analysis, these results show a deviation from constant dark energy at slightly more than the $1\sigma$ level.

\section{Data Availability}

Data from the plots in this paper are available on Zenodo as part of DESI's Data Management
Plan (\url{https://doi.org/10.5281/zenodo.19302052}). The DESI data used in this work is public along the Data Release 1 (details in \url{https://data.desi.lbl.gov/doc/releases/dr1/}).

\acknowledgments

This work has passed internal review by the DESI collaboration.
The authors acknowledge Martin White and Shi-Fan Chen for useful discussions. 

PB acknowledges support from the Department of Energy under contract DE‐SC0019193. AA, HN, DG and IG acknowledge financial support from SECIHTI grant CBF2023-2024-162 and DGAPA-PAPIIT IA101825. AA also acknowledges  DGAPA-PAPIIT IG102123. GN, AA and DG acknowledge financial support from SECIHTI grant CBF-2025-I-2795. GN and DG acknowledge support from DAIP-UG, as well as access to computational resources provided by the DCI-UG DataLab.

This material is based upon work supported by the U.S. Department of Energy (DOE), Office of Science, Office of High-Energy Physics, under Contract No. DE–AC02–05CH11231, and by the National Energy Research Scientific Computing Center, a DOE Office of Science User Facility under the same contract. Additional support for DESI was provided by the U.S. National Science Foundation (NSF), Division of Astronomical Sciences under Contract No. AST-0950945 to the NSF’s National Optical-Infrared Astronomy Research Laboratory; the Science and Technology Facilities Council of the United Kingdom; the Gordon and Betty Moore Foundation; the Heising-Simons Foundation; the French Alternative Energies and Atomic Energy Commission (CEA); the National Council of Humanities, Science and Technology of Mexico (CONAHCYT); the Ministry of Science, Innovation and Universities of Spain (MICIU/AEI/10.13039/501100011033), and by the DESI Member Institutions: \url{https://www.desi.lbl.gov/collaborating-institutions}. Any opinions, findings, and conclusions or recommendations expressed in this material are those of the author(s) and do not necessarily reflect the views of the U. S. National Science Foundation, the U. S. Department of Energy, or any of the listed funding agencies.

The authors are honored to be permitted to conduct scientific research on I'oligam Du'ag (Kitt Peak), a mountain with particular significance to the Tohono O’odham Nation.

\appendix

\section{Influence of the quadrupole bispectrum} \label{app:B202}

In this appendix we explore the impact of the bispectrum quadrupole on the LRG2 mocks. We fit the data using the $\ell = 0,2$ multipoles of the power spectrum over the range $0 < k < 0.301 \ihMpc$ and the bispectrum monopole $B_{000}$ over the range $0 < k < 0.12 \ihMpc$, using the \texttt{FolpsD} model. For mocks, we have already shown that including the bispectrum quadrupole without damping (\texttt{EFT-Pk+Bk}) can bias the recovered parameters, and we therefore restrict its use to $k_{\max} = 0.03 \ihMpc$ in that case.

For the bispectrum quadrupole $B_{202}$, we consider three cases: \emph{i)} not including it; \emph{ii)} including it up to $k_{\max} = 0.03 \ihMpc$, which corresponds to only two data points; and \emph{iii)} including it up to $k_{\max} = 0.08 \ihMpc$, which is the baseline choice used in \cref{sec:mockresults}.

\begin{figure*}
	\begin{center}
	\includegraphics[width=3.3 in]{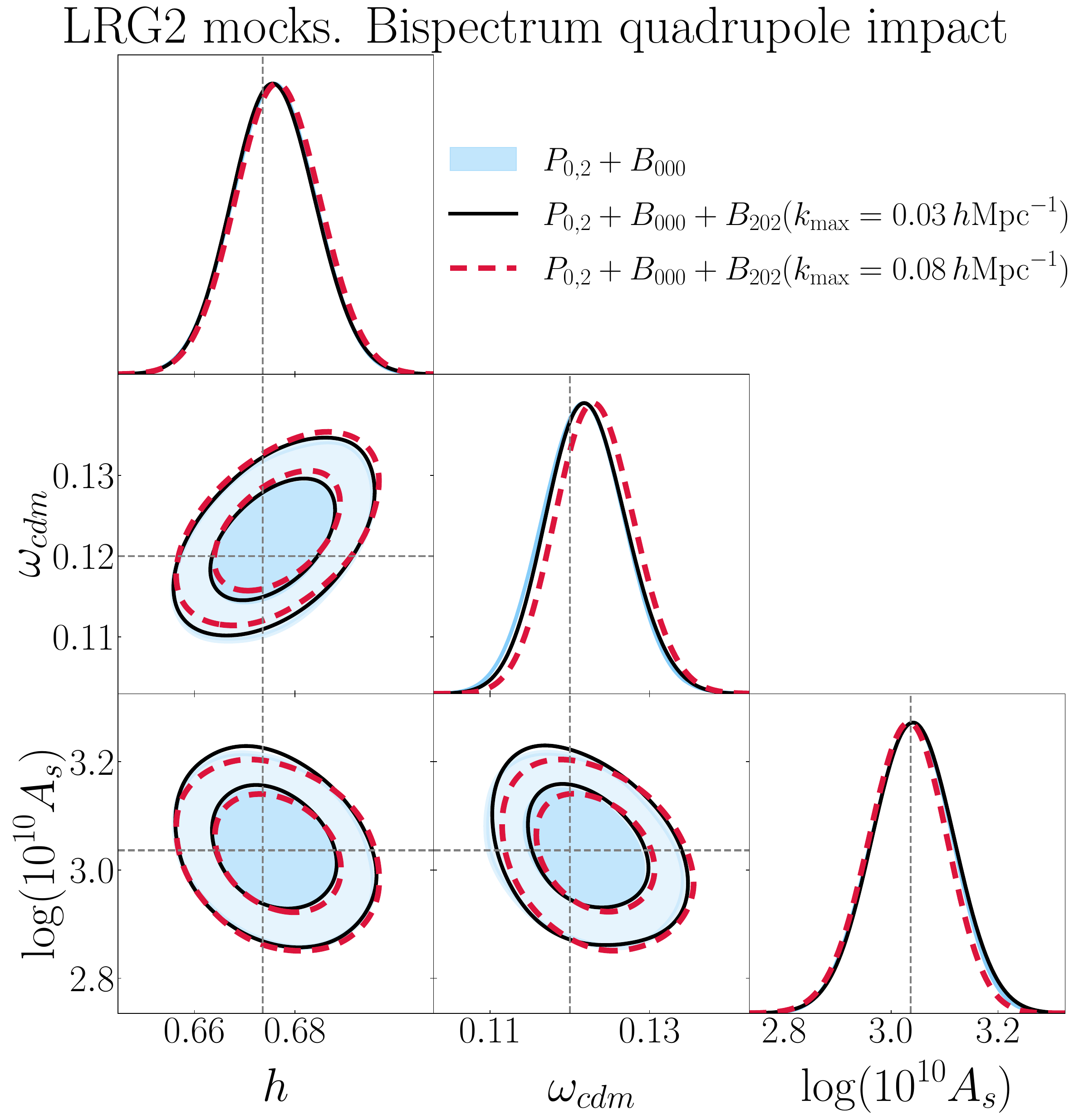}
	\includegraphics[width=2.7 in]{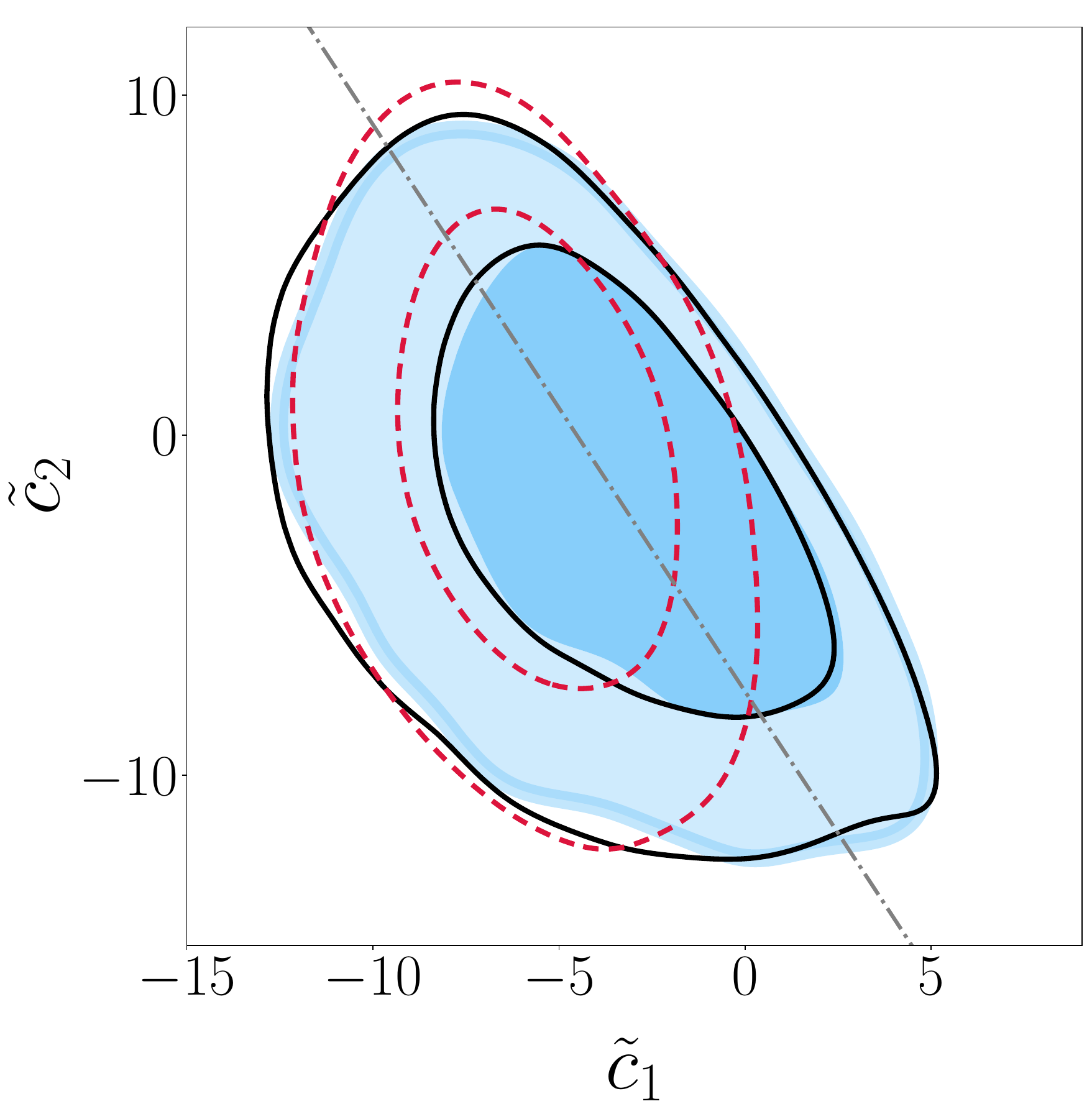}
    \caption{Impact of including the bispectrum quadrupole on parameter constraints for the LRG2 mocks. 
\textit{Left Panel:} Marginalized posterior distributions for the cosmological parameters 
$h,\ \omega_{\rm cdm},\ \text{and}\, \log(10^{10}A_s)$. 
\textit{Right Panel:} Constraints on the NLO counterterms
$(\tilde{c}_1,\ \tilde{c}_2)$. The dashed line indicates the degeneracy between the NLO counterterms
expected when fitting the monopole alone.}\label{fig:B202influence}
	\end{center}
\end{figure*}

\Cref{fig:B202influence} summarizes our results. The corner plot in the left panel shows that the differences between the posteriors are small. In terms of constraining power, we find a modest improvement in the FoM of $3.8\%$ when going from not including the bispectrum quadrupole (case \emph{i}) to the baseline choice (case \emph{iii}). The FoB increases from 0.39 to 0.67 between these two cases, corresponding to a $71\%$ increase. The only parameter showing a noticeable change is $\omega_{cdm}$, whose posterior shifts toward slightly higher values. For case \emph{iii}, the true simulation value lies $0.66\sigma$ from the posterior mean, compared to $0.34\sigma$ in case \emph{i}. The posteriors of $A_s$ and $h$ are essentially unchanged.

In the right panel of \cref{fig:B202influence} we show the two-dimensional posteriors of the NLO counterterms. We also plot the straight line $c_1/3 + c_2/5 = \mathrm{constant}$, which approximates the degeneracy direction imposed by the bispectrum monopole alone. For cases \emph{i} and \emph{ii}, the contours closely follow this degeneracy. In contrast, for case \emph{iii}, where the bispectrum quadrupole has a larger impact, the degeneracy direction becomes mildly tilted.

We conclude that, although the bispectrum quadrupole has a measurable effect on the posteriors, its overall impact is small. In the presence of damping models, this suggests that the parameter $c_2$ is redundant even when including the bispectrum quadrupole.


\section{Extra constraining power from second order biases}\label{app:b2bsinfo}

In this appendix we explore the additional constraining power provided by the damping factor and by the bispectrum. To do this, we construct the probability
\begin{equation}
P(\tilde{b}_2,\tilde{b}_s) = \exp \left( -\Delta \mathbf{b}^\text{T} \,\text{C}^{-1}_{b_2,b_s} \Delta \mathbf{b}\right),
\end{equation}
with the vector difference 
\begin{equation} \label{filter}
\Delta \mathbf{b} = (\tilde{b}_2,\tilde{b}_s) - (\tilde{b}_{2, {\rm mean}}^\text{\texttt{model}},\tilde{b}_{s, {\rm mean}}^\text{\texttt{model}}).
\end{equation}

The ``mean'' labels correspond to the best-fit posterior values from the analysis with a given model (either \texttt{FolpsD-Pk} or \texttt{EFT-Pk+Bk}), and $\text{C}_{b_2,b_s}$ denotes their covariance.

\begin{figure*}
	\begin{center}
	\includegraphics[width=3 in]{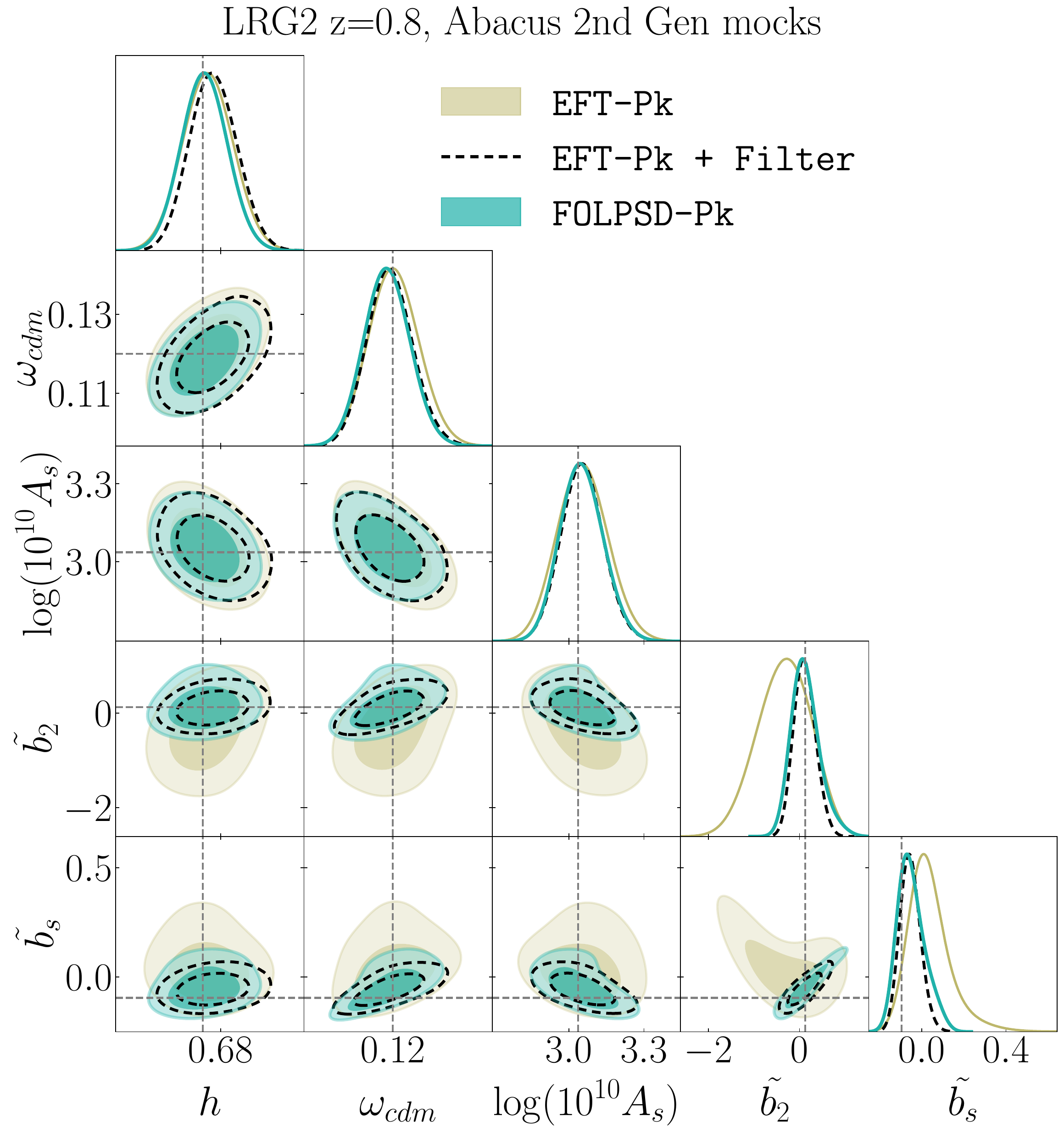}
	\includegraphics[width=3 in]{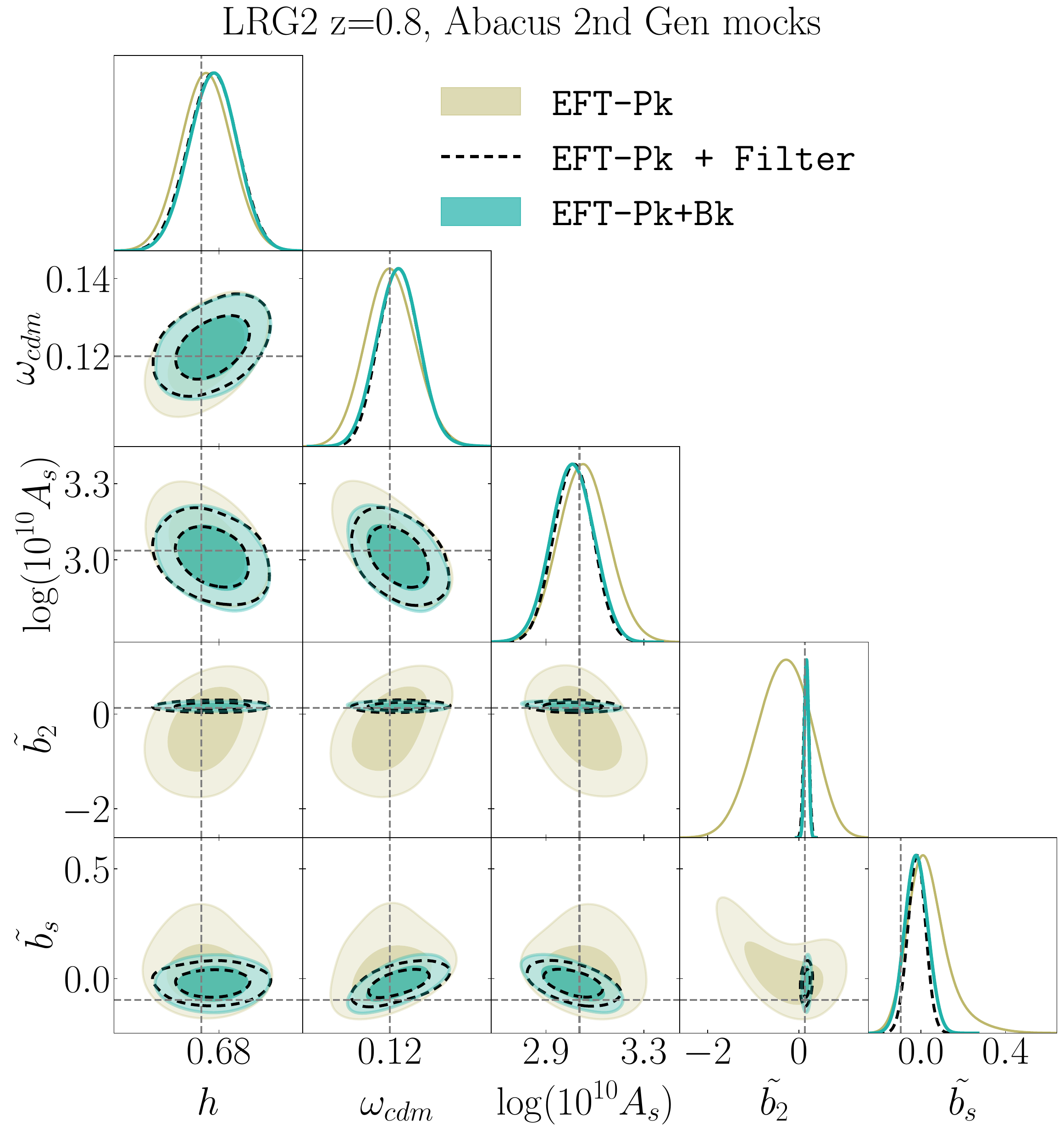}
    \caption{Filtering of the \texttt{EFT-Pk} chains from the LRG2 analysis of \cref{sec:mockresults} 
using the two-dimensional posterior distribution in the $(\tilde{b}_2,\tilde{b}_s)$ parameter space. 
\textit{Left panel}: filtering based on the \texttt{FolpsD-Pk} model. 
\textit{Right panel}: filtering based on the \texttt{EFT-Pk+Bk} model.
    }\label{fig:filter}
	\end{center}
\end{figure*}

We proceed to filter the chains of the \texttt{EFT-Pk}, LRG2-mock fit obtained in \cref{sec:mockresults} by retaining each step based on its $\tilde{b}_2$ and $\tilde{b}_s$ values, with a probability given by \cref{filter}. In the left panel of \cref{fig:filter}, we show the results of this analysis. The posterior distributions from \texttt{EFT-Pk} are shown with light-green contours, while the \texttt{FolpsD-Pk} ones are depicted in light green. The filtered chains, shown with dashed black lines, match the \texttt{FolpsD-Pk} results very well, supporting our claim that the additional constraining power in the cosmological parameters arises mainly from a better fit to $\tilde{b}_2$ and $\tilde{b}_s$, which then translates into a tighter constraint on $b_1$ and ultimately on $A_s$. 
Hence, one expects that the ability to improve constraints with \texttt{FolpsD} would depend on the degeneracies between the linear bias and the second-order local and tidal biases, which are different for different tracers. On the other hand, by examining \cref{fig:Pk_kmax}, the additional constraining power on $\tilde{b}_2$ and $\tilde{b}_s$ arises from extending the analysis to higher wavenumbers.  Therefore, we conclude that for the conservative scale cuts considered in \cref{sec:mockresults}, \texttt{FolpsD} actually obtains its extra information on cosmological parameters mainly from small scales, although indirectly, through improved constraints on the nuisance parameters.

We now perform the filtering of the \texttt{EFT-Pk} chains using the posterior distributions from the \texttt{EFT-Pk+Bk} analysis, as we did for \texttt{FolpsD-Pk}. For this, we replace the mean posterior values and covariance obtained from \texttt{FolpsD-Pk} with those from \texttt{EFT-Pk+Bk} in \cref{filter}. The resulting contours are shown in the right panel of \cref{fig:filter}, strongly suggesting that the additional constraining power on the cosmological parameters when including the bispectrum arises indirectly through a better fit to the bias parameters $b_2$ and $b_s$.

\section{Neutrino masses}\label{app:neutrinos}

\begin{figure*}
	\begin{center}
	\includegraphics[width=4 in]{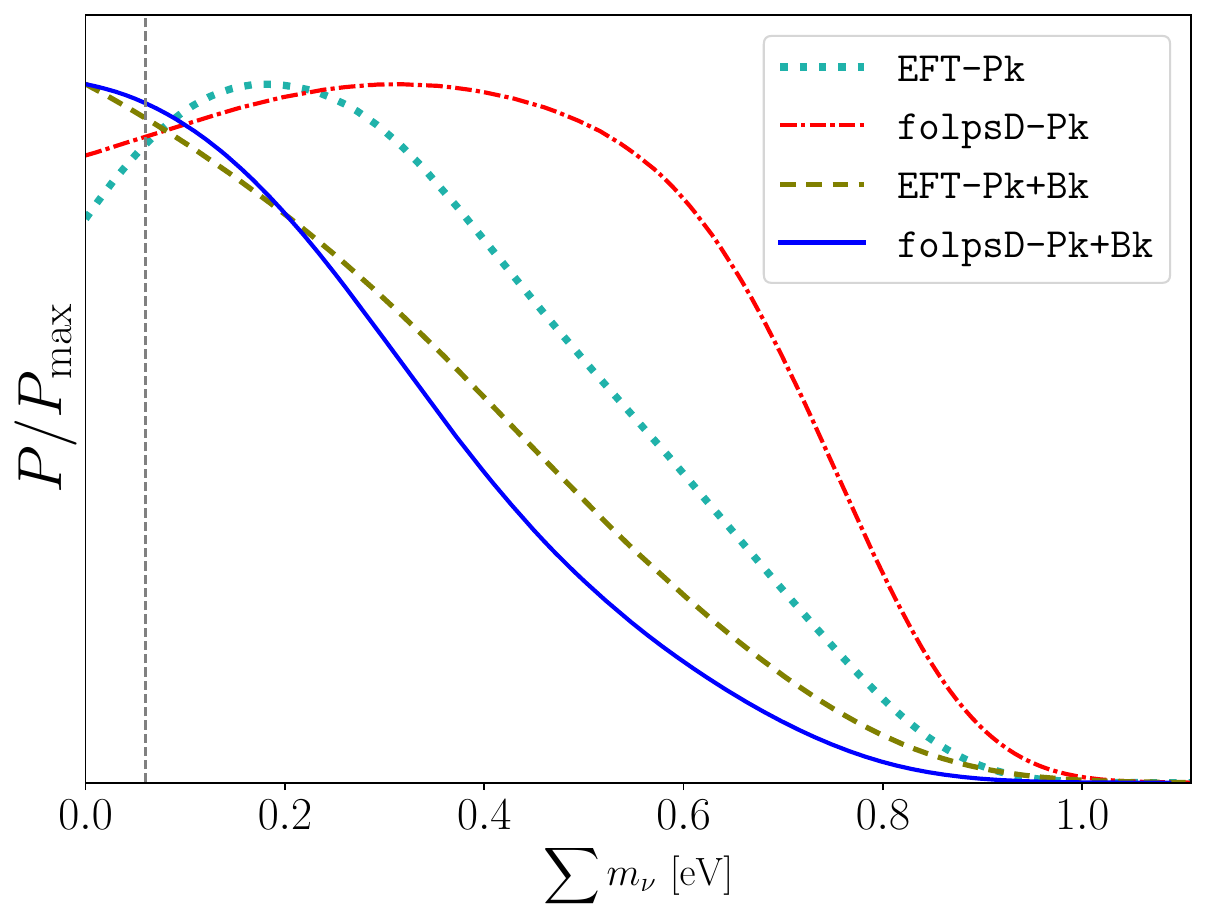}
    \caption{Constraints to the neutrino mass from the galaxy power spectrum of LRG2 Abacus second-generation mocks.}\label{fig:neutrinomasses}
	\end{center}
\end{figure*}

\begin{figure*}
	\begin{center}
	\includegraphics[width=4 in]{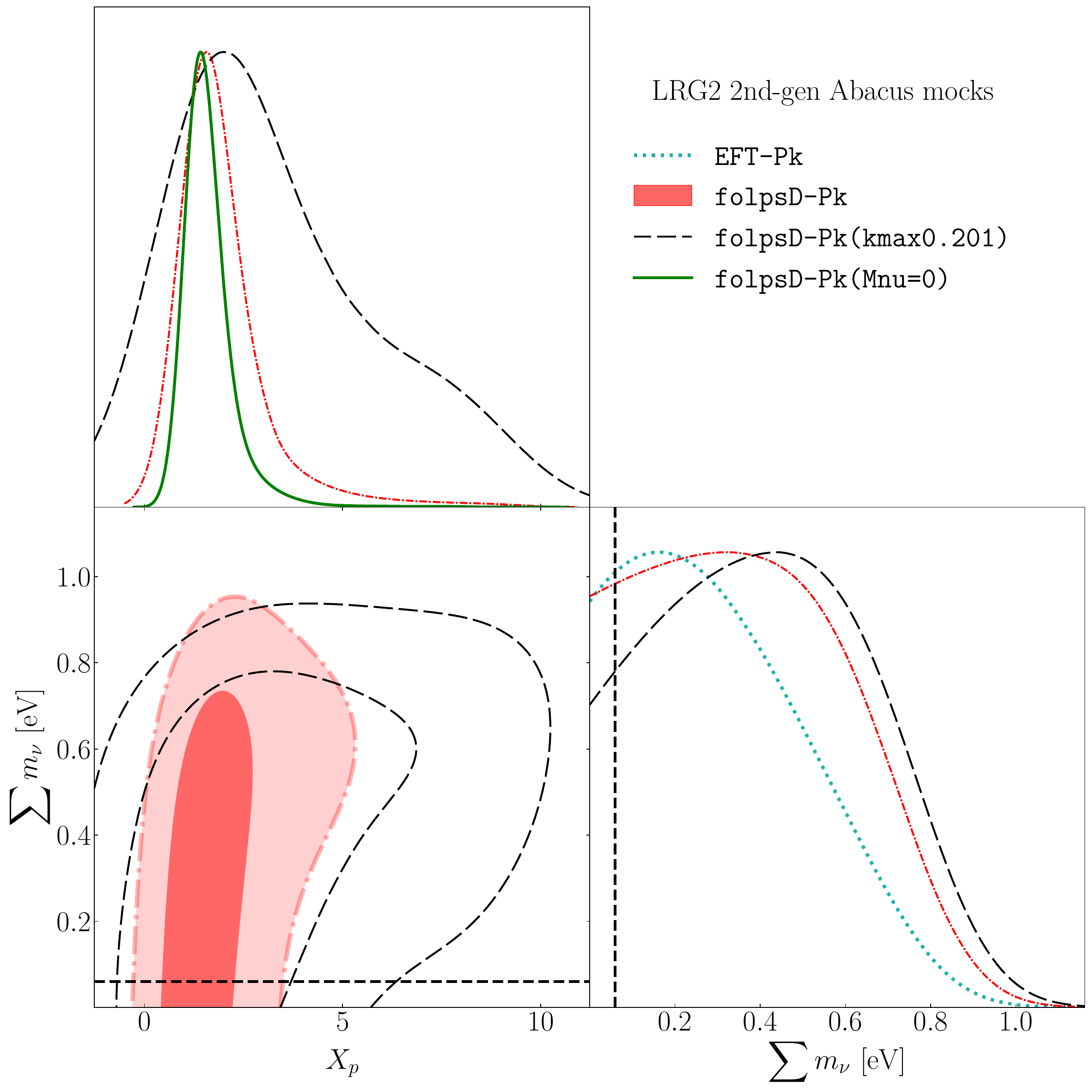}
    \caption{Degeneracies between the neutrino mass and the power-spectrum damping parameter $X_p$ for \texttt{FolpsD-Pk}, comparing the baseline scale cut ($k_{\max} = 0.301 \ihMpc$) with a more conservative choice of $k_{\max} = 0.201 \ihMpc$. For comparison, we also show the constraint on $X_p$ for the baseline model with fixed $\sum m_\nu = 0.06\,\mathrm{eV}$.}\label{fig:neutrinomasses_tri}
	\end{center}
\end{figure*}

Recent DESI-DR1 FS and DESI-DR2 BAO analyses \cite{DESI.DR2.BAO.cosmo,Y3.cpe-s2.Elbers.2025} have put very tight constraints on the sum of the neutrino masses. They have become even tighter with the addition of recent data from the Data Release 6 of the Atacama Cosmology Telescope (ACT) \cite{2025JCAP...11..062L,2025JCAP...11..063C,2025JCAP...11..061N} (see also \cite{2025PhRvD.112h3529G}) and of the South Pole Telescope E-mode polarization analysis \cite{2025arXiv250620707C}, that have reached $\sum m_\nu < 0.048$ eV at 95\% level in the context of $\Lambda$CDM model. This bound is smaller than the minimum masses allowed by observations of neutrino flavor oscillations \cite{Esteban:2018azc}, that impose a bound of $\sum m_\nu > 0.06$ eV at 90\% level. When added the freedom of an evolving dark energy background evolution, the constrain loosen considerably, but their posteriors still peak at zero mass when supernovae data are combined with CMB and BAO+FS data, and remain dominated by the $\sum m_\nu > 0$ prior \cite{Y3.cpe-s2.Elbers.2025}.

Originally, the code \folps\ was developed to constrain neutrino masses. So, it is built on the fkpt theory for kernels beyond EdS \cite{Aviles:2020cax,Aviles:2021que,Rodriguez-Meza:2023rga}. In reference \cite{KP5s3-Noriega}, this method has shown to provide constraints around $14\%$ tighter than the standard EdS using Abacus first-generation mocks.

\subsection{Neutrino masses from LRG2 mocks}

We first show results from fitting the neutrino masses to the LRG2 Abacus second-generation mocks. We vary the neutrino mass with only one massive state ($N_\text{ur}= 2.0328$) over the interval $[0,5]$ in eV units, together with the cosmological and nuisance parameters we used in \cref{sec:mockresults} and shown in \cref{table:parameters_priors}.

The one-dimensional marginalized posterior distributions are shown in \cref{fig:neutrinomasses}. The 95\% c.l. constraints are:
\begin{equation}
\sum m_\nu  \, < \quad
\left\{
\begin{array}{cl}
\quad 0.705 \,\,\text{eV} & \quad(\texttt{EFT-Pk})           \\[4pt]
\quad 0.768 \,\,\text{eV}& \quad(\texttt{FolpsD-Pk}  )      \\[4pt]
\quad 0.675 \,\,\text{eV}& \quad(\texttt{EFT-Pk+Bk} )       \\[4pt]
\quad 0.610 \,\,\text{eV}& \quad(\texttt{FolpsD-Pk+Bk})
\end{array}
\right.
\end{equation}
Among the four models considered, \texttt{FolpsD-Pk+Bk} provides the most stringent constraints. However, the expected hierarchy between models is not preserved, as \texttt{FolpsD-Pk} yields weaker constraints than \texttt{EFT-Pk}. This behavior is driven by additional parameter degeneracies involving the damping parameter $X_p$. As shown in \cite{Noriega:2024lzo}, most of the neutrino information in full-shape analyses comes from the suppression of the BAO wiggles rather than from the broadband power suppression. Since the phenomenological damping applied to the power spectrum also suppresses these oscillations, it naturally introduces strong degeneracies with $X_p$, thereby reducing the constraining power on the neutrino mass.

\Cref{fig:neutrinomasses_tri} illustrates the degeneracy between $X_p$ and $\sum m_\nu$ for \texttt{FolpsD-Pk} using the baseline scale cut ($k_{\max} = 0.301 \ihMpc$). We also show the case of \texttt{FolpsD-Pk} using the same scale cut adopted for the EFT model ($k_{\max} = 0.201 \ihMpc$). In the latter case, the constraints on the neutrino mass are even weaker than for the baseline \texttt{FolpsD-Pk} setup, driven an enhanced degeneracy between $X_p$ and $\sum m_\nu$. The neutrino mass constraint for this case is
\begin{equation}
  \sum m_\nu<  0.813 \,\,\text{eV} \quad(\texttt{FolpsD-Pk(kmax0201)}  ).
\end{equation}

For comparison, we include in \cref{fig:neutrinomasses_tri} results from the \texttt{EFT-Pk} model, as well as the constraint on $X_p$ obtained from the \texttt{FolpsD-Pk} baseline analysis with fixed $\sum m_\nu = 0.06\,\text{eV}$. In this latter case, $X_p$ is better constrained, confirming that the loss of neutrino sensitivity in the damping-based model arises from its degeneracy with the neutrino mass. Overall, this analysis shows that neutrino mass constraints obtained using damping models can be degraded relative to standard EFT due to degeneracies between the damping parameters and $\sum m_\nu$.
The inclusion of the bispectrum, however, may be partially breaking these degeneracies. 

\begin{figure}
	\begin{center}
	\includegraphics[width=4.2 in]{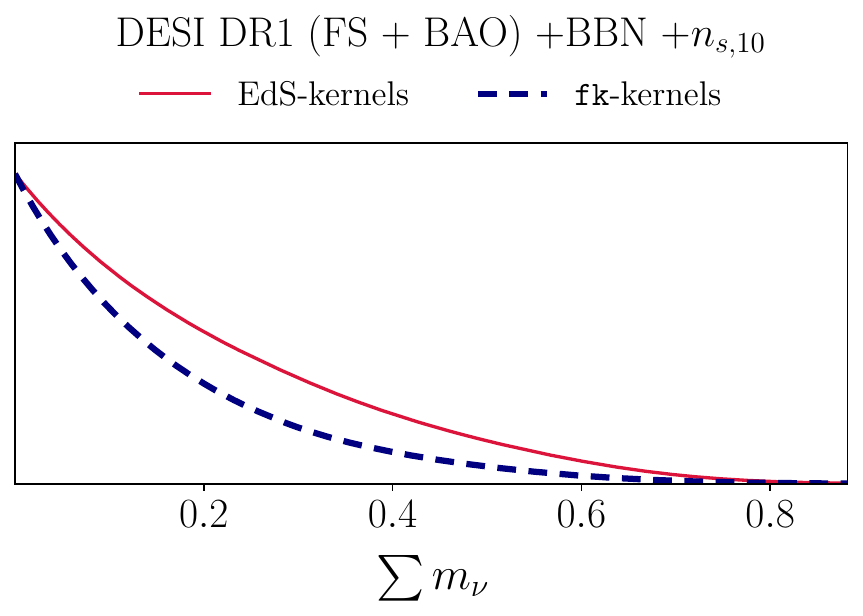} \\[10pt]
	\includegraphics[width=4.2 in]{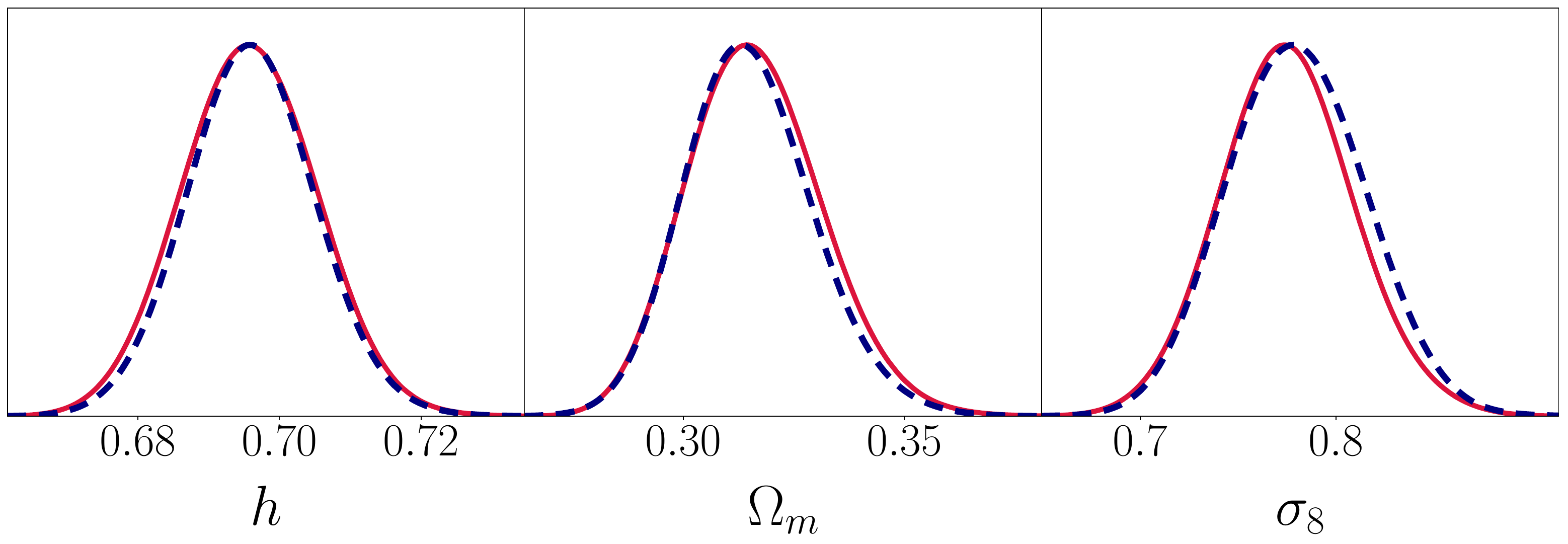}
    \caption{Constraints on the sum of neutrino masses using EFT theory with EdS and \code{fk} kernels. The latter are 15\% more constrictive than EdS. }\label{fig:neutrino_kernels}
	\end{center}
\end{figure}

\subsection{Neutrino masses from DESI DR1 full-shape data: fk vs EdS kernels}\label{app:subneutrinos}

Although for small neutrino masses ($\sim 0.1\,\text{eV}$), EdS kernels are expected to provide very accurate results, deviations can arise for larger neutrino masses. These differences can become relevant when estimating the errors on the neutrino mass inferred from the LSS DESI data, which obtain the neutrino mass signal mainly from the suppression of the power spectrum alone.

We also fit DESI DR1 data using the \texttt{EFT} modeling, including both BAO and full-shape measurements. We impose the BBN prior on $\omega_b$ and the $n_s$ prior on the primordial spectral index, as listed in \cref{table:parameters_priors}. The top panel of \cref{fig:neutrino_kernels} shows the one-dimensional marginalized posterior distribution of  the sum of neutrino masses.  Theire 95\% c.l. constraints are

\begin{align}
\text{DESI(BAO+FS) + BBN + $n_{s10}$}\;: \nonumber\\[6pt]
\text{EdS kernels}: \quad  \sum m_\nu &< 0.548 \,\text{eV}  \quad (95\% \,\,\text{c.l.}), \\[3pt]
\text{fk kernels}: \quad  \sum m_\nu &< 0.464 \,\text{eV}  \quad (95\% \,\,\text{c.l.}). \\[-2pt] &\nonumber
\end{align}
The improve of constraining power is 15\% when going from EdS to \code{fk} kernels, which is consistent with the 14\% improvement obtained for mocks in \cite{KP5s3-Noriega}.
The bottom panel of \cref{fig:neutrino_kernels} shows that the rest of the varied cosmological parameters are not affected by the choice of the kernels. The reason is that the growth factor, $f(k)$, is sensitive to the neutrino mass, and not to the rest of cosmological parameters.  

In this analysis we are only interested in the difference between kernels \code{fk} and EdS. Hence, we have used a different parameter basis and priors than in the rest of this work. We fitted directly the EFT counterterms ($\alpha_0$ and $\alpha_2$) and the shot-noise parameters ($\alpha_0^\text{shot}$ and $\alpha_2^\text{shot}$) defined in \cref{PLOctr,Psn}. We also used the bias parameters $b_1$, $b_2$, $b_{s^2}$, and $b_{3nl}$ from \cite{McDonald:2009dh}. We adopted uninformative priors for all nuisance parameters.

\section{$Z_{1,2}$ kernels beyond EdS}\label{app:Zkernels}

The code \folps\ uses the approximation fkpt to compute kernels beyond EdS. These has been computed in different works \cite{Rodriguez-Meza:2023rga,Aviles:2020wme,Aviles:2021que}, for $F_n$ and $G_n$ kernels. To do so for $Z_n$ kernels is straightforward and is presented in this appendix. 

We start with the relation between density fields in redshift, $\delta_s$, and real spaces, $\delta$:

\begin{align}
 &(2\pi)^3 \dD(\vk) + \delta_s(\vk) = \int d^3x \big[1+\delta(\vx)\big] e^{- i\vk \cdot (\vx + \vu(\vx) )},
 \end{align}
where $\vu(\vx) = \vhn (\vec v(\vx)\cdot \vhn/a H)$ is the peculiar velocity along the LoS in Hubble constant units.
Expanding the exponential and performing the Fourier integral when possible
\begin{align} \label{dsdk}
\delta_s(\vk) &= \delta(\vk) - i k_i u_i(\vk) - i k_i \int d^3x \, e^{- i\vk \cdot \vx }u_i(\vx)\delta(\vx) 
-\frac{1}{2} k_i k_j\int d^3x \, e^{- i\vk \cdot \vx }u_i(\vx)u_j(\vx).
\end{align}
In the presence of massive neutrinos, or in some modified gravity theories, the linear growth is scale dependent, hence $f=f(k,t)$. Thus, we define the $\theta$ field as
\begin{equation}
    \theta(\vk) = - \frac{i k_i v_i(\vk)}{a H f_0}, 
\end{equation}
where $f_0 \equiv f(k=0,t)$. At linear order, velocity a density field are related as
\begin{equation}
    \theta^{(1)}(\vk,t) = \frac{f(k,t)}{f_0(t)}\delta^{(1)}(\vk,t),
\end{equation}
and the linear velocity kernel is $G_1(\vk) = f(k)/f_0$. 
Since we are focusing only in non-rotational fields, 
\begin{equation}
u_i(\vk) 
= i f_0 \frac{\mu}{k} \theta(\vk) \,\hat{n}_i. 
\end{equation}
Equation \eqref{dsdk} then reads\footnote{We use the shorthand notation 
\begin{equation}
    \ikk = \int \Dk{k_1}\Dk{k_2} \dD(\vk_1+\vk_2-\vk).
\end{equation}
}
\begin{align} \label{dsdr}
\delta_s(\vk) =& \delta(\vk) + f_0 \mu^2  \theta(\vk)  
+  \ikk k \mu \frac{\mu_1}{k_1} \theta(\vk_1)\delta(\vk_2) +  \ikk  \frac{k^2\mu^2}{2} \frac{\mu_1}{k_1} \frac{\mu_2}{k_2} \,\theta(\vk_1)\theta(\vk_2)
\end{align}
To linear order, one obtains
\begin{align} \label{dsdr_1}
\delta_s^{(1)}(\vk) &=(1+ f(k) \mu^2 )  \delta(\vk)  \equiv Z_1(k)\delta(\vk)
\end{align}
To second order, one finds
\begin{align} \label{dsdr_2}
\delta_s^{(2)}(\vk) &= \ikk Z_2(\vk_1,\vk_2)\delta^{(1)}(\vk_1)\delta^{(1)}(\vk_2),
\end{align}
with 
\begin{align}
 Z_2(\vk_1,\vk_2) =& F_2(\vk_1,\vk_2) + f_0 \mu^2 G_2(\vk_1,\vk_2)  \nonumber\\
 &+ \frac{k\mu}{2} \left( \frac{\mu_1}{k_1} f(k_1)+ \frac{\mu_2}{k_2} f(k_2)\right)  + \frac{k^2\mu^2}{2} \frac{\mu_1}{k_1} \frac{\mu_2}{k_2}f(k_1)f(k_2),
\end{align}
where $\vk=\vk_1+\vk_2$ and $k\mu=k_1\mu_1 + k_2\mu_2$. Using this last equality, we can rewrite the $Z_2$ kernel in the most common form
\begin{align}
 Z_2(\vk_1,\vk_2) =& F_2(\vk_1,\vk_2) + f_0 \mu^2 G_2(\vk_1,\vk_2)  \nonumber\\
 &+ \frac{k\mu}{2} \Big[ \frac{\mu_1}{k_1} f(k_1) \big(1+f(k_2) \mu_2^2 \big) + \frac{\mu_2}{k_2} f(k_2) \big(1+f(k_1) \mu_1^2\big) \Big]. 
\end{align}
The reader should notice that the bispectrum beyond EdS kernels has also been studied in the past (e.~g.~\cite{Aviles:2023fqx,Bose:2019wuz,Pal:2025hpl,Pal:2025zep}).

\bibliographystyle{JHEP}
\bibliography{references, DESI_supporting_papers}


\section{Author Affiliations}
\label{sec:affiliations}

\noindent \hangindent=.5cm $^{1}${Leinweber Center for Theoretical Physics, University of Michigan, 450 Church Street, Ann Arbor, Michigan 48109-1040, USA}

\noindent \hangindent=.5cm $^{2}${University of Michigan, 500 S. State Street, Ann Arbor, MI 48109, USA}

\noindent \hangindent=.5cm $^{3}${Instituto de Ciencias F\'{\i}sicas, Universidad Nacional Aut\'onoma de M\'exico, Av. Universidad s/n, Cuernavaca, Morelos, C.~P.~62210, M\'exico}

\noindent \hangindent=.5cm $^{4}${Instituto Avanzado de Cosmolog\'{\i}a A.~C., San Marcos 11 - Atenas 202. Magdalena Contreras. Ciudad de M\'{e}xico C.~P.~10720, M\'{e}xico}

\noindent \hangindent=.5cm $^{5}${Institute for Astronomy, University of Edinburgh, Royal Observatory, Blackford Hill, Edinburgh EH9 3HJ, UK}

\noindent \hangindent=.5cm $^{6}${Departamento de F\'{\i}sica, DCI-Campus Le\'{o}n, Universidad de Guanajuato, Loma del Bosque 103, Le\'{o}n, Guanajuato C.~P.~37150, M\'{e}xico}

\noindent \hangindent=.5cm $^{7}${IRFU, CEA, Universit\'{e} Paris-Saclay, F-91191 Gif-sur-Yvette, France}

\noindent \hangindent=.5cm $^{8}${Kavli Institute for Cosmology, University of Cambridge, Madingley Road, Cambridge CB3 0HA, UK}

\noindent \hangindent=.5cm $^{9}${Max Planck Institute for Extraterrestrial Physics, Gie\ss enbachstra\ss e 1, 85748 Garching, Germany}

\noindent \hangindent=.5cm $^{10}${University Observatory, Faculty of Physics, Ludwig-Maximilians-Universit\"{a}t, Scheinerstr. 1, 81677 M\"{u}nchen, Germany}

\noindent \hangindent=.5cm $^{11}${Department of Physics, University of Michigan, 450 Church Street, Ann Arbor, MI 48109, USA}

\noindent \hangindent=.5cm $^{12}${Instituto de Estudios Astrof\'isicos, Facultad de Ingenier\'ia y Ciencias, Universidad Diego Portales, Av. Ej\'ercito Libertador 441, Santiago, Chile}

\noindent \hangindent=.5cm $^{13}${Steward Observatory, University of Arizona, 933 N. Cherry Avenue, Tucson, AZ 85721, USA}

\noindent \hangindent=.5cm $^{14}${Lawrence Berkeley National Laboratory, 1 Cyclotron Road, Berkeley, CA 94720, USA}

\noindent \hangindent=.5cm $^{15}${Department of Physics, Boston University, 590 Commonwealth Avenue, Boston, MA 02215 USA}

\noindent \hangindent=.5cm $^{16}${Dipartimento di Fisica ``Aldo Pontremoli'', Universit\`a degli Studi di Milano, Via Celoria 16, I-20133 Milano, Italy}

\noindent \hangindent=.5cm $^{17}${INAF-Osservatorio Astronomico di Brera, Via Brera 28, 20122 Milano, Italy}

\noindent \hangindent=.5cm $^{18}${Department of Physics \& Astronomy, University College London, Gower Street, London, WC1E 6BT, UK}

\noindent \hangindent=.5cm $^{19}${Instituto de F\'{\i}sica, Universidad Nacional Aut\'{o}noma de M\'{e}xico,  Circuito de la Investigaci\'{o}n Cient\'{\i}fica, Ciudad Universitaria, Cd. de M\'{e}xico  C.~P.~04510,  M\'{e}xico}

\noindent \hangindent=.5cm $^{20}${Department of Astronomy \& Astrophysics, University of Toronto, Toronto, ON M5S 3H4, Canada}

\noindent \hangindent=.5cm $^{21}${Department of Physics \& Astronomy and Pittsburgh Particle Physics, Astrophysics, and Cosmology Center (PITT PACC), University of Pittsburgh, 3941 O'Hara Street, Pittsburgh, PA 15260, USA}

\noindent \hangindent=.5cm $^{22}${University of California, Berkeley, 110 Sproul Hall \#5800 Berkeley, CA 94720, USA}

\noindent \hangindent=.5cm $^{23}${Instituci\'{o} Catalana de Recerca i Estudis Avan\c{c}ats, Passeig de Llu\'{\i}s Companys, 23, 08010 Barcelona, Spain}

\noindent \hangindent=.5cm $^{24}${Institut de F\'{i}sica d’Altes Energies (IFAE), The Barcelona Institute of Science and Technology, Edifici Cn, Campus UAB, 08193, Bellaterra (Barcelona), Spain}

\noindent \hangindent=.5cm $^{25}${Departamento de F\'isica, Universidad de los Andes, Cra. 1 No. 18A-10, Edificio Ip, CP 111711, Bogot\'a, Colombia}

\noindent \hangindent=.5cm $^{26}${Observatorio Astron\'omico, Universidad de los Andes, Cra. 1 No. 18A-10, Edificio H, CP 111711 Bogot\'a, Colombia}

\noindent \hangindent=.5cm $^{27}${Institut d'Estudis Espacials de Catalunya (IEEC), c/ Esteve Terradas 1, Edifici RDIT, Campus PMT-UPC, 08860 Castelldefels, Spain}

\noindent \hangindent=.5cm $^{28}${Institute of Cosmology and Gravitation, University of Portsmouth, Dennis Sciama Building, Portsmouth, PO1 3FX, UK}

\noindent \hangindent=.5cm $^{29}${Institute of Space Sciences, ICE-CSIC, Campus UAB, Carrer de Can Magrans s/n, 08913 Bellaterra, Barcelona, Spain}

\noindent \hangindent=.5cm $^{30}${University of Virginia, Department of Astronomy, Charlottesville, VA 22904, USA}

\noindent \hangindent=.5cm $^{31}${Fermi National Accelerator Laboratory, PO Box 500, Batavia, IL 60510, USA}

\noindent \hangindent=.5cm $^{32}${Department of Astronomy, University of Texas at Austin, 2515 Speedway, TX 78712, USA}

\noindent \hangindent=.5cm $^{33}${Institut d'Astrophysique de Paris. 98 bis boulevard Arago. 75014 Paris, France}

\noindent \hangindent=.5cm $^{34}${Center for Cosmology and AstroParticle Physics, The Ohio State University, 191 West Woodruff Avenue, Columbus, OH 43210, USA}

\noindent \hangindent=.5cm $^{35}${Department of Physics, The Ohio State University, 191 West Woodruff Avenue, Columbus, OH 43210, USA}

\noindent \hangindent=.5cm $^{36}${The Ohio State University, Columbus, 43210 OH, USA}

\noindent \hangindent=.5cm $^{37}${School of Mathematics and Physics, University of Queensland, Brisbane, QLD 4072, Australia}

\noindent \hangindent=.5cm $^{38}${Department of Physics, The University of Texas at Dallas, 800 W. Campbell Rd., Richardson, TX 75080, USA}

\noindent \hangindent=.5cm $^{39}${NSF NOIRLab, 950 N. Cherry Ave., Tucson, AZ 85719, USA}

\noindent \hangindent=.5cm $^{40}${Department of Physics and Astronomy, University of California, Irvine, 92697, USA}

\noindent \hangindent=.5cm $^{41}${Sorbonne Universit\'{e}, CNRS/IN2P3, Laboratoire de Physique Nucl\'{e}aire et de Hautes Energies (LPNHE), FR-75005 Paris, France}

\noindent \hangindent=.5cm $^{42}${Departament de F\'{i}sica, Serra H\'{u}nter, Universitat Aut\`{o}noma de Barcelona, 08193 Bellaterra (Barcelona), Spain}

\noindent \hangindent=.5cm $^{43}${Department of Physics and Astronomy, University of Waterloo, 200 University Ave W, Waterloo, ON N2L 3G1, Canada}

\noindent \hangindent=.5cm $^{44}${Perimeter Institute for Theoretical Physics, 31 Caroline St. North, Waterloo, ON N2L 2Y5, Canada}

\noindent \hangindent=.5cm $^{45}${Waterloo Centre for Astrophysics, University of Waterloo, 200 University Ave W, Waterloo, ON N2L 3G1, Canada}

\noindent \hangindent=.5cm $^{46}${Instituto de Astrof\'{i}sica de Andaluc\'{i}a (CSIC), Glorieta de la Astronom\'{i}a, s/n, E-18008 Granada, Spain}

\noindent \hangindent=.5cm $^{47}${Departament de F\'isica, EEBE, Universitat Polit\`ecnica de Catalunya, c/Eduard Maristany 10, 08930 Barcelona, Spain}

\noindent \hangindent=.5cm $^{48}${Department of Physics and Astronomy, Sejong University, 209 Neungdong-ro, Gwangjin-gu, Seoul 05006, Republic of Korea}

\noindent \hangindent=.5cm $^{49}${Abastumani Astrophysical Observatory, Tbilisi, GE-0179, Georgia}

\noindent \hangindent=.5cm $^{50}${Department of Physics, Kansas State University, 116 Cardwell Hall, Manhattan, KS 66506, USA}

\noindent \hangindent=.5cm $^{51}${Faculty of Natural Sciences and Medicine, Ilia State University, 0194 Tbilisi, Georgia}

\noindent \hangindent=.5cm $^{52}${CIEMAT, Avenida Complutense 40, E-28040 Madrid, Spain}

\noindent \hangindent=.5cm $^{53}${Department of Physics \& Astronomy, Ohio University, 139 University Terrace, Athens, OH 45701, USA}

\end{document}